\documentclass[twoside,12pt]{elsarticle}

\usepackage{amssymb} 
\usepackage{amsmath} 
\usepackage[utf8]{inputenc} 
\usepackage{booktabs}
\usepackage{siunitx}
\usepackage{xspace}
\usepackage{acronym}
\usepackage{todonotes}
\usepackage{dcolumn}
\usepackage{afterpage}

\def\Journal#1#2#3#4{{#1} {#2} (#4) #3\xspace}

\def\PRO{{\em Prog.~Theor.~Phys.}}
\def\NPB{{\em Nucl.~Phys.} B}

\def\PLB{{\em Phys.~Lett.} B}

\def\PL{{\em Phys.~Lett.}}
\def\PRL{{\em Phys.~Rev.~Lett.}}
\def\PREV{{\em Phys.~Rev.}}
\def\PREP{{\em Phys.~Rep.}}

\def\PRD{{\em Phys.~Rev.} D}

\def\PRB{{\em Phys.~Rev.} B}
\def\PR{{\em Phys.~Rev.}}

\def\RMP{{\em Rev.~Mod.~Phys.}}

\def\INTA{{\em Int.~J.~Mod.~Phys.} A}
\def\EPJC{{\em Eur.~Phys.~J.} C}
\def\NIMA{{\em Nucl.~Instrum.~Meth.} A}
\def\JHEP{{\em J.~High~Energy~Phys.}}
\def\JINST{{\em JINST}}
\def\PPNP{{\em Prog.~Part.~Nucl.~Phys.}}

\newcommand{\be}{\begin{equation}}
\newcommand{\ee}{\end{equation}}
\newcommand{\bea}{\begin{eqnarray}}
\newcommand{\eea}{\end{eqnarray}}

\newcommand{\jpsi}{\ensuremath{J/\psi}\xspace}
\newcommand{\nubar}{\ensuremath{{\overline{\nu}}}\xspace}
\newcommand{\qbar}{\ensuremath{{\overline{q}}}\xspace}
\newcommand{\qqbar}{\ensuremath{q\qbar}\xspace}
\newcommand{\tbar}{\ensuremath{{\overline{t}}}\xspace}
\newcommand{\ttbar}{\ensuremath{{t\tbar}}\xspace}
\newcommand{\Tbar}{\ensuremath{{\overline{T}}}\xspace}
\newcommand{\TTbar}{\ensuremath{{T\Tbar}}\xspace}
\newcommand{\bbar}{\ensuremath{{\overline{b}}}\xspace}
\newcommand{\bbbar}{\ensuremath{{b\bbar}}\xspace}
\newcommand{\cbar}{\ensuremath{{\overline{c}}}\xspace}
\newcommand{\ccbar}{\ensuremath{{c\cbar}}\xspace}
\newcommand{\epem}{\ensuremath{{e^{+}e^{-}}}\xspace}
\newcommand{\mumu}{\ensuremath{{\mu^{+}\mu^{-}}}\xspace}
\newcommand{\tautau}{\ensuremath{{\tau^{+}\tau^{-}}}\xspace}
\newcommand{\lplm}{\ensuremath{{\ell^{+}\ell^{-}}}\xspace}
\newcommand{\pp}{\ensuremath{pp}\xspace}
\newcommand{\ppbar}{\ensuremath{{p\overline{p}}}\xspace}
\newcommand{\Bd}{\ensuremath{B_{d}^{0}}\xspace}
\newcommand{\Bdbar}{\ensuremath{\overline{B}_{d}^{0}}\xspace}
\newcommand{\BdBdbar}{\ensuremath{{\Bd\Bdbar}}\xspace}
\newcommand{\ttW}{\ensuremath{{\ttbar W}}\xspace}
\newcommand{\ttg}{\ensuremath{{\ttbar \gamma}}\xspace}
\newcommand{\ttZ}{\ensuremath{{\ttbar Z}}\xspace}
\newcommand{\ttH}{\ensuremath{{\ttbar H}}\xspace}
\newcommand{\Hbb}{\ensuremath{{H\to\bbbar}}\xspace}
\newcommand{\Hgg}{\ensuremath{{H\to\gamma\gamma}}\xspace}
\newcommand{\HZZ}{\ensuremath{{H\to ZZ}}\xspace}
\newcommand{\HWW}{\ensuremath{{H\to W^+W^-}}\xspace}
\newcommand{\Htt}{\ensuremath{{H\to\tautau}}\xspace}
\newcommand{\px}{\ensuremath{p_{x}}\xspace}
\newcommand{\py}{\ensuremath{p_{y}}\xspace}
\newcommand{\pz}{\ensuremath{p_{z}}\xspace}
\newcommand{\pt}{\ensuremath{p_{T}}\xspace}
\newcommand{\met}{\ensuremath{E_{T}^\mathrm{miss}}\xspace}
\newcommand{\metvec}{\ensuremath{\vec{p}_{T}^\mathrm{~miss}}\xspace}
\newcommand{\HT}{\ensuremath{H_{T}}\xspace}
\newcommand{\ST}{\ensuremath{S_{T}}\xspace}
\newcommand{\sqrts}{\ensuremath{\sqrt{s}}\xspace}
\newcommand{\mW}{\ensuremath{m_{W}}\xspace}
\newcommand{\mZ}{\ensuremath{m_{Z}}\xspace}

\newcommand{\mt}{\ensuremath{m_{t}}\xspace}
\newcommand{\mtbar}{\ensuremath{m_{\tbar}}\xspace}
\newcommand{\mtt}{\ensuremath{m_{\ttbar}}\xspace}
\newcommand{\mtpole}{\ensuremath{\mt^\mathrm{pole}}\xspace}

\newcommand{\mT}{\ensuremath{m_{T}}\xspace}
\newcommand{\mb}{\ensuremath{m_{b}}\xspace}
\newcommand{\yt}{\ensuremath{y_{t}}\xspace}
\newcommand{\muf}{\ensuremath{\mu_{F}}\xspace}
\newcommand{\mur}{\ensuremath{\mu_{R}}\xspace}
\newcommand{\alphaS}{\ensuremath{\alpha_{S}}\xspace}
\newcommand{\LamQCD}{\ensuremath{\Lambda_\mathrm{QCD}}\xspace}
\newcommand{\MSbar}{\ensuremath{\mathrm{\overline{MS}}}\xspace}
\newcommand{\GF}{\ensuremath{G_{F}}\xspace}

\newcommand{\sigtt}{\ensuremath{\sigma_{\ttbar}}\xspace}
\newcommand{\Lint}{\ensuremath{\int\mathcal{L}\,\mathrm{d}t}\xspace}
\newcommand{\Vtb}{\ensuremath{V_{tb}}\xspace}
\newcommand{\Vtbsq}{\ensuremath{\left|\Vtb\right|^2}\xspace}
\newcommand{\VminusA}{\ensuremath{V\! -\! A}\xspace}
\newcommand{\Lxy}{\ensuremath{{L_{xy}}}\xspace}
\newcommand{\epsb}{\ensuremath{{\epsilon_{b}}}\xspace}
\newcommand{\SFb}{\ensuremath{{\mathrm{SF}_{b}}}\xspace}
\newcommand{\Afb}{\ensuremath{{A_\mathrm{FB}}}\xspace}
\newcommand{\Ac}{\ensuremath{{A_{C}}}\xspace}
\newcommand{\cosths}{\ensuremath{\cos\theta^{*}}\xspace}
\newcommand{\Rtch}{\ensuremath{R_{t\text{-ch}}}\xspace}
\newcommand{\jetq}{\ensuremath{\mathrm{JetQ}}\xspace}

\newcommand{\eg}{e.\,g.\ }
\newcommand{\ie}{i.\,e.\ }
\newcommand{\spps}{\ensuremath{\mathrm{S\ppbar S}}\xspace}
\newcommand{\xsec}{cross section\xspace}
\newcommand{\ak}{anti-$k_{t}$\xspace}
\newcommand{\st}{single top-quark\xspace}
\newcommand{\sch}{$s$-channel\xspace}
\newcommand{\tch}{$t$-channel\xspace}
\newcommand{\uch}{$u$-channel\xspace}
\newcommand{\Wtch}{$Wt$-channel\xspace}
\newcommand{\wjets}{$W$+jets\xspace}
\newcommand{\zjets}{$Z$+jets\xspace}
\newcommand{\vjets}{$W/Z$+jets\xspace}
\newcommand{\bjet}{$b$-jet\xspace}
\newcommand{\bhad}{$b$~hadron\xspace}
\newcommand{\btag}{$b$-tag\xspace}
\newcommand{\btagging}{$b$-tagging\xspace}
\newcommand{\btagged}{$b$-tagged\xspace}

\newcommand{\invpb}{\per\pico\barn}
\newcommand{\invfb}{\per\femto\barn}
\newcommand{\invab}{\per\atto\barn}

\newacro{sm}[SM]{standard model}
\newacro{bsm}[BSM]{beyond standard model}
\newacro{fcnc}[FCNC]{flavor-changing neutral current}
\newacro{fnal}[FNAL]{Fermi National Accelerator Laboratory}
\newacro{slac}[SLAC]{Stanford Linear Accelerator Center}
\newacro{bnl}[BNL]{Brookhaven National Laboratory}
\newacro{desy}[DESY]{Deutsches Elektronen-Synchrotron}
\newacro{kek}[KEK]{High Energy Accelerator Research Organization}
\newacro{cern}[CERN]{European Organization for Nuclear Research}
\newacro{met}[MET]{missing transverse momentum}
\newacro{qed}[QED]{quantum electrodynamics}
\newacro{qcd}[QCD]{quantum chromodynamics}
\newacro{pdf}[PDF]{parton distribution function}
\newacro{mc}[MC]{Monte Carlo}
\newacro{lo}[LO]{leading order}
\newacro{nlo}[NLO]{next-to-leading order}
\newacro{nnlo}[NNLO]{next-to-next-to-leading order}
\newacro{n3lo}[N$^3$LO]{next-to-next-to-next-to-leading order}
\newacro{lhc}[LHC]{Large Hadron Collider}
\newacro{hllhc}[HL-LHC]{High-Luminosity LHC}
\newacro{ll}[LL]{leading logarithmic}
\newacro{nll}[NLL]{next-to-leading logarithmic}
\newacro{nnll}[NNLL]{next-to-next-to-leading logarithmic}
\newacro{5fs}[5FS]{five-flavor scheme}
\newacro{4fs}[4FS]{four-flavor scheme}
\newacro{ml}[ML]{maximum-likelihood}
\newacro{nn}[ANN]{artificial neural network}
\newacro{bdt}[BDT]{boosted decision tree}
\newacro{mem}[MEM]{matrix-element method}
\newacro{lhctopwg}[LHC\emph{top}WG]{LHC Top Physics Working Group}
\newacro{ckm}[CKM]{Cabibbo-Kobayashi-Maskawa}
\newacro{mcfm}[MCFM]{Monte Carlo for Femtobarn Processes}
\newacro{jes}[JES]{jet-energy scale}
\newacro{jer}[JER]{jet-energy resolution}
\newacro{vlq}[VLQ]{vector-like quark}
\newacro{cl}[CL]{confidence level}
\newacro{dm}[DM]{dark matter}
\newacro{eft}[EFT]{effective field theory}
\newacro{ilc}[ILC]{International Linear Collider}
\newacro{clic}[CLIC]{Compact Linear Collider}
\newacro{cepc}[CEPS]{Circular Electron Positron Collider}
\newacro{kit}[KIT]{Karlsruhe Institute of Technology}

\begin{document}

\begin{frontmatter}

  \title{Top-Quark Physics: Status and Prospects}
  \author{Ulrich Husemann}
  \ead{ulrich.husemann@kit.edu}
  
  \address{Institut f\"ur Experimentelle Kernphysik \\
    Karlsruhe Institute of Technology, Germany}

  \begin{abstract}
  After the discovery of the top quark more than 20~years ago, its
  properties have been studied in great detail both in production and
  in decay.  Increasingly sophisticated experimental results from the
  Fermilab Tevatron and from Run~1 and Run~2 of the LHC at CERN are
  complemented by very precise theoretical predictions in the
  framework of the standard model of particle physics and beyond. In
  this article the current status of top-quark physics is reviewed,
  focusing on experimental results, and a perspective of top-quark
  physics at the LHC and at future colliders is given.
\end{abstract}

\end{frontmatter}
{\small\tableofcontents} 
\clearpage

\section{Introduction}
\label{sec:introduction}

\subsection{Overview}


Particle physics has recently celebrated the 20th anniversary of the
discovery of the top quark. Over the last two decades the most massive
particle of the \ac{sm} of particle physics has been studied in great
detail, both at the Tevatron collider at \ac{fnal} and at the \ac{lhc}
at the \ac{cern}. The results are documented in more than
200~publications by the Tevatron and \ac{lhc} experiments as well as
in many preliminary results, presentations at conferences and
workshops, etc.

The goal of this review is two-fold: it is intended as an introduction
to the field, and at the same time it aims to convey the current state
of the art in top-quark physics. While the focus of the review is on
experimental results, a glimpse of the many achievements in related
developments in particle physics phenomenology is also given. The
introductory part is based on master-level lectures on top-quark
physics given at \ac{kit} and assumes some previous knowledge usually
taught in introductory lectures on experimental and theoretical
particle physics. In the later parts of the review, a variety of
recent results on top-quark physics will be introduced. The focus is
on the basic physics and measurement ideas, leaving out many of the
details which experimental physicists have spent most of their time
on. For a given physics question, the analysis methods, as well as the
sensitivities, of the different experiments are often very
similar. Therefore the numbers and figures quoted in the review should
be taken as illustrative examples.

There has been a substantial number of review articles on top-quark
physics published in recent years. The review
articles~\cite{Cristinziani:2016vif,Boos:2015,Kroninger:2015oma,Chierici:2014eqa,Deliot:2014uua,Gerber:2014xea,Jung:2014iqa,JABEEN:2013mva,BARBERIS:2013wba,Schilling:2012dx}
are general overviews of top-quark physics from an experimental point
of view, sometimes restricted to just Tevatron or \ac{lhc}
results. In~\cite{delDuca:2015gca, Bernreuther:2008ju} the theoretical
and phenomenological aspects of top-quark physics are
discussed. Further review articles deal with more specialized topics,
for example single top-quark
production~\cite{Wagner-Kuhr:2016zbg,Giammanco:2015bxk,Boos:2012hi},
the top-quark mass~\cite{Cortiana:2015rca}, \ttbar production
asymmetries~\cite{Aguilar-Saavedra:2014kpa}, or top-quark physics at
the HERA $ep$ collider~\cite{Behnke:2015qja}.

This review is structured as follows: The remainder of this chapter is
dedicated to a brief historical introduction to top-quark physics.  In
Section~\ref{sec:theory} the basic concepts of top-quark physics are
introduced. The most important experimental techniques employed to
study top quarks are discussed in Section~\ref{sec:techniques}. Some
readers may want to skip these introductory chapters and jump directly
to the discussion of recent top-quark physics results starting in
Section~\ref{sec:production}. In this chapter measurements of
top-quark production in various production and decay channels are
introduced. One of the most important measurements in top-quark
physics is the determination of the top-quark mass, which will be
discussed in Section~\ref{sec:mass}. Further production and decay
properties of the top quark, including those expected in \ac{bsm}
physics scenarios, have been studied in great detail, as shown in
Section~\ref{sec:properties}. The review is completed with a look at
the prospects for top-quark physics at future collider experiments in
Section~\ref{sec:outlook}.

\subsection{Historical Remarks}
\label{sec:history}
The discovery of the top quark by the CDF and D0 collaborations at the
Tevatron in 1995~\cite{Abe:1995hr,Abachi:1995iq} marks the end of a
long quest for the sixth and last quark of the \ac{sm} and at the same
time the beginning of a long quest to understand the top quark's
properties and its role in the \ac{sm} and beyond.

\subsubsection{The Road to the Top}
In the original quark model by Gell-Mann~\cite{GellMann:1964nj} and
Zweig~\cite{Zweig:1981pd}, based on the approximate $SU(3)$ symmetry of
the mass spectrum of light mesons and
baryons~\cite{GellMann:1962xb,Ne'eman:1961cd}, hadrons consist of the
three lightest quarks: up, down, and strange. It was realized by
Cabibbo in 1963 that electroweak currents that change the strangeness
quantum number of a hadron by one unit ($\Delta S=1$) show a different
coupling strength than currents with
$\Delta S=0$~\cite{Cabibbo:1963yz}.  In modern particle physics
language this means that the physical quarks (mass eigenstates) and
the quarks that participate in the electroweak interaction (flavor
eigenstates) are not aligned, a phenomenon called flavor mixing. A
fourth quark, the charm quark, was postulated by Glashow, Iliopoulos,
and Maiani in 1970 to explain the strong suppression of \ac{fcnc}
processes such as $K^0\to\mumu$ by the destructive interference of
scattering amplitudes with up and charm quarks (``GIM
mechanism'')~\cite{Glashow:1970gm}. The charm quark was discovered by
interpreting the \jpsi resonance observed in experiments at
\ac{bnl}~\cite{Aubert:1974js} and \ac{slac}~\cite{Augustin:1974xw} as
a \ccbar bound state. This discovery completed the second generation
of quarks. In both quark generations a quark with a third component of
the weak isospin of $I_3=+1/2$ and a charge of $Q=+2/3$ in units of
the elementary charge $e$ (``up-type quark'') and a quark with
$I_3=-1/2$ and charge $Q=-1/3$ (``down-type quark'') form a weak
isospin doublet.

The 1964 experiment by Christenson, Cronin, Fitch, and Turlay used
neutral kaon decays to show that the weak interaction is not invariant
under the combined discrete symmetry operation of charge conjugation
$C$ and parity $P$ (``CP
violation'')~\cite{Christenson:1964fg}. Kobayashi and Maskawa realized
in 1973 that what is known now as the electroweak sector of the
\ac{sm} provides a mechanism for CP violation through flavor mixing
only if there are at least three generations of
quarks~\cite{Kobayashi:1973fv}.  The charged lepton of the third
generation, the tau lepton, was discovered at \ac{slac} in
1975~\cite{Perl:1975bf}, shortly followed by the discovery of the
$\Upsilon$ resonances at \ac{fnal} in 1977~\cite{Herb:1977ek},
interpreted as bound states of a third-generation quark, the bottom
quark, and its antiparticle (\bbbar).

The open question at the time was if the bottom quark is a weak
isospin singlet or is part of another doublet. To shed light on this
question, the quantum numbers of the bottom quark were determined in
\epem collision experiments at \ac{desy}. The cross section for the
production and hadronic decay of the $\Upsilon(1S)$ resonance in \epem
collisions is proportional to the partial width of the $\Upsilon(1S)$
for decays to electrons, $\Gamma_{ee}$, which can be related to the
bottom-quark charge. The experimental results of the PLUTO
collaboration from 1978 favored a charge of
$Q=-1/3$~\cite{Berger:1978dm}. Measurements of the angular
distribution of \bhad{}s produced in \epem collision supported the
quantum numbers $I_3=-1/2$ and $Q=-1/3$ for the bottom
quark~\cite{Bartel:1984rg}, strongly suggesting that the bottom quark
is the down-type quark of the third generation whose $I_3=+1/2$
isospin partner was yet to be discovered.

An isospin partner for the bottom quark is also well-motivated
theoretically by the chiral anomaly. In quantum field theories,
anomalies are symmetries of the Lagrangian that are absent in the full
theory including quantum corrections. For a four-dimensional chiral
gauge theory to be renormalizable, the chiral anomaly, generated by
the non-conservation of gauge currents \eg in triangle diagrams, must
be absent. In the \ac{sm} the chiral anomaly is ``accidentally''
canceled because there is the same number of quark and lepton flavors,
and the number of color charges is three. Hence, to avoid chiral
anomalies the third generation of quarks should be a weak isospin
doublet, consisting of the bottom quark and its isospin partner, the
top quark.

In the early 1980s it seemed natural to search for top quarks with
masses similar to the bottom quark mass, of the order of
\SI{10}{GeV}~\footnote{In this article, natural units with $\hbar=c=1$
  are used throughout. Hence energy, momentum, and mass are measured
  in units of \si{GeV}, and the units for time and length are~\si{\per
    \GeV}.}. Direct searches for the process $\epem\to\ttbar$ were
conducted at the \epem colliders PEP (\ac{slac}, center-of-mass energy
$\sqrts\lesssim\SI{30}{GeV}$), PETRA (\ac{desy},
$\sqrts\lesssim\SI{45}{GeV}$), TRISTAN (\ac{kek},
$\sqrts\lesssim\SI{64}{GeV}$), SLC (\ac{slac},
$\sqrts\approx\SI{91}{GeV}$), and LEP~1 (\ac{cern},
$\sqrts\approx\SI{91}{GeV}$). In absence of a signal, lower limits on
the top-quark mass of up to $\mt>\SI{45}{GeV}$ at 95\% \ac{cl} were
placed, see e.g.~\cite{Boos:2015} for references and further details.

At the \spps proton-antiproton collider at \ac{cern} with
$\sqrts=\SI{540}{GeV}$, top quarks could manifest themselves in decays
of real $W$ bosons, $W^+\to t\bbar$,~\footnote{Charge-conjugated
  decays are implied, unless stated otherwise.} if their mass is below
\SI{70}{GeV}. In 1984, the UA1 experiment claimed a ``clear signal''
compatible with a $W$ boson decaying into a 40-GeV top
quark~\cite{Arnison:1984iw}. From today's perspective, the ``signal''
was most likely caused by an underestimation of the background from
$W$-boson production in association with jets, for which no adequate
simulation tools existed at the time. At the end of their data-taking
the \spps experiments UA1 and UA2 were only able to provide lower
limits on the top-quark mass up to approximately $\mt>\SI{70}{GeV}$ at
95\% \ac{cl}, see e.g.~\cite{Boos:2015} for references.

\begin{figure}[t]
  \centering
  \includegraphics[width=0.48\textwidth]{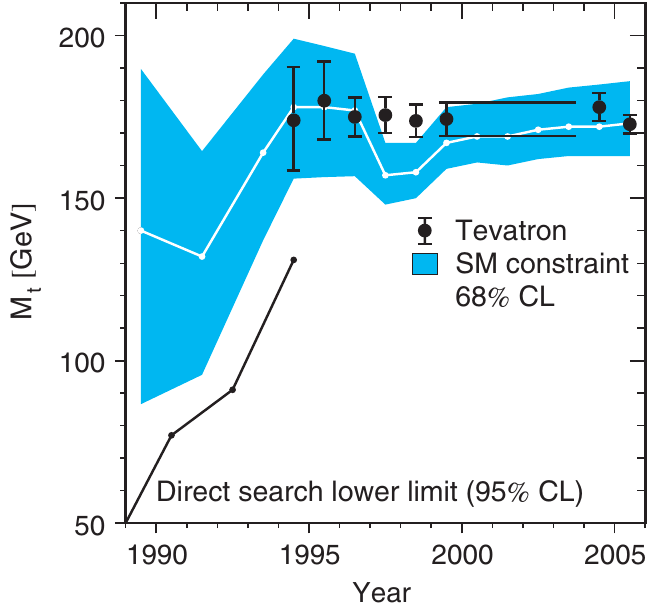}\hspace{3mm}
  \includegraphics[width=0.48\textwidth]{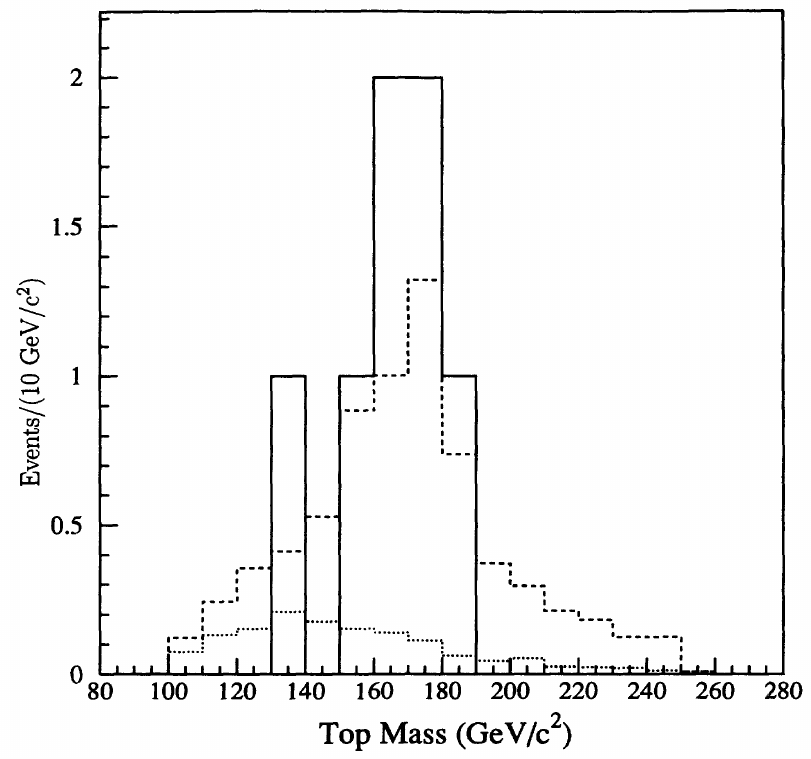}
  \caption{Comparison of direct and indirect determinations of the
    top-quark mass as a function of time~\cite{ALEPH:2005ab}
    (left). Reconstructed top-quark mass distribution in data (solid
    histogram) and \ac{mc} simulation (dashed histograms) based on
    seven candidate events recorded by the CDF experiment between
    August of~1992 and May of~1993~\cite{Abe:1994st} (right).}
  \label{fig:mt_limits}
\end{figure}

In parallel, indirect hints of a large top-quark mass came from the
observation of $\BdBdbar$ flavor oscillations with the ARGUS
experiment at \ac{desy}~\cite{Prentice:1987ap}, in the process
$\epem\to\Upsilon(4S)\to\BdBdbar$. The \ac{cern} UA1 experiment had
reported a three-standard deviation excess of same-sign muon pairs in
\ppbar collisions earlier~\cite{Albajar:1986it} that can be
interpreted as evidence for $\BdBdbar$ oscillations.  The oscillation
frequency depends on the mass difference $\Delta m_d$ between the two
\Bd-meson mass eigenstates, which in turn is a function of the
top-quark mass \mt. The large oscillation frequency observed by ARGUS
pointed to top-quark masses well above \SI{50}{GeV}. Additional
indirect constraints on \mt were derived from the combined analysis of
electroweak precision data obtained at the ``$Z$ boson factories''
LEP~1 and SLC. Some of the radiative corrections to the masses of the
$W$ and $Z$ bosons are proportional to $\mt^2$ and further electroweak
observables are sensitive to \mt as well. From the LEP experiments
alone, a value of
$\mt = \SI[parse-numbers=false]{173^{+12}_{-13}~^{+18}_{-20}}{GeV}$
was quoted before the discovery of the top quark~\cite{LEPEW:1994aa},
where the first uncertainty interval comes from the experimental
uncertainties propagated through the combined analysis and the second
uncertainty interval corresponds to the lack of knowledge of the Higgs
boson mass in the 1990s. The limits on \mt as a function of time are
compared to direct measurements at the Tevatron in
Fig.~\ref{fig:mt_limits} (left).

\subsubsection{Tevatron Run I: Discovery and First Measurements}
At the Tevatron \ppbar collider at \ac{fnal}, the center-of-mass
energy of \SI{1.8}{TeV} marked a significant increase compared to the
\spps, such that the top quark became directly accessible in collider
experiments for the first time. The first Tevatron collisions were
recorded by the CDF experiment in 1985. In subsequent years, CDF
improved the lower limit on the top-quark mass to $\mt>\SI{91}{GeV}$
at 95\% \ac{cl}~\cite{Abe:1991ue}. Starting in 1992, the Tevatron
commenced its Run~I with the two experiments CDF and D0 taking data
with improved detectors. For ``heavy'' top quark masses above
approximately \SI{85}{GeV} the decay $t\to W^+ b$ is allowed
kinematically, which was reflected in modified search strategies at
the Tevatron. First indications of an excess of collision events above
the background expectation compatible with \ttbar production showed up
in the following years~\cite{Abe:1994st,Abachi:1995ms}, see
Fig.~\ref{fig:mt_limits}~(right). The discovery of the top quark was
announced publicly in a joint seminar of the CDF and D0 collaborations
at \ac{fnal} on March 2, 1995, and published in the journal Physical
Review Letters the following day~\cite{Abe:1995hr,Abachi:1995iq}.  A
popular account of the top-quark discovery can be found
in~\cite{Liss:1997vk}.

Typical Tevatron Run-I top physics analyses used between
\SI{100}{\invpb} and \SI{125}{\invpb} of integrated
luminosity\footnote{In the following, luminosity figures are given
  per experiment, unless noted otherwise.},  equivalent to tens to
hundreds of \ttbar pairs available for analysis, depending on the
decay channel. Top-quark physics highlights of Run~I included
measurements of the \ttbar production cross
section~\cite{Affolder:2001wd,Abazov:2002gy}, the top-quark
mass~\cite{Azzi:2004rc} and various other properties, such as the
$W$-boson polarization in the decay
$t \to W^+ b$~\cite{Affolder:1999mp,Abazov:2004ym} and first searches
for physics beyond the \ac{sm} with top quarks, \eg for decays of a
hypothetical heavy resonance $Z'\to\ttbar$~\cite{Affolder:2000eu} or
top-quark decays into charged Higgs bosons, $t\to H^+b$, which occur in
models with an extended Higgs sector compared to the
\ac{sm}~\cite{Affolder:1999au,Abazov:2001md}. Top-quark physics at
Tevatron Run~I also pioneered various novel data analysis techniques,
such as the \ac{mem} to determine the top-quark
mass~\cite{Kondo:1988yd,Abazov:2004cs}.

\subsubsection{Tevatron Run II: Is the Top Really the Sixth Quark of
  the Standard Model?}
\label{sec:runii}
Tevatron Run~II started in 2001 with an increased center-of-mass
energy of \SI{1.96}{TeV} and significant upgrades to the CDF and D0
detectors. Until the end of Tevatron data-taking on September~30,
2011, the Tevatron delivered a total of \SI{12}{\invfb} of integrated
luminosity each to CDF and D0. Typical top-quark analyses were thus
performed on data samples of several hundreds to thousands of events
containing top quarks. Using the Run-II datasets, the Tevatron
experiments have addressed a broad range of questions in top-quark
physics, from inclusive production \xsec{}s and precise measurements
of the top-quark mass and couplings to a variety of searches for
\ac{bsm} physics with top quarks. At the time of writing this review,
many ``legacy'' publications using the full Run-II datasets have
already been published, others are being finalized.  Three highlights
of the Run-II top-quark physics program are briefly sketched in the
following.

The precision achievable in measurements of the top-quark mass at the
Tevatron was limited both by the dataset size and by uncertainties in
the reconstruction of jet energies.  Further refinements to the
\ac{mem} and novel concepts to constrain uncertainties in the
jet-energy scale from the data itself (``in-situ calibration'')
during Tevatron Run~II lead to a significant reduction of the
uncertainty on the top-quark mass, culminating in the current single
most precise Tevatron measurement of the top-quark mass, performed by
the D0 collaboration, which has a relative uncertainty of only
0.43\%~\cite{Abazov:2014dpa,Abazov:2015spa}.

The \ac{sm} predicts the electroweak production of single top quarks
in addition to the dominant \ttbar pair production, which is a
\ac{qcd} process. Single top-quark production was observed for the
first time by CDF and D0 in
2009~\cite{Abazov:2009ii,Aaltonen:2009jj}. Sophisticated multivariate
analysis methods were necessary to separate the small signal from an
overwhelming background. The development and validation of these
methods also paved the way for Higgs-boson searches at the Tevatron
and similar methods are being employed in top-quark and Higgs-boson
physics as well as in searches for \ac{bsm} physics at the \ac{lhc}.

In \ttbar production a small forward-backward asymmetry between the
top quark and antiquark is expected~\cite{Kuhn:1998jr}. The Tevatron
Run~II results on the \ttbar production asymmetry gained considerable
interest. The first results already indicated asymmetry values larger
than expected from \ac{qcd}~\cite{Abazov:2007ab,Aaltonen:2008hc}. By
2011, with about half of the Run~II datasets analyzed, the CDF
experiment observed discrepancies between the data and \ac{nlo}
\ac{qcd} expectations at the level of three standard deviations for
\ttbar invariant masses above
\SI{450}{GeV}~\cite{Aaltonen:2011kc}. These observations triggered a
plethora of publications from the theory community as well as an
extensive measurement program. However, after a full suite of
measurements and improved \ac{sm} predictions including \ac{nnlo}
\ac{qcd} corrections, no strong hints of BSM physics in \ttbar
production asymmetries remain.


\subsubsection{LHC Run 1: From Re-Discovery to a Top Factory}
The start of \ac{lhc} data taking at $\sqrts=\SI{7}{TeV}$ in 2010 was
also the beginning of a new era in top-quark physics. With
approximately 3.5 times higher center-of-mass energy compared to the
Tevatron, \xsec{}s for top-quark production are expected to be more
than 20 times higher than at the Tevatron. Already after the first
three years of data-taking, the datasets recorded by the ATLAS and CMS
experiments contained about a million top-quark events, rendering the
\ac{lhc} the first ``top-quark factory.''

The goal for the first months at the \ac{lhc} was to ``rediscover the
\ac{sm},'' \ie to identify and measure the basic properties of all
known \ac{sm} particles, including the top quark. First measurements
of the \ttbar production \xsec by the CMS and ATLAS collaborations
became available in the second half of
2010~\cite{Khachatryan:2010ez,Aad:2010ey} using the first
\SI{3}{\invpb} of $pp$ collisions data. With the full 2010 dataset of
about \SI{35}{\invpb} the precision on the \ttbar production \xsec
already approached the precision achieved at the Tevatron. Data-taking
at the \ac{lhc} commenced in 2011, with another \SI{5}{\invfb} of data
recorded at $\sqrt{s}=\SI{7}{TeV}$. For the 2012 data-taking run of
the ATLAS and CMS experiments, the center-of-mass energy of the
\ac{lhc} was increased to \SI{8}{TeV} and a dataset of \SI{20}{\invfb}
was recorded.  With these \ac{lhc} Run-1 datasets, a wide variety of
precision measurements of top-quark properties and searches for
\ac{bsm} physics with top quarks were performed. Some of the analyses of 
the Run~1 datasets are still being finalized at the time of writing
this review. Results on top-quark production have also been obtained
using data taken with the LHCb experiment during \ac{lhc} Run~1.


\subsubsection{LHC Run 2: Towards Ultimate Precision}
After a two-year shutdown for maintenance of the \ac{lhc} machine and
experiments (``Long Shutdown 1''), the \ac{lhc} was restarted in early
2015 (``Run 2''). The center-of-mass energy was further increased to
\SI{13}{TeV}, which boosted typical top-quark cross sections by a
factor of about three compared to Run~1. ATLAS and CMS have recorded
$pp$ collision data equivalent to a luminosity of about
\SI{3.5}{\invfb} in 2015 and approximately \SI{40}{\invfb} in 2016. At
this integrated luminosity, the Run-2 top-quark datasets are already
about five times as large as the Run-1 datasets. The \ac{lhc} design
instantaneous luminosity of \SI{1e34}{\cm^{-2} s^{-1}} was reached and
exceeded by 50\% in~2016. Again, \xsec measurements were the first
top-physics results based on the \ac{lhc} Run-2 datasets that were
published. Many further results on top-quark properties and searches
for \ac{bsm} physics keep appearing while this review is being
written.

\subsection{Working Groups Across Experiments and Combination of
  Results}
\label{sec:combination}
While first and foremost, the Tevatron and \ac{lhc} experiments
publish experimental results based on their own datasets and methods,
there are also collaborative efforts across the experiments. 
The statistical combination of measurements aims at reducing the
statistical and systematic uncertainty of a result. This requires good
understanding of how systematic uncertainties are defined in each
experiment and how they are correlated across the experiments.

Both at the Tevatron and at the \ac{lhc} working groups have formed to
define guidelines for the combination of physics measurements. The
guidelines may include recommendations on the treatment of systematic
uncertainties and their correlations, ``reference \xsec{}s'' for
signal and background processes considered in top-quark physics, and
agreements on how to present the results of measurements.

At the Tevatron, the Top Subgroup of the Tevatron Electroweak Working
Group~\cite{TevTop} has provided combinations of CDF and D0
measurements on the top-quark mass and the \ttbar production
\xsec. Similarly, the \ac{lhctopwg}~\cite{LHCTopWG} has developed
recommendations on systematic uncertainties, as well as compilations,
comparisons, and combinations of ATLAS and CMS measurements.


\section{Top-Quark Physics at Hadron Colliders}
\label{sec:theory}

The basic tool for top-quark physics is a high-energy particle
collider. The dominant \ttbar production process is accessible both at
lepton and at hadron colliders, provided the center-of-mass energy of
the collisions is above the production threshold of twice the
top-quark mass and the luminosity is large enough to acquire datasets
with a sufficient number of \ttbar pairs. Until now only hadron
colliders have provided sufficient center-of-mass energy and
luminosity for top quarks to be produced. Therefore the discussion of
top-quark physics in this chapter is focused on hadron colliders. A
brief account of the top-quark physics prospects at future lepton
colliders will be given in Section~\ref{sec:future}.  

This chapter starts with brief overviews of hadron collider kinematics
and physics at large momentum transfer, often called ``high-\pt
physics,'' as well as simulation tools for hadron collider
physics. The discussion of basic hadron collider physics is followed
by a brief account of the production mechanisms and decay channels of
top quarks as well as the most important properties of the top quark
expected in the \ac{sm}.

\subsection{Hadron Collider Kinematics}
Most experiments at circular colliders utilize a right-handed
coordinate system with the $z$ axis pointing along the counterclockwise beam
direction, the $y$ axis pointing upwards, and the $x$ axis pointing
towards the center of the collider ring. The coordinates are often
expressed in a cylindrical coordinate system that reflects the
symmetry of the detector, with the distance to the beam axis $\rho$,
the angle $\theta$ from the $z$ axis (``polar angle'') and the angle
$\phi$ from the $x$ axis in the $xy$ plane (``azimuthal angle''),
perpendicular to the beam axis.

In hadron colliders the particles participating in the fundamental
collision processes are the partons within the hadrons. The
$z$~components of the colliding partons' momenta, \pz, are unknown in
a given collision event, only their probability distribution is
known. This is accounted for by choosing kinematic variables which are
insensitive to the lack of knowledge about \pz. The velocity of a
particle along the $z$ direction, $\beta_z = \pz/E$, is often
expressed in terms of the rapidity $y$:
\begin{equation}
  y\equiv\tanh^{-1}\beta_z 
  = \tanh^{-1}\left(\frac{\pz}{E}\right) 
  = \frac{1}{2}\ln\left(\frac{E+\pz}{E-\pz}\right).
\end{equation}
It can be shown that rapidity distributions, \eg the number of
particles per unit rapidity, $\mathrm{d}{N}/\mathrm{d}y$, are
invariant under Lorentz boosts along the $z$~direction. In the limit
of momenta much larger than the mass of a particle, the rapidity
converges to pseudorapidity:
\begin{equation}
  \lim_{|\vec{p}|\gg m} y \equiv \eta = -\ln\tan\left(\frac{\theta}{2}\right). 
\end{equation}
The pseudorapidity of a particle is a purely geometrical quantity, it
only depends on the polar angle $\theta$, but not on the particle's
mass. Another class of kinematic variables often used at hadron
colliders are transverse quantities, such as the transverse momentum
$\pt \equiv \sqrt{\px^2 + \py^2}$, with \px and \py being the $x$ and $y$
components of the particle momentum. Transverse quantities are
invariant under Lorentz boosts along the beam direction by
construction.

The initial-state particles of a hadron-hadron collision are collinear
to the $z$ axis to very good approximation. Momentum conservation in
the $xy$ plane requires that the vectorial sum of the transverse
momenta of all final-state particles is (approximately) zero as
well. This constraint can be used to indirectly detect weakly
interacting particle that do not leave a signal in a hadron collider
detector, such as neutrinos. The corresponding observable is the
missing transverse momentum $\metvec$, defined as the negative
vectorial sum of all reconstructed particle momenta in a collision
event. Its absolute value \met is often called missing transverse
energy (MET). For a single undetected particle, \met is equivalent to
the \pt of that particle; however, the particle's \pz remains
undetermined. Experimentally \acs{met} reconstruction is challenging,
because the observable depends on all other particles in the detector
and their calibration and is prone to misreconstruction.

\subsection{High-\pt Physics at Hadron Colliders}
\label{sec:highpt}
For many \ac{qcd} processes at hadron colliders, the physics effects
at short distances---or equivalently at high energies---and at large
distances, \ie small energies, can be factorized. The \xsec for such a
process can be expressed as a \xsec for the high-energy (``hard'')
parton-parton scattering process weighted by \acp{pdf} of the partons
participating in the scattering processes, integrated over all parton
momenta and summed over all parton types.  The hard scattering \xsec
is process-specific and can be computed in perturbative \ac{qcd},
while the \acp{pdf} are universal and can be measured independently of
the hard process. The factorization formula for the \xsec reads
\begin{equation}
\sigma = 
\sum_{jk}^{\mathrm{partons}}\int_0^1 \mathrm{d}x_j\,\mathrm{d}x_k\,
f_j(x_j,\muf^2)\,f_k(x_k,\muf^2) \,
\hat\sigma\left(x_j x_k s, \muf,\alphaS(\mur)\right).
\end{equation}
The \ac{pdf}s $f_i(x_i,\muf^2)$ are universal functions that describe
the probability to find a parton $i$ with a given longitudinal
momentum fraction $x_i$ when the hadron is probed at a momentum
transfer of \muf. This introduces a new energy scale \muf to the
process, called the factorization scale, which can be viewed as the
energy scale that separates physics processes at short distances from
those at long distances. The \ac{pdf} absorbs all long-distance
effects in the initial state that would lead to infrared and/or collinear
divergences\footnote{Infrared divergences occur if massless particles
  with vanishing momenta are radiated from other particles. Massless
  particles radiated at very small angles lead to collinear
  divergences.} in collider observables if treated in perturbative
\ac{qcd}.  The hard scattering \xsec $\hat\sigma$ is a function of the
partonic center-of-mass energy squared $\hat s = x_j x_k s$ ($s$ being
the $pp$ center-of-mass energy squared), the factorization scale and
the strong coupling constant \alphaS. As $\hat\sigma$ is computed in
perturbation theory, the renormalization procedure to treat
ultraviolet divergences\footnote{In perturbation theory, ultraviolet
  divergences occur if particle momenta in virtual (``loop'')
  corrections approach infinity.}  results in an additional energy
scale, the renormalization scale~\mur. The default choice of energy
scale to compute $\ttbar$ pair production is the top-quark mass:
$\mur=\muf=\mt$.

In the above discussion, only the partonic final state has been
considered. However, \ac{qcd} color confinement requires the final
state particles to be color-neutral. The process of converting colored
partons into hadrons, called hadronization, cannot be computed in
\ac{qcd} perturbation theory. Instead phenomenological models are
employed, as implemented in \acf{mc} event generators, see
Section~\ref{sec:mc}. As the hadronization probability is unity, the
\xsec $\sigma$ remains unchanged.

\subsection{Monte-Carlo Simulation Tools for Top-Quark Physics}
\label{sec:mc}
To compare calculations of hadron-hadron collisions to experimental
data software tools based on the \acf{mc} method are employed. The
output of these tools is simulated collision events that resemble
experimental data both with respect to the physics processes involved
in the hadron-hadron scattering and the interactions of the final state
particles with the particle detector. The following discussion is
restricted to the simulation of the physics processes in \ac{mc} event
generators.

\ac{mc} event generators follow the factorization approach discussed
in Section~\ref{sec:highpt}. The \acp{pdf}, which are required to
describe the structure of the colliding hadrons, have been derived
from a set of measurements sensitive to the hadron structure, \eg deep
inelastic $ep$ scattering and jet production.  Recent \ac{pdf} sets
are available from several research groups; examples include
NNPDF3.0~\cite{Ball:2014uwa}, CT14~\cite{Dulat:2015mca},
MMHT2014~\cite{Harland-Lang:2014zoa},
HERAPDF2.0~\cite{Abramowicz:2015mha}, and
ABMP2016~\cite{Alekhin:2016uxn}. Technically \acp{pdf} can be accessed
conveniently via a common interface provided by the
LHAPDF~\cite{Buckley:2014ana} program library. The hard scattering
\xsec may be implemented at different orders in \ac{qcd} perturbation
theory. General-purpose \ac{mc} event generators of the first
generation typically included only $2\to 1$ and $2\to 2$ processes at
\ac{lo}. Starting in the early 2000s, also $2\to n$ processes (with
$n \lesssim 6$) and \ac{nlo} event generators became available. This
increased the precision of \ac{mc} predictions significantly, as \eg
the emission of additional partons or real and virtual \ac{nlo}
corrections were included in simulated \ttbar events. In parallel
automated \ac{mc} event generators were introduced, first at \ac{lo},
then also at \ac{nlo}. These generators are able to automatically
compute the full set of contributions to the hard process given the
Feynman rules of the underlying theory (both \ac{sm} and \ac{bsm}).

The process of turning partons into hadrons cannot be treated
perturbatively and relies on models. The process can be separated into
two steps, parton shower and hadronization, both of which are
implemented in general-purpose \ac{mc} event generator packages. The
parton shower is a probabilistic method to model the fragmentation of
partons that effectively resums soft and collinear radiation off the
partons, typically to \ac{ll} order precision in the corresponding
observables, see also Section~\ref{sec:ttbarprod}. Various specialized
matching and merging techniques are available to consistently
interface \ac{nlo} and $2\to n$ event generators to the parton shower
without double-counting parton emissions due to higher order processes
and the parton shower. Hadronization is described with models, the
most popular being based on the Lund string
model~\cite{Andersson:1983ia} and the cluster
model~\cite{Webber:1983if}.

Current \ac{mc} event generators used in top-quark physics at the LHC
include the \ac{nlo} generator {\sc Powheg
  v2}~\cite{Nason:2004rx,Frixione:2007vw,Alioli:2010xd,Jezo:2016ujg} and the
automated \ac{lo} and \ac{nlo} generator {\sc
  MadGraph5\_aMC@NLO}~\cite{Frixione:2002ik,Alwall:2014hca}. Both are
typically interfaced to {\sc
  Pythia8}~\cite{Sjostrand:2006za,Sjostrand:2014zea} or {\sc
  Herwig7}~\cite{Bellm:2015jjp} for the parton shower. Other popular
\ac{mc} generator choices include the \ac{lo} generator for $2\to n$
processes {\sc Alpgen}~\cite{Mangano:2002ea} and {\sc
  Sherpa}~\cite{Gleisberg:2008ta,Hoeche:2014qda}, which includes
\ac{lo} and \ac{nlo} matrix elements as well as its own parton shower.

Oftentimes calculations of the inclusive production \xsec for signal
and background processes include higher-order corrections and are thus
more precise than current \ac{mc} event generators. As will be
discussed in Sections~\ref{sec:ttbarprod} and~\ref{sec:singletopprod},
\xsec{}s for \ttbar and \st production are available up to \ac{nnlo}
accuracy~\cite{Czakon:2013goa,Brucherseifer:2014ama,Berger:2016oht}.
For such processes the normalization of the \ac{mc} sample is
corrected with a scale factor to match the more precise calculation,
ignoring the effect that higher-order corrections may have on the
shapes of kinematic observables.


\subsection{The Top Quark in the Standard Model}
\label{sec:topsm}

\subsubsection{Quantum Numbers and Decays}
\label{sec:decays}
In the \ac{sm} the top quark has the following properties: The top
quark is a fundamental fermion with spin $s=1/2$. It carries an
electric charge of $Q=2/3$ and is a color triplet. It forms a weak
isospin doublet together with the bottom quark, where the top quark is
the up-type quark with the third component of the weak isospin
$I_3 = +1/2$.

\paragraph{Decays}
The top quark decays via the electroweak charged-current process
$t \to W^+ q$, where $q$ is a down-type quark. The part of the \ac{sm}
Lagrangian density describing this interaction reads
\begin{equation}
\mathcal{L}_{Wtb} = - \frac{g}{\sqrt{2}}\,V_{tq}\, \qbar\,\gamma^\mu\,
 \frac{1}{2}\left( 1 - \gamma_5 \right) t\, W_\mu^-
+\text{h.c.},
\label{eq:Wtbsm}
\end{equation}
where $g$ is the electroweak coupling constant, $V_{tq}$ is the
element of the \ac{ckm} matrix responsible for $t\to q$ transitions,
\qbar is the adjoint spinor of the down-type quark, and $t$ is the
spinor of the top quark. The \VminusA Dirac structure of the $Wtb$
vertex, $\gamma^\mu( 1 - \gamma_5 )/2$, reflects the experimental fact
that $W$ bosons only couple to left-handed quarks and right-handed
antiquarks.  At \ac{lo} the total decay width of the top quark is
given by
\begin{equation}
  \Gamma_{t}^{\mathrm{LO}} = \frac{G_F}{8\pi\sqrt{2}}\, \mt^3
  \left( 1 - \frac{m_W^2}{\mt^2}\right)^2
  \left( 1 + 2\, \frac{m_W^2}{\mt^2}\right) \approx \SI{1.5}{GeV},
  \label{eq:width}
\end{equation}
where \mt is the top-quark mass, \mW is the $W$-boson mass, and
$\GF=\sqrt{2}\, g^2/(8\,\mW^2)$ is the Fermi constant. The
comparatively small mass of the bottom quark has been neglected in
Eq.~(\ref{eq:width}). The decay width has been computed in a fully
differential way including \ac{nnlo} \ac{qcd} corrections and \ac{nlo}
electroweak corrections, which reduce the top-quark decay width by
approximately 10\% compared to the \ac{lo}
prediction~\cite{Gao:2012ja,Brucherseifer:2013iv}.  The partial decay
width for the decay channel $t \to W^+ q$, $\Gamma(t\to Wq)$, is
proportional to the \ac{ckm} matrix element
$|V_{tq}|^2$. Experimentally, the relation
$|\Vtb| \gg |V_{ts}| > |V_{td}|$ holds, such that the
``\ac{ckm}-allowed'' decay $t \to W^+ b$ is by far the dominant decay
mode, with a branching fraction
$\mathcal{B}(t\to Wb)\equiv\Gamma(t\to Wb)/\sum_{q=d,s,b}\Gamma(t\to
Wq) = 0.998$ expected for a unitary \ac{ckm} matrix for three quark
generations~\cite{Olive:2016xmw}. The inverse of the total decay
width, the top-quark mean lifetime, is $\tau_t=\SI{5e-25}{s}$. This
value is shorter than the typical time scale of hadronization, which
can be estimated from the inverse of the energy scale \LamQCD at which
\ac{qcd} becomes non-perturbative:
$1/\LamQCD \approx 1/(\SI{200}{MeV}) \approx \SI{3e-24}{s}$. This
leads to two important consequences: Top quarks decay before
hadronization and do not form bound states such as top mesons
($t\qbar$) or toponium (\ttbar). The top-quark spin polarization and
the correlation between spins are largely preserved and can therefore
be computed and observed more easily than for other quarks. The
fraction of polarization transferred to the decay products, often
called the ``spin analyzing power'' $\kappa$, is different for the
different decay products. The value for the $W^+$ boson is
$\kappa=0.39$ and for the $b$ quark $\kappa=-0.39$, and the value for
the neutrino or the up-type quark of the $W^+$-boson decay is
$\kappa=-0.3$. The charged lepton or the down-type quark from the
$W^+$-boson decay assume the value $\kappa=1$. In an ensemble of 100\%
polarized top quarks the charged lepton will be emitted parallel to
the top-quark spin with the highest
probability~\cite{Bernreuther:2008ju}, making charged leptons the most
attractive top-quark decay products to study polarization effects.

\paragraph{$W$-Boson Polarization}
The \ac{sm} top-quark decay is governed by an electroweak \VminusA
interaction; therefore the $W$ boson in the final state is
polarized. This renders top quarks the only \ac{sm} source of
polarized $W$ bosons. At \ac{lo} the \ac{sm} predicts the following
fractions of left-handed polarization ($F_L$), longitudinal
polarization ($F_0$), and right-handed polarization ($F_R$):
\begin{eqnarray}
F_L = \frac{2\mW^2}{\mt^2 + 2\mW^2}\approx 0.3, \quad
F_0 = \frac{\mt^2}{\mt^2 + 2\mW^2}\approx 0.7,\\
F_R = \frac{\mb^2}{\mt^2} 
\frac{2\mW^2}{(1-\mt^2/\mW^2)^2 (\mt^2 + 2\mW^2)} \approx 0,\nonumber
\end{eqnarray}
where $F_L + F_0 + F_R = 1$. The most precise \ac{sm} prediction of
the polarization fractions includes \ac{nnlo} \ac{qcd}
corrections~\cite{Czarnecki:2010gb}.

The large value of $F_0\approx 0.7$ is related to the
Brout-Englert-Higgs mechanism~\cite{Higgs:1964pj, Higgs:1964ia,
  Englert:1964et, Nambu:1961tp, Guralnik:1964eu}, which is responsible
for the $W$ boson's longitudinal degree of freedom and hence its
mass. To conserve momentum and angular momentum, a right-handed $W$
boson can only be produced together with a positive-helicity bottom
quark. Due to the comparatively low bottom quark mass \mb, the
left-handed bottom quarks produced in top-quark decays dominantly
carry negative helicity, hence the fraction $F_R$ of right-handed
$W$~bosons is close to zero.

\paragraph{Classification of Decays}
Experimentally collision events containing \ttbar pairs are classified
by the decay of the $W^+$ and the $W^-$ boson from the \ttbar
decay. $W^+$ ($W^-$) bosons decay into hadronic final states $q\qbar'$
with a branching fraction of approximately~2/3 and into a charged
lepton $\ell^+$ ($\ell^-)$ and its corresponding (anti)neutrino
$\nu_\ell$ ($\nubar_\ell$) with a branching fraction of
approximately~1/9. This results in the following classification scheme
for \ttbar decay channels:
\begin{itemize}
\item Fully hadronic (also: all-hadronic, all-jets) channel: \\
  $\ttbar \to W^+b\,W^-\bbar \to q\qbar'b\,q''\qbar'''\bbar,$
\item Single-lepton (also: lepton+jets, semileptonic) channel: \\
  $\ttbar \to W^+b\,W^-\bbar \to \ell^+\nu_\ell b\,q\qbar'\bbar 
  \,\text{ and }\, \ttbar \to W^+b\,W^-\bbar \to q\qbar' b\,\ell^-\nubar_\ell \bbar,$
\item Dilepton channel:
  $\ttbar \to W^+b\,W^-\bbar \to \ell^+\nu_\ell b\, \ell'^-\nubar_{\ell'}\bbar.$
\end{itemize}
The fully hadronic channel has the largest branching fraction of
$(2/3)^2 \approx 0.45$ but also suffers from the largest
background. The single-lepton channel with its moderate branching
fraction of $2\times2/3\times(2\times1/9)\approx0.29$ has moderate
backgrounds, while the dilepton channel has the smallest branching
fraction of only $(2\times 1/9)^2\approx0.05$, but only very small
backgrounds\footnote{The factors $2\times 1/9$ (instead of
  $3\times 1/9$) in the single-lepton and dilepton channels are
  introduced because only electrons and muons are considered charged
  leptons in the above classification scheme, while tau leptons are
  treated separately due to their many leptonic and hadronic decay modes
  and large hadronic backgrounds.}. A more detailed discussion of the
background processes most relevant to \ttbar production follows in
Section~\ref{sec:event_selection}.

\subsubsection{Mass}
\label{sec:mass_theory}
In the \ac{sm}, the mass of fermions is generated by their Yukawa
coupling to the Higgs boson, linking the left-handed and right-handed
components of their spinors. The corresponding part of the \ac{sm}
Lagrangian for top quarks reads:
\begin{equation}
\mathcal{L}_{\mathrm{Yukawa},t} = 
-\yt\,\frac{v}{\sqrt{2}} \left(\tbar_L t_R + \tbar_R t_L\right) = 
-\yt\,\frac{v}{\sqrt{2}}\, \ttbar = -\mt\, \ttbar,
\end{equation}
where $y_t$ is the Yukawa coupling constant of the top quark, $v$ is
the vacuum expectation value of the Higgs field, $t_L$ and $t_R$ are
the left-handed and the right-handed components of the top-quark
spinor $t$, and \mt is the top-quark mass. It is worth noting
that---unlike for any other fundamental fermion---the numerical value
of \yt is unity to good approximation. This may just be a numerical
coincidence, but is often interpreted as a hint of the special role
that the top quark could play in \ac{bsm} physics.

The mass of the top quark is not a uniquely defined quantity. In
\ac{qcd} perturbation theory quark masses are renormalized and thus
become energy-scale dependent. The pole mass (also ``on-shell mass'')
\mtpole is a seemingly obvious choice to define the top-quark mass. As
the top quark does not hadronize it can be considered an unstable
``free'' fermion and its pole mass is defined as the real part of the
(renormalized) top-quark propagator's pole. However, this definition
is only unique in a given fixed order of \ac{qcd} perturbation
theory. Moreover, as quarks cannot be observed as free particles due
to \ac{qcd} confinement, the full quark propagator does not contain a
pole. It can also be shown that certain radiative corrections that
have to be considered to all orders (``infrared renormalon'') are hard
to control and lead to irreducible intrinsic uncertainties of the
\mtpole definition. Another open question is if the mass parameter
used in \ac{mc} generators can be identified with \mtpole, as \ac{mc}
generators use an energy cut in the parton shower as well as a
hadronisation model, both of which cannot be easily mimicked by a
perturbative calculation.

The class of scale-dependent ``short-distance masses,'' such as the
mass definition in the modified minimal subtraction renormalization
scheme (``\MSbar mass''), do not contain non-perturbative ambiguities.
Short-distance masses can be converted to pole mass in a given order
of perturbation theory in a unique way, which however comes with
uncertainties due to the truncation of the perturbative
series~\cite{Marquard:2015qpa,Marquard:2016dcn}. The inclusive
intrinsic uncertainty of the top-quark pole mass due to renormalon
effects has recently been demonstrated to be only
\SI{70}{MeV}~\cite{Beneke:2016cbu}, much smaller than other
uncertainties occuring in measurements of the top-quark mass at hadron
colliders. On the other hand, attempts to calibrate the top-quark mass
used in \ac{mc} generators to a short-distance mass for \epem initial
states show larger uncertainties of the order of
\SI{300}{MeV}~\cite{Butenschoen:2016lpz}. The debate on the ultimate
precision achievable in top-quark mass measurements is ongoing.

\subsection{Top Quark-Antiquark Pair Production}
\label{sec:ttbarprod}
\begin{figure}[t]
  \centering
  \begin{minipage}[b]{0.35\textwidth}
    \centering
    \includegraphics[width=\textwidth]{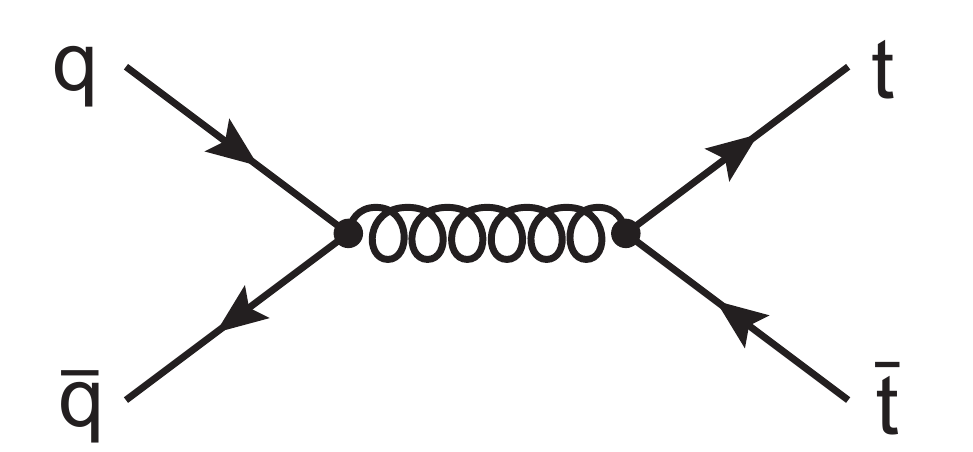}\\[3mm]
    \includegraphics[width=0.85\textwidth]{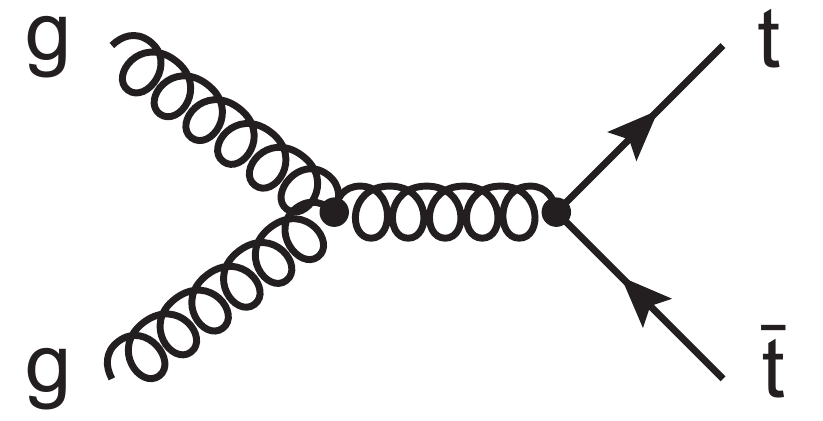}
  \end{minipage}
  \hspace{5mm}
  \begin{minipage}[b]{0.3\textwidth}
    \includegraphics[width=\textwidth]{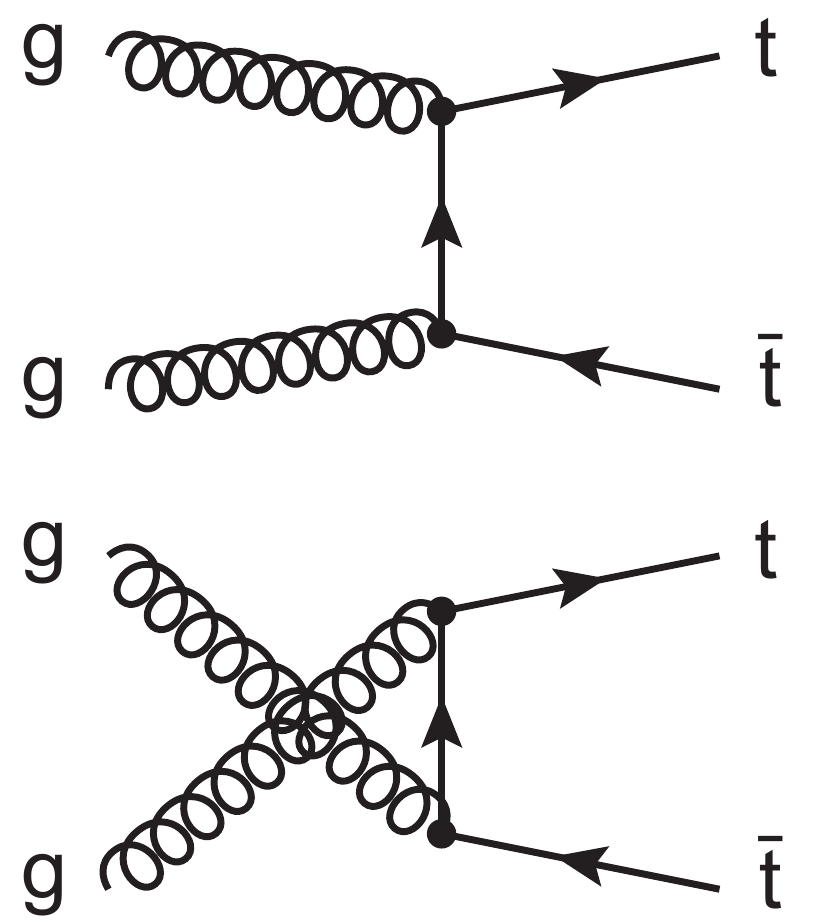}
  \end{minipage}

  \caption{Feynman diagrams for \ttbar production in \ac{qcd} at
    \ac{lo}: \qqbar annihilation (top left), $gg$ fusion in the \sch
    (bottom left), $gg$ fusion in the $t$-channel (top right), and
    $gg$ fusion in the $u$-channel (bottom right). Feynman diagrams
    created with {\sc JaxoDraw}~\cite{Binosi:2003yf}.}
  \label{fig:ttbarprod}
\end{figure}

The most abundant production process for \ttbar pairs at hadron
colliders is \ac{qcd} pair production. At parton level two \ac{lo}
processes with \xsec{}s proportional to $\alphaS^2$ contribute that
lead to \ttbar final states (see Fig.~\ref{fig:ttbarprod}):
gluon-gluon ($gg$) fusion in the $s$-, $t$-, and \uch\footnote{The
  Lorentz-invariant kinematic variables $s$, $t$, and $u$ are called
  Mandelstam variables. In scattering processes they denote the
  ``direction'' of the momentum transfer by a virtual particle.} and
quark-antiquark (\qqbar) annihilation. The relative fractions of $gg$
and \qqbar initiated processes depends on the \acp{pdf} of the
initial-state hadrons and the center-of-mass energy of the collisions.
In \pp collisions, \qqbar annihilation can take place between valence
quarks or sea quarks and sea antiquarks, while in \ppbar collisions,
valence quarks from the proton can annihilate with valence antiquarks
from the antiproton. This makes \qqbar annihilation more likely in
\ppbar collisions at the Tevatron compared to \pp collisions at the
\ac{lhc}.  The center-of-mass energy \sqrts of the collisions
determines at which momentum fraction$~x$ the partons in the initial
state hadrons are probed: to produce a \ttbar pair at rest, the
partonic center-of-mass energy $\sqrt{x_1 x_2 s}$ must be equal to
twice the top-quark mass.  For larger \sqrts, smaller~$x$ values and
larger momentum transfers get relevant, and it becomes increasingly
likely to probe a gluon inside the hadrons.  At the \ac{lhc} at
$\sqrts=\SI{13}{TeV}$, \ttbar production is dominated by $gg$ fusion
(approximately 90\%), while only 10\% of the \ttbar pairs are produced
via \qqbar annihilation.

At \ac{nlo}, \ttbar production processes with \xsec{}s proportional to
$\alphaS^3$ become relevant. These processes include higher-order
corrections to the \ac{lo} processes with the real emission of gluons
and virtual corrections. In addition new production channels open up:
Processes with $qg$ and $\qbar g$ initial states contribute for the
first time at \ac{nlo}. The ultraviolet divergences occurring in
\ac{nlo} calculations are systematically canceled by
renormalization. This introduces the renormalization scale \mur into
the calculation. Infrared and collinear divergences of the initial
state particles are systematically absorbed in the \acp{pdf},
introducing the factorization scale \muf into the calculation, as
discussed in Section~\ref{sec:highpt}. The inclusive \ttbar production
\xsec is known to \ac{nlo} accuracy since the late
1980s~\cite{Nason:1987xz,Beenakker:1988bq}. The first full \ac{nnlo}
calculation of the inclusive \ttbar production \xsec, \ie including
processes up to $\alphaS^4$, became available in
2013~\cite{Czakon:2013goa}.

The precision of \ttbar \xsec calculations can be further improved by
resumming contributions which may become large in certain areas of
phase space to all orders in \ac{qcd} perturbation theory. These may
include \eg the emission of soft gluons or effects at the kinematic
production threshold, where the velocity of the \ttbar pair
$\beta_\ttbar$ approaches zero, $\hat s \approx 2 \mt$. The leading
contributions at $n$-th order are proportional to
$\alphaS^n \ln(\dots)^{2n}$, hence they are often called \acf{ll}
contributions. Contributions at \ac{nll} order are proportional to
$\alphaS^n \ln(\dots)^{2n-1}$, etc. Logarithmic corrections to the
inclusive \ttbar production \xsec are known to \ac{nnll} order.  The
most precise prediction of the inclusive \ttbar production \xsec to
date (\ac{nnlo}+\ac{nnll}) reaches an uncertainty of less than
4\%~\cite{Czakon:2013goa}. Prior to the full \ac{nnlo} result several
\ac{nlo}+\ac{nnll} calculations were published, often refered to as
``approximate \ac{nnlo},'' as they already included important parts of
the \ac{nnlo} calculations.  Numerical access to the \xsec formulae
for \ttbar production as a function of \alphaS, \mur, and \muf and for
a given \ac{pdf} set is provided by software tools such as {\sc
  top++}~\cite{Czakon:2011xx} an {\sc
  Hathor}~\cite{Aliev:2010zk}. Differential \xsec{}s at approximate
\ac{nnlo} can be obtained from the {\sc DiffTop}
program~\cite{Guzzi:2014wia}.

Additional improvements to the \ac{sm} prediction of the \ttbar \xsec
are obtained by including electroweak corrections proportional to
$\alphaS^2\,\alpha$~\cite{Beenakker:1993yr,Bernreuther:2006vg,Kuhn:2006vh}. Another
approach is to consider the full process
$\pp \to W^+ b\,W^- \bbar + X$, \ie both \ttbar production and decay,
at \ac{nlo}, including all interference effects and kinematic
configurations in which only one or none of the top quarks is on its
mass shell~\cite{Denner:2010jp,Cascioli:2013wga}.

Top quarks and antiquarks produced in \ttbar pair production show only
very small polarization (approximately~1\%, depending on the initial
state and the choice of the quantization
axes~\cite{Bernreuther:2015yna}); however, their spins are
significantly correlated. The quantum-mechanical observable connected
to a spin is its projection to a quantization axis. The magnitude
of the \ttbar spin correlation effect depends on the choice of the
quantization axes (``spin basis''); therefore the spin basis is often
chosen to maximize the size of the effect.  One typical choice is the
beam basis, for which the quantization axis for both the top quark and
antiquark is the beam axis in the laboratory frame. One can show that
in the beam basis, the spins in $\qqbar \to \ttbar$ are 100\%
correlated close to the kinematic threshold, where $\beta_\ttbar$ is
close to zero. In the helicity basis, the quantization axes are the
flight directions of the $t$ and the \tbar in the \ttbar rest frame
and hence the spin projections are equal to the $t$ and \tbar
helicities. In the helicity basis $\qqbar \to \ttbar$ are 100\%
correlated for $\beta_\ttbar \to 1$. The process $gg\to\ttbar$ does
not show 100\% spin correlation for any choice of quantization axes,
as the $t$ and the \tbar carry like helicities for $\beta_\ttbar\to 0$
and opposite helicities for $\beta_\ttbar\to 1$.

\subsection{Single-Top Quark Production}
\label{sec:singletopprod}
\begin{figure}[t]
  \centering
  \includegraphics[width=0.25\textwidth]{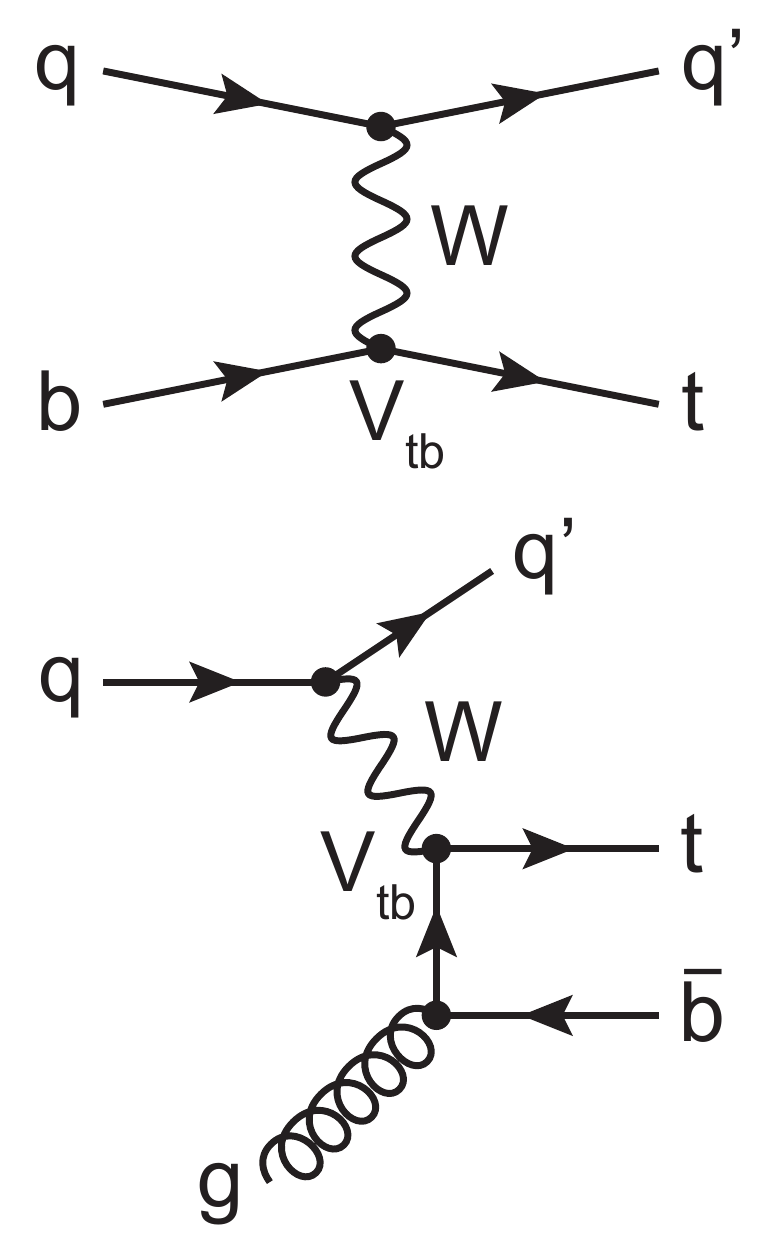}
  \hspace{3mm}
  \includegraphics[width=0.25\textwidth]{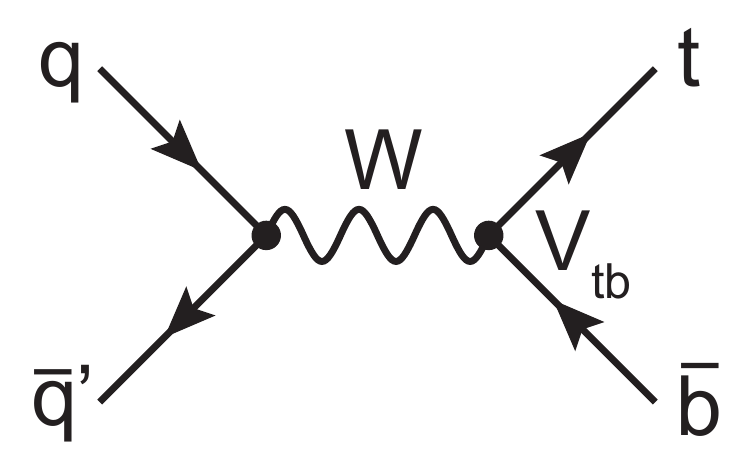}
  \hspace{3mm}
  \includegraphics[width=0.25\textwidth]{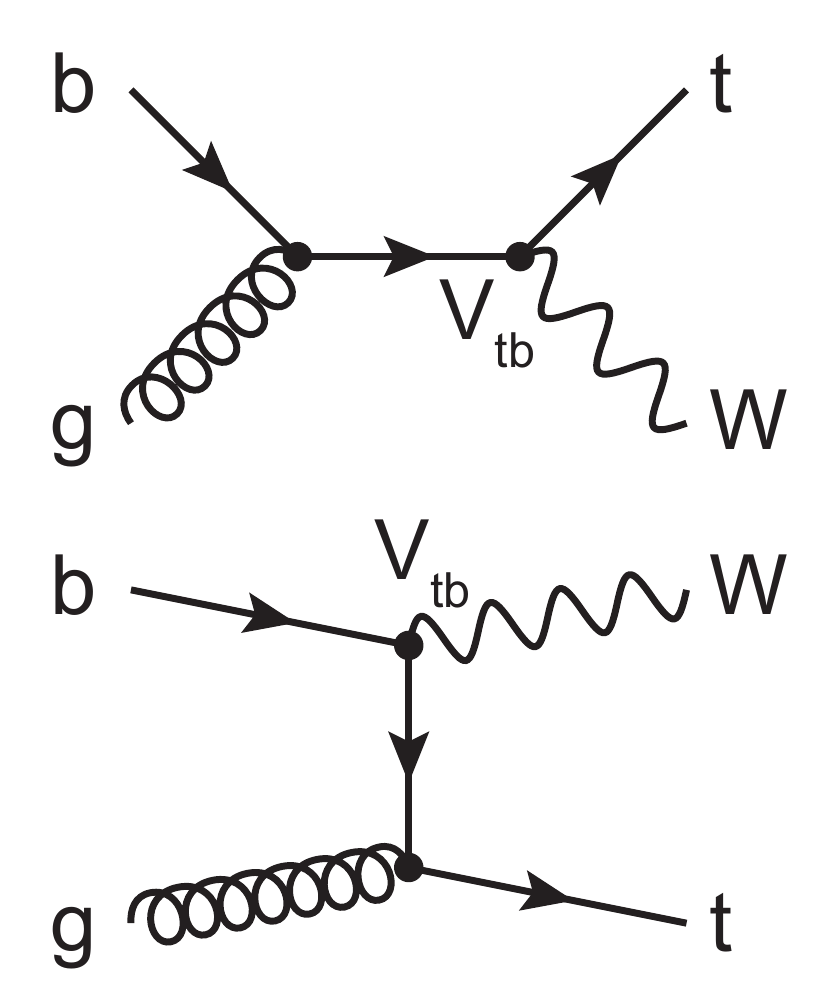}

  \caption{Feynman diagrams for electroweak \st production at \ac{lo}:
    \tch production in the \acl{5fs} and \acl{4fs} (left), \sch
    production (center), and associated $Wt$ production
    (right). Feynman diagrams created with {\sc
      JaxoDraw}~\cite{Binosi:2003yf}.}
  \label{fig:stopprod}
\end{figure}

Top quarks can also be produced singly in electroweak processes; the
inclusive \xsec{} is about two to three times smaller than for strong
\ttbar production. The production processes are classified by the
virtuality of the $W$ boson exchanged in the process. The most
abundant \st production process at the \ac{lhc} is \tch production
(\ac{sm} expectation: 70\% of the total \xsec), followed by the
associated production of a top quark and a real $W$ boson (25\%), and
$s$-channel production (5\%). At the Tevatron, 70\% $t$-channel and
30\% \sch \st production are predicted by the \ac{sm}, the $Wt$
contribution is negligible. \ac{lo} Feynman diagrams of these
processes are displayed in Fig.~\ref{fig:stopprod}. The electroweak
production vertex contains the CKM matrix element \Vtb. This offers
the opportunity to measure \Vtb directly in single top-quark
production. As the $W$ boson only couples to left-handed quarks and
right-handed antiquarks, the top (anti)quarks produced in the above
processes are 100\% polarized.

Single-top production in the \tch is mediated by a space-like virtual
$W$ boson. The process can be calculated in a scheme in which the
initial state $b$ quark originates from flavor excitation in the
proton (\acf{5fs}). The \ac{lo} production process in the \ac{5fs} is
$qb \to q't$. Alternatively, in the \acf{4fs} the \ac{lo} process is
the $2\to3$ process $qg \to q't\bbar$, where the initial-state gluon
splits into a \bbbar pair and one of the $b$ quarks interacts with the
virtual $W$ boson to produce the top quark. The ``spectator'' quark
$q'$ is typically emitted at rather small angles with respect to the
beam axis, resulting in one of the most striking signatures of \tch
\st production, a jet at large $|\eta|$.

Theoretically the \tch process is known differentially to \ac{nnlo},
first calculated assuming stable top
quarks~\cite{Brucherseifer:2014ama}, recently also for production and
decay~\cite{Berger:2016oht}. Earlier calculations of the \tch
production \xsec were performed at approximate
\ac{nnlo}~\cite{Kidonakis:2011wy}. The \ac{nlo} corrections to the
\ac{lo} \tch \xsec are ``accidentally'' small---of the order of a few
percent---and the \ac{nnlo} corrections are of the same order.

In \sch \st production a time-like virtual $W$ boson is
exchanged. The process is known to approximate
\ac{nnlo}~\cite{Kidonakis:2010tc}.  In associated $Wt$ production the
$W$ boson is real. This process is known to approximate \ac{nnlo}
accuracy~\cite{Kidonakis:2010ux}. \ac{nlo} corrections to $Wt$
production in which the intermediate top (anti)quark is on its mass
shell, called ``double resonant'' processes, share the same final
state with \ttbar production. In \ac{mc} generators this overlap can
be consistently removed by either removing all double-resonant
contributions (``diagram removal'') or by local cancellation of
double-resonant contributions via subtraction terms (``diagram
subtraction''). Both methods lead to comparable results. A more
comprehensive way of dealing with the overlap between $Wt$ and \ttbar
production is to consider the full process
$\pp \to W^+ b\,W^- \bbar + X$ at \ac{nlo}, as introduced in
Section~\ref{sec:ttbarprod}. Numerical access to \st production
\xsec{}s at fixed-order \ac{nlo} for all three production channels is
provided as part of the parton-level \ac{mc} generator
\ac{mcfm}~\cite{Campbell:2010ff} and recent versions of {\sc
  Hathor}~\cite{Kant:2014oha}.

\subsection{Summary}
In the \ac{sm}, the properties of the top quark are well defined. The
high-precision computations available for top-quark production and
many top-quark properties enable tests of the \ac{sm} as well as
searches for \ac{bsm} physics when confronted with measurements of
comparable precision. These measurements will be the subject of the
remainder of this review.

\section{Experimental Techniques in Top-Quark Physics}
\label{sec:techniques}

A typical hadron collider detector consists of a tracking detector, an
electromagnetic and a hadron calorimeter, and a muon detector, grouped
around the interaction point like the shells of an onion.  Momenta of
charged particles are determined by tracking their trajectories in
strong magnetic fields. In the calorimeters, particle energies are
determined in a destructive measurement from electromagnetic and
hadronic showers initiated by the particles. The raw signals of the
various subdetectors are further processed to reconstruct basic
analysis objects such as electrons, muons, and jets. The experimental
signatures of events with \ttbar pairs or single top quarks may
contain charged leptons, neutrinos, and jets, initiated from gluons,
light ($u$, $d$, $s$, $c$) quarks, or bottom ($b$) quarks. As a
consequence of this rich mixture of signatures all subdetectors of a
hadron collider detector are required in top-quark physics. Collision
events are selected according to the signatures expected from events
with top quarks such that these events are kept but background
processes with similar signatures are suppressed. Based on the event
selection higher-level data analysis methods are employed to obtain
physics results.

In this chapter some general aspects of the data analysis chain in
top-quark physics are reviewed. Note however that many details of the
analysis chain have to be tailored specifically to a given
measurement. After a sketch of the reconstruction of the basic
analysis objects, aspects of data selection and background suppression
methods are discussed. Techniques to reconstruct top quarks and
methods of statistical data analysis are also presented.

\subsection{Analysis Objects}

\subsubsection{Leptons}
\label{sec:leptons}

\paragraph{Electrons}
Electrons from the decay $W^+\to e^+\nu_e$ are reconstructed in the
tracking detector and electromagnetic calorimeter. The electrons are
expected to have large transverse momenta ($\gtrsim\SI{20}{GeV}$) and
be well isolated from other particles in the event. The isolation
requirement is fulfilled for events in which the sum of track momenta
or the sum of energy deposited in the calorimeter in a cone around the
electron (excluding the energy of the electron and of bremsstrahlung
photons) is below a threshold. A further selection (``electron
identification'') is applied to distinguish electrons from other
particles with similar detector signatures, \eg charged pions. At the
\ac{lhc} the electron identification is typically based on information
on the shapes of energy clusters in the calorimeter, assisted by
tracking information. To achieve optimal separation, this information
is often processed in multivariate methods, which will be introduced
in Section~\ref{sec:multivariate}. The energy scale of electrons is
typically calibrated against the invariant mass of well-known \ac{sm}
particles, such as the quarkonia \jpsi and $\Upsilon$, and the $Z$
boson.

\paragraph{Muons}
Muons are reconstructed in the tracking detector and the muon
detector, but leave only little energy in the calorimeters. Similar to
the electron selection, the selection of muons from the decay
$W^+\to \mu^+\nu_\mu$ is based on their large transverse momenta and
isolation, combined with---typically multivariate---muon
identification criteria.

\paragraph{Efficiency Determination}
The efficiencies for lepton reconstruction and identification can be
determined from correlated pairs of leptons in $\jpsi\to\lplm$ or
$Z\to\lplm$ events using a tag-and-probe method, see
\eg\cite{Chatrchyan:2012xi}. One lepton (``tag'' lepton) is selected
with strict criteria, the other lepton (``probe'' lepton) is selected
with looser criteria. The fraction of selected events in which both
the tag and probe lepton pass the reconstruction or identification
criteria is a measure of the corresponding efficiency.  The
tag-and-probe efficiencies, often determined as a function of the
lepton kinematics (transverse momentum and pseudorapidity), may differ
in data and simulated data. The simulated data are corrected for this
effect by applying appropriate scale factors defined as the ratio of
tag-and-probe efficiencies in data and simulated data.

\subsubsection{Jets, Missing Transverse Momentum, and Particle Flow}
\label{sec:b-tagging}

Jets are reconstructed combining the information of subdetectors,
typically the hadron and electromagnetic calorimeters and the tracking
detector. 

\paragraph{Jet Algorithms}
While at the Tevatron the jet reconstruction in top-quark
physics was mainly based on algorithms that define jets based on
geometric cones, the \ac{lhc} experiments use sequential recombination
jet algorithms, most prominently the \ak
algorithm~\cite{Cacciari:2008gp}. The size of a jet in $\eta$-$\phi$
space is characterized by the radius parameter
\begin{equation}
  R = \sqrt{\Delta\eta^2 + \Delta\phi^2},
\end{equation}
where $\Delta\eta$ ($\Delta\phi$) is the distance from the jet axis in
pseudorapidity (azimuthal angle)\footnote{Sometimes the rapidity $y$
  is used instead of the pseudorapidity $\eta$.}. In top-quark physics
at \ac{lhc} Run~2 the radius parameter of the \ak algorithm is chosen
to be $R=0.4$.  The \ak algorithm fulfills the requirements of
infrared and collinear safety: the same set of jets is reconstructed
in an event if an additional particle with very low momentum or at
very small angle to another particle is added to the event.  

\paragraph{Jet Energy Scale and Resolution}
Due to the non-linear detector response to jets, the \ac{jes} must be
calibrated carefully, typically with a combination of simulation-based
and data-driven methods. The \ac{jes} calibration performed to correct
the jet response in the \ac{lhc} data of Run~1 is discussed in great
detail in~\cite{Aad:2014bia,Chatrchyan:2011ds}. Also data-simulation
discrepancies in the \ac{jer} are corrected for, usually by smearing
the momenta of simulated jets to match the resolution observed for
reconstructed jets in the data.

\paragraph{B-Jets and B-Tagging}
Top-quark decays in the dominant mode $t\to W^+ b$ always produce a
bottom quark in the final state, which subsequently hadronizes into a
\bhad. Jets containing \bhad{}s (``\bjet{}s'') can be identified by
dedicated \btagging algorithms. These algorithms are based on the
distinctive properties of \bhad{}s such as their long lifetime of the
order of picoseconds, their high mass of the order of \SI{5}{GeV}, or
their semileptonic decays $B\to\ell\nu\,X$. Experimental signatures
related to the long lifetimes include secondary vertices with large
displacement from the primary collision vertex or charged-particle
tracks with large impact parameters relative to the primary
vertex. The high mass of \bhad{}s results in ``broader'' jets compared
to jets from light quarks and high relative \pt of the lepton in
semileptonic decays. The lepton from $B\to\ell\nu\,X$ is typically
non-isolated and carries rather low absolute \pt (``soft lepton''). In
addition the fragmentation of $b$ quarks is said to be ``hard'': the
\bhad carries a large fraction of the $b$-quark energy.  Recent
\btagging algorithms at hadron colliders combine the available
information on jets with \bhad{}s in a single multivariate classifier.

The performance of a \btagging algorithm can be quantified by the
probability to correctly identify a jet coming from a $b$ quark as a
\bjet and by the probability to wrongly identify a jet from a
light-flavor quark or a gluon (``mistag'').  A \btagging classifier
can either be used by assigning \btag{}s to all jets that show
classifier values above standardized working points with a fixed
mistag probability or by exploiting the full shape of the classifier's
distribution. As the \btagging and mistag efficiency may be different
in data and simulated data, $b$-tagging algorithms must be calibrated,
such that the simulation can be corrected with scale factors. Datasets
enriched with \ttbar events are well suited for such a calibration, as
they contain two jets with \bhad{}s from the \ttbar decay, see also
Section~\ref{sec:tool}.

\paragraph{Missing Transverse Momentum}
The entire detector is required to reconstruct the \acf{met} caused
for example by the undetected neutrino(s) from leptonic $W$-boson
decays. To calibrate the \ac{met} reconstruction, the calibration of
all other analysis objects must be known.

\paragraph{Particle Flow}
In the CMS experiment, the reconstruction of analysis objects follows
the particle-flow approach~\cite{CMS:2010eua}. For each object type
the optimal combination of subdetectors is chosen to determine its
four-momentum. One benefit of this approach is the improved jet energy
and \ac{met} resolution: The energies of all charged particles in a
jet are inferred from their momenta, which are very precisely measured
by the tracking detectors, and only the neutral hadron energies have
to be reconstructed in the low-resolution hadron calorimeter.

\subsection{Data Selection}
\label{sec:selection}
The \xsec{}s for top-quark production are about nine orders of
magnitude lower than the inelastic $pp$ scattering \xsec. Many other
\ac{sm} processes have \xsec{}s larger than the \ttbar or \st
production \xsec. These processes contribute to the background in a
given production and decay channel if they have similar experimental
signatures. The signal-to-background ratio\footnote{The
  signal-to-background ratio is the ratio of signal and background
  events in a given data sample. Another way of expressing the
  separation of signal and background is the signal purity, defined as
  the fraction of signal events in a sample containing both signal and
  background events.}  for top-quark events in a hadron collider data
sample is improved by a multistage online and off\/line data
selection.

\subsubsection{Preselection}
The online data selection is performed by a multilevel trigger system,
where the first step is usually implemented in custom-made electronics
and later steps are implemented in software on large computing
farms. The main trigger paths\footnote{A trigger path is a combination
  of triggers at different levels to select a specific set of trigger
  objects.}  used in top-quark physics consist of triggers that select
one or more isolated electrons or muons above a threshold in
transverse momentum~\pt. These trigger paths enable the efficient
selection of single-lepton and dilepton \ttbar events as well as \st
events. Further possible trigger paths include combinations of
triggers sensitive to a large number of high-\pt jets and
\ac{met}. The trigger efficiency is determined both in data and
simulated data, for example using a tag-and-probe method, see
Section~\ref{sec:leptons}. Any difference is corrected for with
appropriate scale factors applied to the simulation.

The next step in the data selection is the preselection of
high-quality collision events with all relevant detector parts
operational, a suitable trigger fired and a primary vertex
successfully reconstructed. Events containing signals not from
beam-beam collisions, such as beam halo, cosmic rays or coherent noise
in the detector, are vetoed.

\subsubsection{Event Selection and Major Backgrounds}
\label{sec:event_selection}
A further selection step is required to separate signal events with
top quarks from background events originating from other physics
processes. The signatures of signal events and the most important
background processes depend on the production and decay channel
considered.  The event selection criteria may include cuts on the
minimum (and/or maximum) number of leptons, jets, and \bjet{}s. The
selection may also exploit the specific kinematic properties of these
objects or of global observables in top-quark events, such as
\ac{met}.

In top-quark decay channels containing one or more charged leptons
background processes may be separated in those with real isolated
charged leptons and those in which other objects are misidentified as
charged leptons (``fake leptons''). One prominent example is \ac{qcd}
multijet production: due to the large production \xsec even the small
fraction of jets misidentified as charged leptons contributes to the
background. Therefore the \ac{qcd} multijet background is often hard
to estimate, as will be discussed in~\ref{sec:background}.

\paragraph{Dilepton Channel}
In the \ttbar dilepton decay channel, exactly two isolated high-\pt
leptons with opposite charge signs are selected (\epem, \mumu, or
$e^\pm\mu^\mp$), which strongly suppresses \ac{sm} backgrounds. Events
with a same-flavor lepton pair with an invariant mass around the
$Z$-boson mass, which occur in the associated production of $Z$ bosons
and jets (``\zjets''), are rejected.  One or two \btagged jets may be
required to further suppress background containing light-flavor or
gluon jets.

The most important real-lepton backgrounds for \ttbar dilepton events
with same-flavor lepton pairs are $\gamma^*/Z\to\epem/\mumu$+jets
(``Drell-Yan'') and associated $Wt$ production. For leptons with
different flavors, also background from $Z\to\tautau$+jets production
with leptonic $\tau$ decays becomes relevant. Background processes
with one or more misidentified leptons include the production of
$W$~bosons in association with jets (``\wjets'') and \ac{qcd} multijet
production. Overall the dilepton channel has the smallest branching
fraction but the most favorable signal-to-background ratio of all
\ttbar decay channels.

\paragraph{Single-Lepton Channel}
The most striking feature of \ttbar pairs decaying in the
single-lepton channel is a single isolated high-\pt lepton. The event
selection requires this lepton and at least three or four high-\pt
jets. Further requirements may be a significant amount of \ac{met}
from the neutrino and one or two \btagged jets.

The background level in the single-lepton channel is moderate. The
background is composed of processes with isolated high-\pt leptons,
such as \st production, \wjets and \zjets production, and production
of electroweak boson pairs, $WW$, $WZ$, and $ZZ$. \ac{qcd} multijet
background also contributes to the background in the single-lepton
channel if one of the jets is misidentified as a charged lepton.

\paragraph{Fully Hadronic Channel}
The signature of \ttbar events in the fully hadronic decay channel
consists of six jets, two of which originating from \bhad{}s. Unlike
the channels discussed above, a selection of isolated leptons cannot
be used to suppress background in this channel. Therefore the
fully-hadronic channel suffers from large \ac{qcd} multijet
background, which may be somewhat reduced by requiring two jets to be
\btagged.

\paragraph{Single-Top Production}
In all \st production channels, the top-quark decay
$t\to W^+ b\to \ell^+\nu_\ell\, b$ is considered, requiring a high-\pt
lepton and large \ac{met} as well as a \btagged jet from the top-quark
decay.  The additional signature of \tch \st production is a light jet
with large $|\eta|$ (see Section~\ref{sec:singletopprod}).  In the
\sch an additional \bjet is expected so that a second \btag is usually
required.  The additional $W$ boson in the final state of associated
$Wt$ production is usually required to decay leptonically. Therefore
the characteristic signature of the \Wtch is two oppositely charged
high-\pt leptons---similar to the \ttbar dilepton channel but with
only one \bjet. In all \st production channels, \ttbar production is a
major background. Other backgrounds include \wjets, \zjets, $WW$, $WZ$
and $ZZ$ processes as well as \ac{qcd} multijet events in which jets
are misidentified as leptons. The relative importance of the
backgrounds depends on the channel.

\subsubsection{Background Estimation Techniques}
\label{sec:background}
The background remaining after the event selection can be controlled
with various techniques. While some of the techniques are specific to
certain analyses, there are also some recurring concepts. These will
be discussed in this section.

The level of background from processes that are known to be
well-modeled in the \ac{mc} simulation is estimated directly from
simulated events. In case the inclusive \xsec predicted by the \ac{mc}
simulation does not match the most precise calculations, the simulated
events are often scaled such that their integral matches the number of
events expected from the calculation for a given integrated
luminosity\footnote{This procedure changes the production rates of
  processes but neglects potential differences in the shapes of
  kinematic distributions due to higher-order corrections.}.

Background processes for which the simulation has known deficiencies
or for which it is difficult to populate the relevant parts of phase
space with a sufficient number of simulated events are often estimated
from the data itself. The data is split into a signal-enriched signal
region and one or more signal-depleted control regions. This split can
either be based on the event kinematics or on the analysis-object
selection in the same kinematic region. The background rate is
determined in the control region(s), often by a \ac{ml} fit (see
Section~\ref{sec:likelihood}) to the data in which the signal and the
relevant background rates are free parameters. The background rate is
then translated to the signal region using the \ac{mc} simulation.
Such data-driven or data-assisted procedures to determine the
background result in estimates of the background rate and often also
of further properties, \eg shapes of kinematic distributions.

A background process often estimated from data is \ac{qcd} multijet
production. Due to the large \xsec and the small misidentification
probability of jets as charged leptons, simulations of multijet events
often do not provide a reliable estimate of the background in events
with top quarks. Instead a model of misidentified jets is built from
electron or muon candidates in the data for which one or more of the
lepton identification criteria failed. Events with such lepton
candidates form a disjoint set of events with kinematic properties
that closely resemble those of the \ac{qcd} background events passing
the event selection.  The normalization of the \ac{qcd} background is
obtained from an \ac{ml} fit to control regions, while the shapes of
kinematic distributions are taken from the model in the signal region.

\subsection{Top-Quark Reconstruction}
In many top-quark physics analyses it is desirable to reconstruct the
four-momenta of the top (anti)quarks from the leptons, jets, and
\ac{met} observed in an event. However, a one-to-one correspondence
between parton-level objects such as the top quarks and their decay
products and reconstruction-level objects such as leptons and jets
only exists in a crude \ac{lo} picture of hadronic collisions. Beyond
\ac{lo} this picture is complicated \eg due to additional jets from
gluon radiation with large \pt and/or at large angles with respect to
the original parton.  Apart from this conceptual question, the
top-quark reconstruction faces problems such as underdetermined
kinematics due to neutrinos, and the combinatorics of assigning jets
to partons from the top-quark decay (usually the bottom quarks from
top decays and the light-flavor quarks from hadronic $W$ decays).

\subsubsection{Neutrino Reconstruction}
In single-lepton and dilepton final states, one or two neutrinos from
$W$-boson decays escape the detector undetected. The only kinematic
observable available is \metvec, a two-vector in the transverse plane. In
single-lepton \ttbar events, a single neutrino is the only (real)
source of \ac{met}, however, $p_{z,\nu}$, the $z$ component of its momentum,
remains unknown. Using the $W$-boson mass as a constraint and
neglecting the lepton and neutrino masses, two solutions for \pz can
be obtained by solving the quadratic equation
\begin{equation}
(p_\ell + p_\nu)^2 = p_W^2 = m_W^2,
\end{equation}
where $p_\ell$, $p_\nu$, and $p_W$ are the four-momenta of the charged
lepton, the neutrino, and their parent $W$ boson. For the two
neutrinos in dilepton final states the kinematic system is
underdetermined and additional assumptions have to be made, see
\eg\cite{Sonnenschein:2006ud}.

\subsubsection{Jet-Parton Assignment and Kinematic Fitting}
In an \ac{lo} picture each jet can be assigned uniquely to one parton
from the top-quark decay. However, the correct assignment is
unknown. For example, in a single-lepton \ttbar decay with two bottom
quarks and two light quarks from the hadronic $W$ boson decay, there
are $4!=24$ possible permutations of jet-parton assignments. This
number is reduced to 12~permutations because exchanging the assignment
of the light quarks from the hadronic $W$-boson decay does not change
the event kinematics. The combinatorics can be further reduced if
b-tagged jets are always assigned to the bottom quark or
antiquark.

There are various ways to pick the ``best'' permutation of jet-parton
assignments in an event. A popular method is to construct a figure of
merit based on a $\chi^2$-like variable that compares the invariant
top-quark and $W$-boson masses with their nominal values. The mass of
the semileptonically decaying top quark is reconstructed from the
invariant mass of a charged lepton, a neutrino and a \bjet,
$m_{\ell\nu b}$. The mass of the hadronically decaying $W$ boson is
inferred from the invariant mass of two jets, $m_{jj}$, and the mass
of the hadronically decaying top quark from the three-jet mass
$m_{jjj}$\footnote{Note that the correlation between $m_{jj}$ and
  $m_{jjj}$ is ignored by considering the two observables separately.}.
The individual terms are usually weighted with factors $1/\sigma^2$,
which contain the widths of the invariant mass distributions, for the
semileptonic top-quark decay ($\sigma_{\mt,\mathrm{lep}}$), the
$W$-boson decay ($\sigma_{\mW,\mathrm{had}}$) and the hadronic
top-quark decay ($\sigma_{\mt,\mathrm{had}}$), determined from the
correct jet-parton assignment in \ac{mc}-simulated events:
\begin{equation}
\chi^2 = \frac{(m_{\ell\nu b}^2 - \mt)^2}{\sigma_{\mt,\mathrm{lep}}^2} +
\frac{(m_{jj}^2 - \mW)^2}{\sigma_{\mW,\mathrm{had}}^2} +
\frac{(m_{jjj}^2 - \mt)^2}{\sigma_{\mt,\mathrm{had}}^2}.
\label{eq:chi2}
\end{equation}
The jet-parton assignment can also be performed using machine-learning
techniques that are trained on simulated data to pick the ``best''
permutation according to a more sophisticated figure of merit. As an
alternative to picking the ``best'' permutation of jet-parton
assignments, also all permutations can be considered, weighted by
their probability to be the ``best'' permutation, determined from
\ac{mc} simulations.

A more precise method to reconstruct the kinematics of \ttbar events
is based on the observation that the resolution of certain kinematic
observables is limited and the observables may hence be slightly
mismeasured. In a kinematic fit, the figure of merit for the ``best''
jet-parton permutation is parameterized as a function of those
kinematic observables that can only be reconstructed with limited
resolution. Examples of such observables include the jet energies and
directions and \metvec. Each of these observables is allowed to be
varied within its resolution in the kinematic fit. This way, the
kinematic fit adjusts the event kinematics to find the optimum figure
of merit for a given permutation, before selecting the ``best''
permutation, or weighting all permutations.

\subsubsection{Boosted Top-Quark Reconstruction}
\label{sec:boosted}
\begin{figure}[t]
  \centering
  \includegraphics[width=0.4\textwidth]{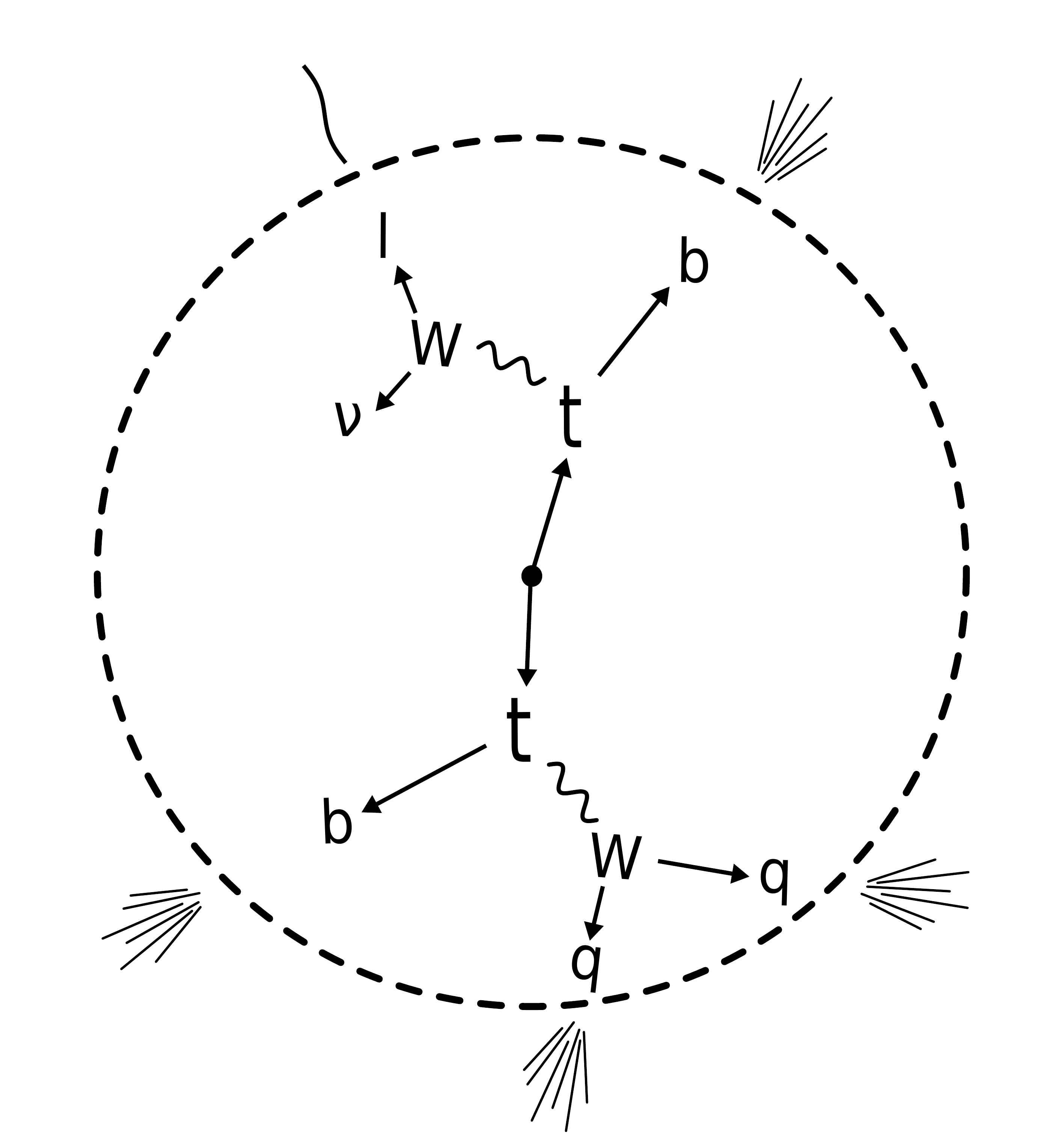}
  \hspace{5mm}
  \includegraphics[width=0.4\textwidth]{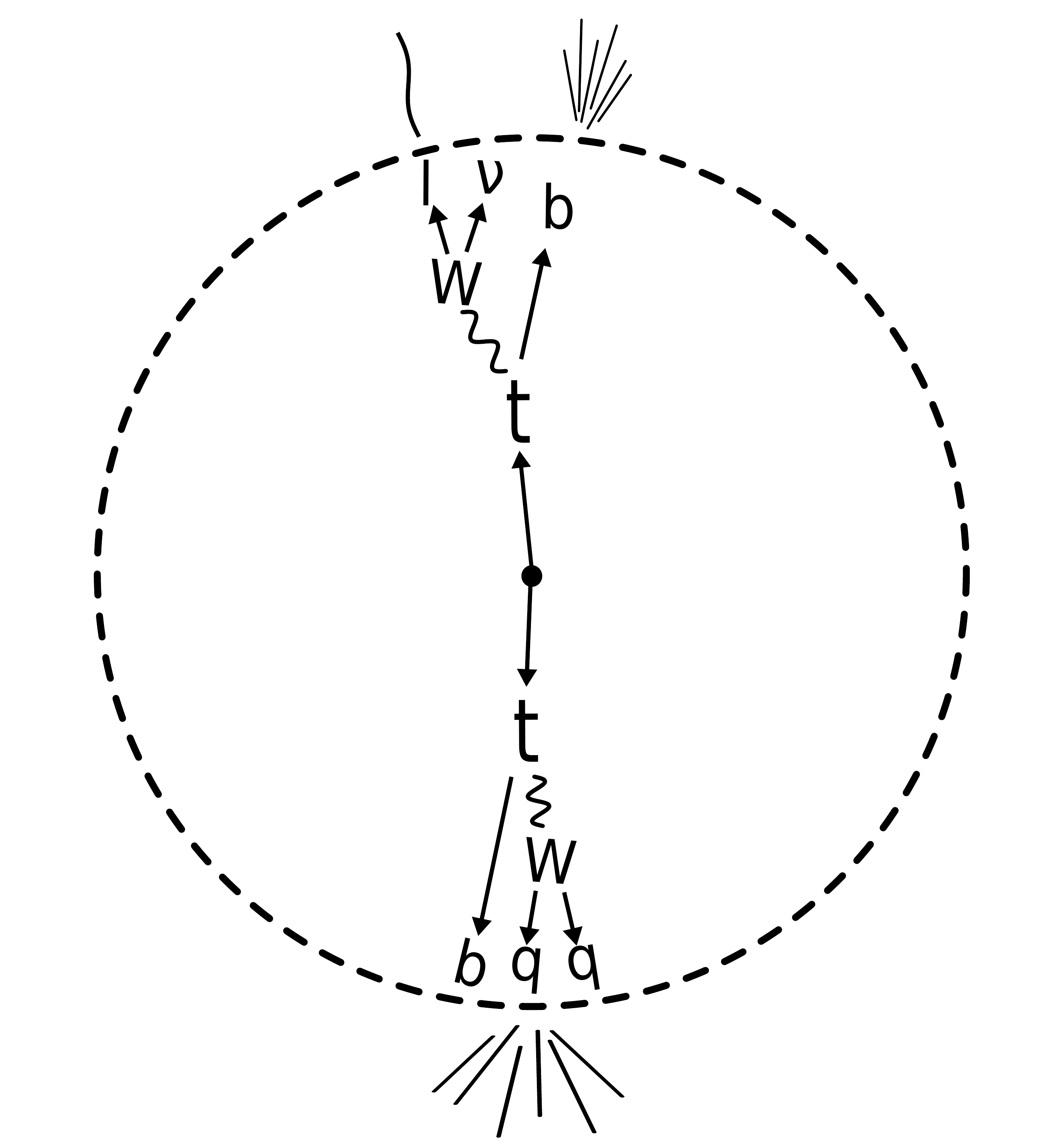}

  \caption{Illustration of resolved (left) and boosted (right) event
    topologies in single-lepton \ttbar decays. Picture courtesy of Shawn
    Williamson.}
  \label{fig:boost}
\end{figure}

Top quarks produced in high-energy collisions, \eg at the current LHC
Run~2, may receive large momenta, either in regular \ac{sm} processes
or by hypothetical high-mass particles decaying to top quarks.  For
top quarks with $\pt\gtrsim\SI{200}{GeV}$ the decay products (jets and
leptons) start becoming collimated, such that they begin to overlap in
$\eta$-$\phi$ space. This is illustrated in Fig.~\ref{fig:boost}. Such
topologies require specialized ``boosted-jet'' reconstruction
algorithms. The boosted-jet reconstruction comes with the additional
benefit that the combinatorial problem of jet-parton assignment is
mitigated, as events with boosted analysis objects contain fewer (but
more complicated) reconstructed objects than events in which all jets
can be resolved.

In the last decade a large number of algorithms was conceived to
analyze boosted-jet topologies. In these algorithms, jets are first
reconstructed with large radius parameters (``fat jets''),
$R = 0.8-1.5$. As a second step, the substructure of the fat jets is
examined. The main classes of jet-substructure algorithms are
jet-declustering algorithms and jet-shape algorithms. The key idea of
declustering algorithms is to undo the last steps of the jet
clustering algorithm to identify those structures in the jet related
to the decay of the mother particle. Grooming techniques such as
pruning or trimming remove soft radiation uncorrelated to the decay
particles. This results in better resolution for the jet mass and
reduced pile-up\footnote{At high-luminosity hadron colliders, the
  hadron-hadron collision of interest is overlaid by other collisions
  in the same bunch crossing as well as the ``afterglow'' of
  collisions from previous bunch crossings in the detector. These are
  jointly referred to as pile-up.} dependence. Jet shape algorithms
such as $n$-subjettiness~\cite{Thaler:2010tr} assign a probability to
a fat jet to stem from $n$ overlapping jets. Combinations of several
techniques can also be used. The exact choice of algorithm depends on
the expected event topologies and typical momentum range of the
boosted objects. Reviews of jet-substructure algorithms can be found
\eg in~\cite{Salam:2009jx,Altheimer:2012mn,Adams:2015hiv}.

\subsection{Statistical Methods}
\label{sec:statistics}
Analysis of top-quark events often relies on advanced statistical
methods, many of which are based on the \acf{ml}
principle. Sophisticated statistical methods are employed in several
analysis steps, in order to maximize the precision of measurements or
the sensitivity of searches for \ac{bsm} physics. Examples include the
selection of physics objects and entire events (see
Section~\ref{sec:selection}), the classification of the selected events
as either signal or background, and the extraction of physics
information such as model parameters from the data.

The methods relevant for top-quark physics are often made available in
the {\sc C++}-based data analysis framework {\sc
  Root}~\cite{Brun:1997pa}, which is very widely used in particle
physics. It is beyond the scope of this review to explain the methods
and tools in detail. The interested reader is referred to text books
such as~\cite{Cowan:1998ji}.

\subsubsection{Maximum-Likelihood Method}
\label{sec:likelihood}
A typical task in physics data analysis is to extract model parameters
from data. The \ac{ml} method solves this task by first constructing
the likelihood function $\mathcal{L(\vec{\mu})}$ as the product of
probability densities $P(\vec{x}_i|\vec{\mu})$ for single measurements
$\vec{x}_i$ given the true parameter vector $\vec{\mu}$:
\begin{equation}
  \mathcal{L(\vec{\mu})} = \prod_i P(\vec{x}_i|\vec{\mu}).
\end{equation}
The \ac{ml} estimator of $\vec{\mu}$ is the maximum of the likelihood
function, usually determined by minimizing
$-\ln \mathcal{L(\vec{\mu})}$.  For discretized (``binned'') data
distributions, $P$ is the Poisson distribution of the number of events
in each bin given the number of events expected from the model. The
model expectation is usually obtained from simulated data and
represented as bins of a template histogram.  The model parameters
$\vec{\mu}$ estimated with the \ac{ml} method are the normalizations
of the histograms, which are in turn proportional to the total number
of events expected from the model. Unbinned data distributions can be
fitted with continuous probability density distributions, \eg
constructed by kernel-density estimates~\cite{Cranmer:2000du}.

In more sophisticated \ac{ml} models used with binned data further
parameters are added that describe the influence of systematic
uncertainties on the normalizations and shapes of the template
histograms. The model parameters are then split into the
``parameter(s) of interest'' $\vec{\beta}$ and additional ``nuisance
parameters'' $\vec{\delta}$.  In a Bayesian approach, a-priori
knowledge, for example from auxiliary measurements, is used to
constrain $\vec{\delta}$.  To obtain an estimate of the parameters of
interest and their uncertainties, the nuisance parameters can be
either profiled or marginalized. Profiling means that the profile
likelihood ratio
\begin{equation}
  \lambda(\vec{\beta}) = 
  \frac{\mathcal{L(\vec{\beta},\hat{\hat{\vec{\delta}}}})}
  {\mathcal{L(\hat{\vec{\beta}},\hat{\vec{\delta}}})}    
\end{equation}
is minimized instead of the original likelihood. The numerator of the
profile likelihood ratio is the minimum of the likelihood function at
a fixed value of $\vec{\beta}$, where the nuisance parameters assume
the values $\hat{\hat{\vec{\delta}}}$, the denominator is the global
minimum of the likelihood function, with parameter values
$\hat{\vec{\beta}}$ and $\hat{\vec{\delta}}$. In the marginalization
approach the likelihood function is integrated numerically, typically
with \ac{mc} methods. The parameters of interest are then extracted
from the projections of the likelihood function on these parameters
(``marginal distributions'').

A frequentist method to deal with systematic uncertainties is to
perform ensenble tests by drawing pseudo-experiments (also: ``toy
experiments'', ``MC experiments''). Many random variations of
distributions are generated and the entire analysis chain is performed
on each variation. The variance of the results is a measure of the
uncertainty.  Examples of software tools used in top-quark physics
that include the above sophisticated \ac{ml} methods are {\sc
  RooFit/RooStats}~\cite{Verkerke:2003ir,Moneta:2010pm} shipped with
{\sc root}, and {\sc theta}~\cite{Ott:2010}.

One way of interpreting measurements of top-quark properties is to
compare the \ac{bsm} physics prediction for an observable with the
corresponding \ac{sm} prediction. The statistical method applied in
the comparison is called hypothesis test. First the null hypothesis
$H_0$ (\eg\ac{sm}) and the alternative hypothesis $H_1$ (\eg\ac{bsm})
are formulated and a test statistic is constructed that is able to
discriminate between $H_0$ and $H_1$. A popular choice of the test
statistic is the ratio of likelihoods for the vector of measurements
$\vec{x}$ given $H_0$ or $H_1$:
\begin{equation}
  r(\vec{x}) = \frac{L(\vec{x}|H_0)}{L(\vec{x}|H_1)}.
\end{equation}
From the observed value of the likelihood ratio $r_\mathrm{obs}$, the
significance for the hypotheses is obtained.

\subsubsection{Multivariate Classification}
\label{sec:multivariate}
The selection of analysis objects and the classification of events as
signal-like or background-like is often performed using methods from
(supervised) machine learning.  Such methods use simulated data to
teach (``train'') an algorithm how to distinguish signal from
background processes based on a non-linear combination of several
input variables. It is important for these methods not to generalize
peculiar features of the simulated data used for the training to the
entire sample (``overtraining''). Among the many methods available in
the statistics literature (see \eg\cite{Hastie:2009}), the most
popular in top-quark physics are \acp{nn} and \acp{bdt}.
Currently the main tool employed in the top-quark
physics community is the Toolkit for Multivariate Data Analysis ({\sc
  tmva})~\cite{Hocker:2007ht} which is shipped with {\sc
  Root}. Alternatives include the {\sc Python} package {\sc
  scikit-learn}~\cite{Pedregosa:2012toh}, and the commercial \ac{nn}
package {\sc NeuroBayes}~\cite{Feindt:2006pm}.

\subsubsection{Matrix-Element Method}
\label{sec:mem}
The entire parton-level kinematics of a physics process is contained
in the squared scattering amplitude of the process, also called the
(hard) matrix element. The \acf{mem} is a method to construct an
event-based likelihood discriminant to separate signal from background
that fully exploits all information in the event by using the squared
matrix element~\cite{Kondo:1988yd}. Currently most \ac{mem}
implementations use matrix elements at \ac{lo} \ac{qcd} perturbation
theory; however, concepts to implement \ac{nlo} corrections into the
\ac{mem} have emerged recently, see
\eg\cite{Alwall:2010cq,Campbell:2012cz,Martini:2015fsa}. The
explanation below follows the review article~\cite{Fiedler:2010sg}.

For an event with a given set of reconstructed kinematic variables
$\vec{x}$ a likelihood function $L(\vec{x}|S)$ is constructed under
the hypothesis that the event is a signal event. Also for one or more
background hypotheses, likelihood functions $L(\vec{x}|B_i)$ are
constructed. These are combined for each event in a likelihood ratio
discriminant, \eg in the form
\begin{equation}
R(\vec{x}) = \frac{L(\vec{x}|S)}{L(\vec{x}|S) + \sum_i c_i L(\vec{x}|B_i)},
\end{equation}
where each background likelihood function can be assigned a different
weight $c_i$.  For a given signal or background hypothesis the
likelihood function is constructed from the sum of cross sections of
all sub-processes that lead to the parton-level final state $y$, with
kinematics $\vec{y}$, that could have lead to the reconstruction-level
final state $x$, with kinematics $\vec{x}$, using the \ac{qcd}
factorization approach (assuming $pp$ collisions):
\begin{equation}
\sigma(pp\to y) = \sum_{jk}^{\mathrm{partons}}
\int_0^1 \mathrm{d}z_j\,\mathrm{d}z_k
f_j(z_j) f_k(z_k)\, \frac{(2\pi)^4}{z_j z_k s} 
\left|M(jk\to y)\right|^2 \mathrm{d}\Phi.
\end{equation}
In the above equation, the sums are over the partons $j$ and $k$ and
the integrals are over their momentum fractions $z_j$ and $z_k$. The
parton distribution functions are denoted $f_i(z_i)$, and the hard
matrix element for the process leading to the parton-level final state
$y$ is $M(jk\to y)$. The Lorentz-invariant phase space measure is
symbolically written as $\mathrm{d}\Phi$. Note that the phase space
integral is numerically expensive as all unobserved variables in each
event (often of the order of 20) have to be integrated over.

To translate from the parton-level final state $y$ to the
reconstruction-level final state $x$, $\sigma(pp\to y)$ is folded with
a transfer function $W(\vec{x}|\vec{y})$:
\begin{equation}
\sigma(pp\to x) = \int \sigma(pp\to y)\, W(\vec{x}|\vec{y})\,\mathrm{d}\vec{y}
\end{equation}
The transfer function accounts for the limited detector resolution and
for the combinatorics of assigning reconstruction-level quantities to
partons and is determined from \ac{mc}-simulated data. The final
likelihood functions $L(\vec{x}|S)$ and $L(\vec{x}|B_i)$ are obtained
by normalizing the cross sections to the (fiducial) cross sections of
the processes.

\subsubsection{Unfolding Techniques}
\label{sec:unfolding}
Physics quantities reconstructed with a collider detector and
theoretical calculations of observables cannot be compared
directly. This problem can be solved in two ways. Either the
theoretical calculations are fed into a detailed simulation of the
detector and hence ``forward-folded'' into detector-related effects
such as limited acceptance and resolution. Alternatively the detector
effects can be removed from the reconstructed quantities by unfolding
techniques. In top-quark physics unfolding is typically applied in
measurements of differential \xsec{}s, see
Section~\ref{sec:differential}.

Mathematically the relation of reconstructed and ``true'' quantities
can be expressed in a Fredholm integral equation:
\begin{equation}
g(\vec{x}) 
= \int R(\vec{x}|\vec{y}) \, f(\vec{y}) \,\mathrm{d} \vec{y} + b(\vec{x})
= \int A(\vec{x}|\vec{y}) \, \epsilon(\vec{y}) \, f(\vec{y})
\,\mathrm{d} \vec{y} + b(\vec{x}),
\label{eq:fredholm}
\end{equation}
where $g(\vec{x})$ is the distribution of the reconstructed quantity
as a function of the set of kinematic variables $\vec{x}$, and
$f(\vec{y})$ is the ``true'' distribution from theory, depending on a
different set of kinematic variables $\vec{y}$. The reponse function
(also: transfer function) $R(\vec{x}|\vec{y})$, which may be written
as the product of an acceptance function $\epsilon(\vec{y})$ and a
resolution function $A(\vec{x}|\vec{y})$, parameterizes the detector
effects. In addition the background distribution $b(\vec{x})$ must be
considered. Unfolding means solving Eq.~(\ref{eq:fredholm}) for
$f(\vec{y})$, which is an ill-posed mathematical problem. The solution
chosen in particle physics analyses starts with discretizing the
distributions in bins of histograms:
\begin{equation}
g_i = \sum_{j=1}^{m} R_{ij} f_j + b_i.
\end{equation}
A straight-forward matrix inversion to solve for $f_j$ is not useful
in a physics analysis, because physics data always contain statistical
fluctuations, which cannot be distinguished from real structure in the
data without further assumptions. This leads to numerical
instabilities in the matrix inversion. Therefore regularization
techniques are applied that assume that distributions of physics
observables are ``smooth.'' Various regularization techniques are
discussed in the literature. Among the most popular in top-quark
physics are Tikhonov regularization, as \eg implemented in the {\sc
  Root} class \texttt{TUnfold}~\cite{Schmitt:2012kp}, and
regularization by singular-value decomposition as in
\texttt{TSVDUnfold}~\cite{Hocker:1995kb}.

Another approach employed in top-quark physics is called fully Bayesian
unfolding~\cite{Choudalakis:2012hz}. In this approach Bayesian
inference is applied to the unfolding problem and the probability
density of a true distribution $f(\vec{y})$ given the reconstructed
distribution $g(\vec{x})$ is obtained from Bayes' theorem:
\begin{equation}
  p(f|g) \propto 
  \mathcal{L}(g|f) \cdot \pi(f),
\end{equation}
where $\mathcal{L}(g|f)$ is the likelihood function of the measured
values $g$ given the true distribution $f$ and $\pi(f)$ is the prior
probability density of $f$. In this method backgrounds and systematic
uncertainties can be included consistently as described in
Section~\ref{sec:likelihood}.

\subsubsection{Statistical Combination of Measurements}
\label{sec:statcomb}
Statistical methods can be used to combine sets of measurements from
the same or from different experiments with the goal of reducing
uncertainties, see Section~\ref{sec:combination}. A simple
prescription for combining a set of measurements would be the weighted
arithmetic mean of the measured values, where the weights are the
inverse of the variance of the values. However, in all realistic
cases of top-quark physics, not only the statistical and systematic
uncertainties of the individual measurements must be considered, but
also their correlations.

Information on all uncertainties and their correlations is available
if the measurements are interpreted using the same \ac{ml} model. In
such a combination on the level of likelihood functions the model
parameters and their uncertainties are estimated from all data in a
consistent way. However, such an approach requires a large degree of
coordination between the individual measurements and may thus not
always be feasible, in particular when combining measurements from
different experiments. In this case the combination is often performed
on the level of measured values instead of likelihood functions, with
a reasonable guess on their covariance matrix. A popular combination
method for this purpose is called BLUE (best linear unbiased
estimator)~\cite{Lyons:1988rp}.

\section{Top-Quark Production}
\label{sec:production}

The measurement of the production \xsec of \ttbar pairs and single top
quarks constitutes a test of the \ac{sm} description of heavy quark
production. The level of understanding of top-quark production
increases with increasingly precise measurements and theoretical
calculations of the production processes. In this chapter, recent
top-quark production \xsec{} results from the Tevatron and the
\ac{lhc} will be reviewed, illustrating current experimental methods
and their precision. The presentation includes inclusive and
differential \ttbar and \st production \xsec{}s, and \xsec{}s for the
associated production of \ttbar plus ``something else'' ($\ttbar+X$),
such as jets, \acl{met}, photons, $W$ and $Z$ bosons, as well as Higgs
bosons.

\subsection{Observables and Measurement Techniques}

\paragraph{Inclusive Cross Section} 
The most inclusive observable to measure particle production is the
inclusive (also: total) production \xsec. The inclusive \xsec is a
measure of the production probability in the full kinematic phase
space of the production processes\footnote{In this context, phase
  space is understood as the space of all possible final state
  configurations in top-quark events with all possible four-momenta
  consistent all conservation laws.}. The first \xsec
measurements performed in the top-quark sector, both at the Tevatron
and the \ac{lhc}, were inclusive \ttbar \xsec{}s \sigtt. The
observables and techniques discussed in this section apply to other
production processes as well.

All \xsec measurements start with a basic selection of candidate
events to suppress background while retaining a large fraction of
signal events in the data sample, see Section~\ref{sec:selection}.
After the event selection the simplest way to extract \sigtt is to
perform a ``counting experiment'':
\begin{equation}
  \sigtt = \frac{N_\mathrm{top} - N_\mathrm{bkg}}{\Lint \cdot
    \epsilon},
  \label{eq:xsec}
\end{equation}
where $N_\mathrm{top}$ and $N_\mathrm{bkg}$ are the number of
top-quark events and background events, \Lint is the integrated
luminosity and $\epsilon$ the efficiency to detect top-quark events in
the full phase space. While counting experiments are simple and
robust, the need for absolute predictions of the signal efficiency and
background level limits their precision.  More precise \xsec results
can be obtained by exploiting the kinematic properties of the final
state particles. The shapes of kinematic distributions are determined
for the signal and all background processes and stored in discretized
form in template histograms. The sum of template histograms for the
signal and background processes is then fitted to the data using
\acf{ml} methods as described in Section~\ref{sec:likelihood}. Fitting
kinematic distributions in signal-enriched and background-enriched
regions simultaneously allows for better constraints on the background
level, resulting in reduced statistical uncertainty of the result.  At
the same time additional systematic uncertainties arise due to the
limited knowledge of the shapes of kinematic distributions. In more
sophisticated fitting procedures, also the shapes of kinematic
distributions are allowed to vary within their uncertainties. The
top-quark production \xsec and its uncertainties may then be
determined either from a multi-parameter profile likelihood ratio fit,
or the uncertainties are estimated using pseudo-experiments.

\paragraph{Fiducial Cross Section}
The need to know the absolute efficiency makes measurements of
inclusive \xsec{}s model-dependent.  The efficiency $\epsilon$ in
Eq.~(\ref{eq:xsec}) may be factorized into the detector acceptance and
the detection efficiency of final state particles within the detector
acceptance. While the detection efficiency can be calibrated using
data to high accuracy (see Section~\ref{sec:selection}), a
determination of the detector acceptance, \ie the ratio of detectable
events to all events, requires an (often large) extrapolation to the
full phase space of the \ttbar final state. The extrapolation is
usually performed using simulated data samples and hence depends on
the \ac{mc} model on which the simulation is based. This
model dependence can be reduced by measuring the cross section in a
restricted (``fiducial'') region of the phase space that closely
resembles the detector acceptance. Typical phase space requirements
include the detector's pseudorapidity range and a minimum transverse
momentum of analysis objects. The fiducial phase space is usually
defined in the \ac{mc} simulation on the particle level, after the
particles hadronize but before they decay. The fiducial cross section
can then be extrapolated to the full phase space by employing the
predictions of different \ac{mc} models.

\paragraph{Differential Cross Section}
Differential \xsec{}s are \xsec{}s as a function of one or more
kinematic observables. They allow more detailed insights into the
\ttbar production mechanism. In recent years, many differential
\xsec{} measurements have been performed in a fiducial region of phase
space to reduce their model dependence. Measurements of the
distribution of kinematic observables can be translated into
differential \xsec{}s using the unfolding techniques, such as those
described in Section~\ref{sec:unfolding}. After all detector-related
effects are removed by unfolding, differential \xsec{}s from different
experiments, if performed in the same fiducial phase space, can be
directly compared among each other. Differential \xsec{}s can also be
compared with predictions from \ac{mc} event generators (for
particle-level and parton-level measurements) or from theoretical
calculations (only for parton-level measurements). For comparisons at
particle level, the software framework {\sc
  Rivet}~\cite{Buckley:2010ar} is often used.

\subsection{Inclusive \ttbar Production}
\label{sec:inclusive}

\begin{figure}[t]
  \centering
  \includegraphics[width=\textwidth]{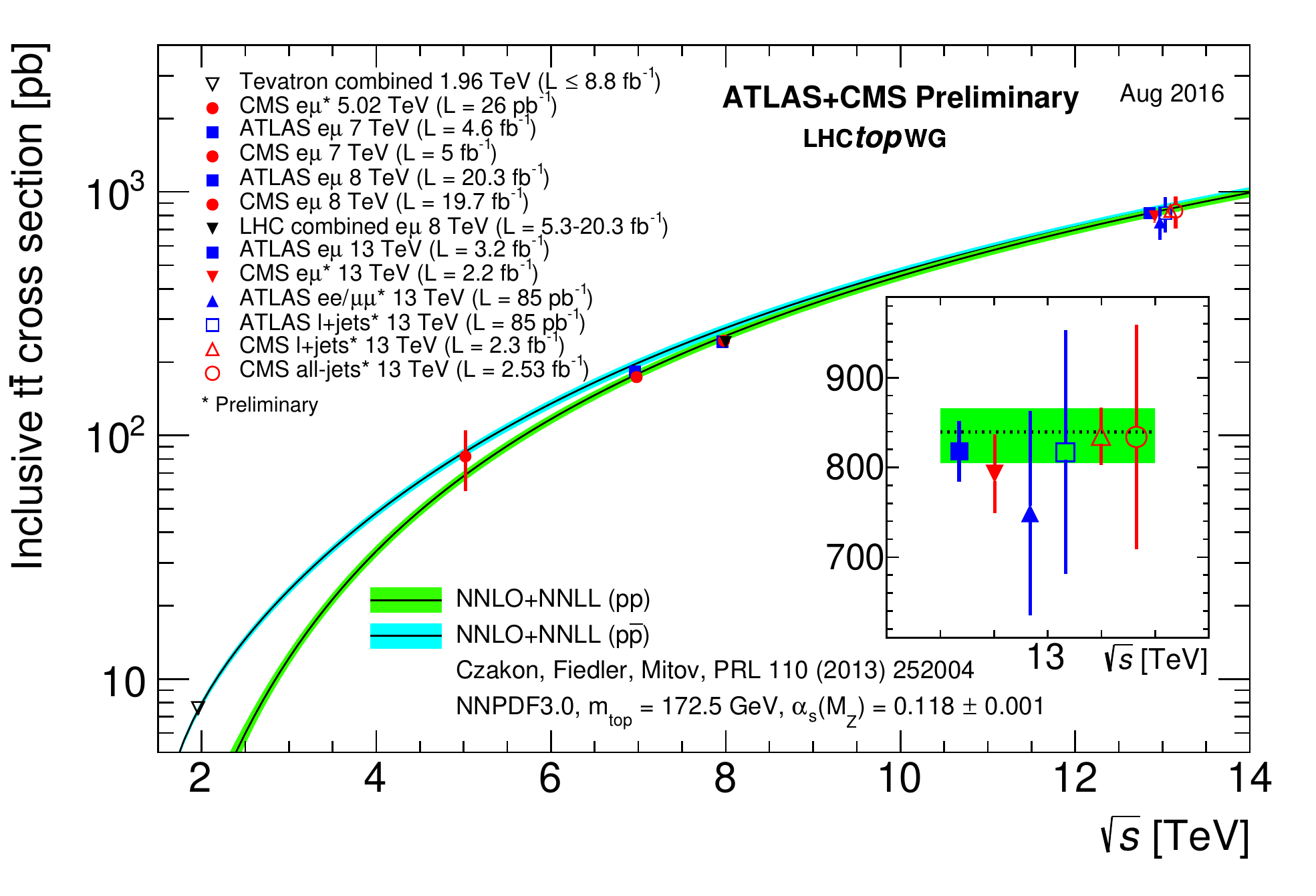}
  \caption{Compilation of measurements and \ac{sm} predictions of
    the inclusive \ttbar \xsec as a function of the center-of-mass
    energy for \ppbar collisions at the Tevatron and $pp$ collisions
    at the \ac{lhc}~\cite{LHCTopSummary}.}
  \label{fig:lhctopxsec}
\end{figure}

The inclusive \ttbar production \xsec has been measured for \ppbar
initial states at the Tevatron and for $pp$ initial states at the
\ac{lhc} as well as for various center-of-mass energies: \SI{1.8}{TeV}
and \SI{1.96}{TeV} for \ppbar collisions, and
\SI{5.02}{TeV}~\footnote{In November 2015 the \ac{lhc} delivered $pp$
  collisions at a center-of-mass energy of~\SI{5}{TeV} as part of its
  heavy-ion program. The inclusive \ttbar \xsec was measured using a
  data sample corresponding to an integrated luminosity
  of~\SI{26}{\per\pico\barn}~\cite{CMS:2016pqu}.}, \SI{7}{TeV},
\SI{8}{TeV}, and \SI{13}{TeV} for $pp$ collisions. These measurements
test the theoretical understanding of the dependence of the \ttbar
production \xsec on the initial state and the center-of-mass
energy. The most precise measurements of the \ttbar production \xsec
so far have been performed in the dilepton and single-lepton decay
channels. A summary compiled by the \ac{lhctopwg} is displayed in
Fig.~\ref{fig:lhctopxsec}, showing excellent agreement between
measurements and the most precise \ac{sm} predictions to date at
\ac{nnlo} with \ac{nnll} resummation~\cite{Czakon:2013goa}.

\paragraph{Tevatron Results}
At the Tevatron, a combination of inclusive \ttbar production \xsec
measurements from the CDF and D0 experiments has been performed,
resulting in a combined value of
\begin{equation}
  \sigma_\ttbar = \left( 7.60 \pm 0.20\,\mathrm{(stat)} \pm 0.36\,\mathrm{(syst)}\right)\,\si{pb},
\end{equation}
assuming a top-quark mass of
$\mt =\SI{172.5}{GeV}$~\cite{Aaltonen:2013wca}\footnote{As will be
  discussed in Section~\ref{sec:xsecmass}, the \ttbar production \xsec
  decreases steeply with increasing \mt. The detection efficiency is a
  function of \mt as well.  Therefore \ttbar \xsec measurements are
  quoted at a fixed value of \mt, usually the default value used in
  the \ac{mc} simulation. Sometimes the \mt dependence of the \xsec is
  quoted as well.}.  The combination achieves a precision of 5.4\%,
which is dominated by systematic uncertainties due to limitations in
signal and detector modeling, as well as the uncertainty of the
Tevatron luminosity.  The most precise individual measurements at the
Tevatron were obtained in the single-lepton
channel~\cite{Aaltonen:2010ic,Abazov:2016ekt}, while the precision in
the dilepton channel was limited by the rather small Tevatron
dataset~\cite{Aaltonen:2013tsa,Abazov:2016ekt}. The D0 experiment also
pioneered the extraction of the top-quark mass from the \ttbar \xsec,
as will be discussed in Section~\ref{sec:xsecmass}.

\paragraph{LHC: High-Precision Results}
Most recent \ac{lhc} measurements of the inclusive \ttbar \xsec are
first reported as fiducial cross sections and then extrapolated to the
full phase space with several \ac{mc} models. Given the large \ttbar
data samples recorded at the \ac{lhc} the precision of these
measurements is limited by systematic uncertainties. The most precise
\ac{lhc} measurements to date are summarized in
Table~\ref{table:xsec}.  The smallest systematic uncertainties in
\ac{lhc} Run~1 have been achieved in the $e\mu$ channel with its low
background level and small expected number of jets~\cite{Aad:2014kva,
  Khachatryan:2016mqs}. For example, the CMS
measurement~\cite{Khachatryan:2016mqs} is based on a simultaneous
binned profile likelihood ratio fit in 12~different event categories
in the $e\mu$ dilepton channel. The categories are defined by the
number of \btagged jets and the number of additional non-\btagged jets
in the events. In the categories with additional jets the fit is
applied to the \pt distribution of the non-\btagged jet with the
lowest \pt, while in the categories without additional jets a counting
experiment is performed by fitting the total event yield.

Major experimental uncertainties on the ATLAS and CMS high-precision
measurements originate from the determination of the
luminosity\footnote{At the level of precision achieved in measurements
  of the \ttbar production \xsec at the \ac{lhc} also the uncertainty
  on the beam energy becomes relevant. So far, this is only taken into
  account by ATLAS.}, the trigger and lepton identification
efficiencies, and the estimation of background from $\gamma^*/Z$+jets
events. Another class of uncertainties, particularly important when
fiducial \xsec{}s are extrapolated to the full phase space, is due to
the \ttbar modeling in the \ac{mc} simulation.  Adding all
uncertainties in quadrature, both ATLAS and CMS arrive at very similar
total uncertainties below 4\% for \ac{lhc} Run~1, comparable to the
current uncertainty of the most precise \ac{sm} prediction of less
than~4\%~\cite{Czakon:2013goa}.

In 2014, the ATLAS and CMS inclusive \ttbar \xsec measurements
available at $\sqrt{s}=\SI{8}{TeV}$ at the time were combined, to
arrive at an even smaller total uncertainty of
3.5\%~\cite{CMS:2014gta}. The current uncertainty on \ttbar \xsec{}s
measured with \ac{lhc} Run~2 data~\cite{Aaboud:2016pbd,CMS:2016mqs} is
slightly higher than the uncertainties obtained in Run~1, but is
expected to improve with larger data samples and better understanding
of systematic uncertainties.

\begin{table}
  \centering
  \small
  \caption{Summary of most precise inclusive \ttbar \xsec
    measurements from the ATLAS~\cite{Aad:2014kva, Aaboud:2016pbd} and
    CMS~\cite{Khachatryan:2016mqs,CMS:2016mqs}
    experiments together with their relative statistical and systematic
    uncertainties. 
    The systematic uncertainties are separated into uncertainties
    originating from experimental and
    theoretical sources as well as luminosity (and beam energy) 
    uncertainties.}
  \vspace{1mm}

  \begin{tabular}{lcccccc}
    \toprule
    Measurement & $\sigma_\ttbar$ (\si{pb}) & stat (\%) &
    exp (\%) & th (\%) & lumi (\%) & Ref.\\
    \midrule
    ATLAS $e\mu$ \SI{7}{TeV} & 182.9 & 1.7 & 2.3 & 2.0 & 1.8 & \cite{Aad:2014kva} \\
    CMS $e\mu$ \SI{7}{TeV} & 173.6 & 1.2 & \multicolumn{2}{c}{$^{+2.6}_{-2.3}$} & 2.2 & \cite{Khachatryan:2016mqs}\\
    \midrule
    ATLAS $e\mu$ \SI{8}{TeV} & 242.4 & 0.7 & 2.3 & 3.1 & 1.7 & \cite{Aad:2014kva} \\
    CMS $e\mu$ \SI{8}{TeV} & 244.9 & 0.6 &
                                           \multicolumn{2}{c}{$^{+2.6}_{-2.2}$} & 2.6 & \cite{Khachatryan:2016mqs}\\
    \midrule
    ATLAS $e\mu$ \SI{13}{TeV}     & 818   & 1.0 & \multicolumn{2}{c}{3.3} & 2.7 & \cite{Aaboud:2016pbd}\\
    CMS $e/\mu$+jets \SI{13}{TeV} & 834.6 & 0.3 & \multicolumn{2}{c}{2.7} & 2.7 & \cite{CMS:2016mqs}\\ 
    \bottomrule
  \end{tabular}
  \label{table:xsec}
\end{table} 

\paragraph{LHC: Further Results}
The inclusive \ttbar \xsec{} has also been established in other decay
channels and found to be consistent with the high-precision channels
with electrons and muons. This constitutes a check of the \ttbar
production mechanism, because the \xsec{} could be influenced by
\ac{bsm} physics differently in different channels.  The additional
\ttbar decay channels in which the \ttbar \xsec has been determined
include final states with one or two tau
leptons~\cite{Abazov:2010pa,Collaboration:2012hz,Aaltonen:2014hua,Aad:2012mza,Chatrchyan:2012vs,Aad:2012vip},
which are sensitive to charged Higgs boson production, and fully
hadronic \ttbar
decays~\cite{Abazov:2009ss,Aaltonen:2010pe,Chatrchyan:2013ual,Khachatryan:2015fwh,CMS:2016rtp}.
Inclusive \ttbar \xsec measurements have also been pursued for top
quarks with large transverse momenta (``boosted top quarks''),
together with differential \xsec measurements, as will be discussed in
Section~\ref{sec:differential}.

Top quark-antiquark production has become accessible also in $pp$
collisions at the LHCb experiment, where the process has been
established with a significance of 4.9 standard deviations 
using a dataset of approximately \SI{2}{\invfb} at
$\sqrts=\SI{8}{TeV}$~\cite{Aaij:2016vsy}. While classic collider
experiments at the Tevatron and the \ac{lhc} cover the ``central''
kinematic region of $|\eta|<2.5$, the forward-spectrometer design of
LHCb leads to coverage of the complementary kinematic region of
forward pseudorapidities, $2.0 \lesssim \eta \lesssim 4.5$. The
fiducial \ttbar production \xsec in this kinematic region is measured
along with the \xsec{}s for associated $W+\bbbar$ and $W+\ccbar$
production. The signature is a high-\pt electron or muon and two
heavy-flavor tagged jets, and \ttbar, $W+\bbbar$, and $W+\ccbar$
candidate events are separated using a multivariate discriminant. The
measured fiducial \xsec{}s are in agreement with \ac{sm} predictions
at \ac{nlo}.

\subsection{Differential \ttbar Production Cross Section}
\label{sec:differential}

The Tevatron and the \ac{lhc} experiments have published differential
\ttbar \xsec{} measurements as a function of various kinematic
properties of analysis objects. Kinematic observables may be separated
in quantities that can be measured directly and reconstructed
quantities that have to be inferred from the quantities measured
directly. Some differential \xsec results are presented as normalized
to the inclusive or fiducial \ttbar \xsec determined from the same
measurement. In this way normalization uncertainties, \eg the
luminosity uncertainty, cancel and the sensitivity of the measurement
to the shapes of kinematic distributions is improved.

\paragraph{Kinematic Observables}
The directly measured observables, \eg the kinematic distributions of
leptons and jets with \bhad{}s, are corrected back to the level of
stable particles, which are accessible in \ac{mc} generators, in a
fiducial region of phase space. The fiducial region is usually defined
by the detector acceptance in \pt and $\eta$ for leptons and
jets. Particle-level charged leptons are taken as the generated
leptons and sometimes ``dressed'' with soft photons from \ac{qed}
radiation. Particle-level jets are jets clustered from stable
generated particles except neutrinos with the same \ak algorithm
applied to reconstructed particles. On particle-level \btagging is
mimicked by adding ``ghost \bhad{}s''~\cite{Cacciari:2007fd} with
negligible momenta to the list of final-state particles before the jet
algorithm is applied and declaring jets in which one or more ghost
\bhad{}s are found as \bjet{}s. The particle-level observables can be
compared to the output of \ac{mc} event generators, for example using
{\sc Rivet}~\cite{Buckley:2010ar}, to test how well a given \ac{mc}
generator models the observables.

The kinematics of the top quarks and antiquarks or of the \ttbar
system are defined only on the level of partons rather than
particles. The partons are considered before decay, but after gluon
and photon radiation. Parton-level observables have the advantage that
they can be compared with theoretical calculations directly.  While it
is reasonable to assume a very good correspondence between the
particle and the parton level, strictly speaking there is no
unambiguous way to translate particle-level results to the parton
level.  Therefore particle-level pseudo-observables (``pseudo-top'')
have been agreed upon in the \ac{lhctopwg}, where care has been taken
to define the reconstructed quantities in a theoretically safe and
unambiguous way, see \eg\cite{Aad:2015eia,CMS:2015tpz}. Differential
\xsec{}s can also be determined as a function of event-level
quantities such as momentum sums which do not require the
reconstruction of the top quark and antiquark from their decay
products.

At the level of stable top quarks, differential \xsec measurements can
be compared with \ac{sm} predictions directly. These predictions are
available at various levels of precision that go beyond the precision
available in current \ac{mc} event generators: approximate \ac{nnlo}
with \ac{nnll} resummation~\cite{Ahrens:2010zv}, approximate
\ac{nnlo}~\cite{Guzzi:2014wia,Broggio:2014yca} and approximate
\ac{n3lo}~\cite{Kidonakis:2014pja}. Recently also differential
distributions at full~\ac{nnlo} precision became
available~\cite{Czakon:2015owf,Czakon:2016dgf}.

\begin{figure}[!t]
  \centering
  \includegraphics[width=0.45\textwidth]{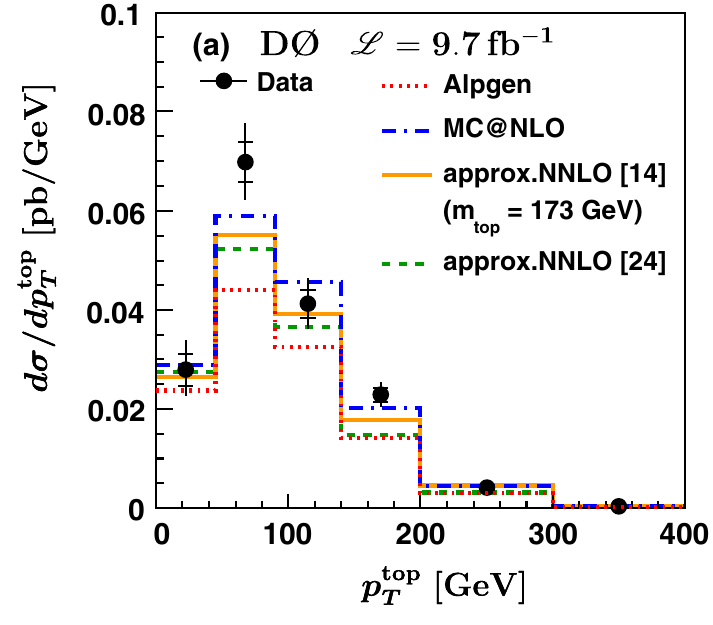}\hspace{10mm}
  \includegraphics[width=0.36\textwidth]{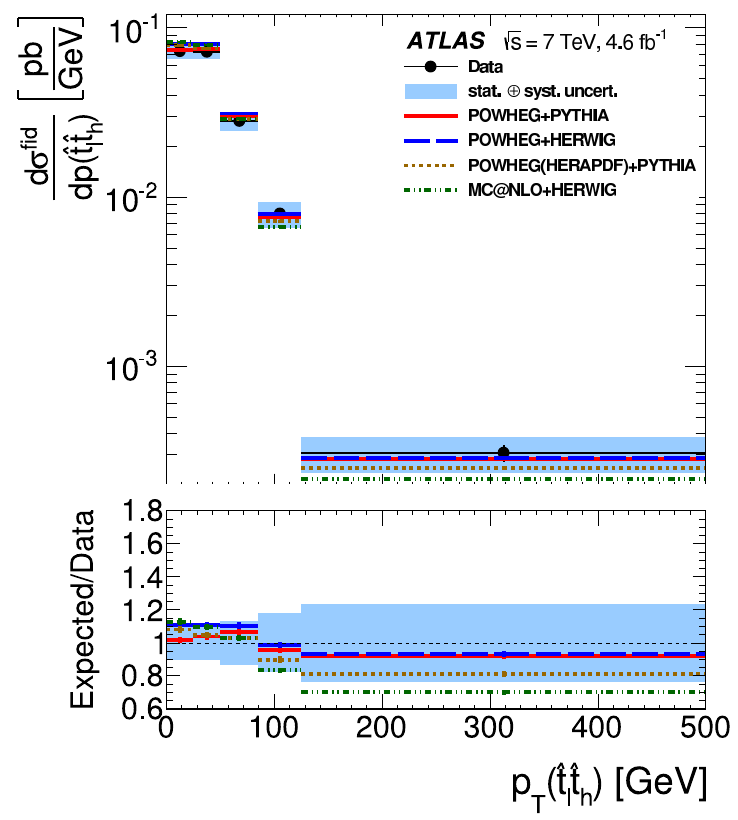}\\[3mm]
  \includegraphics[width=0.45\textwidth]{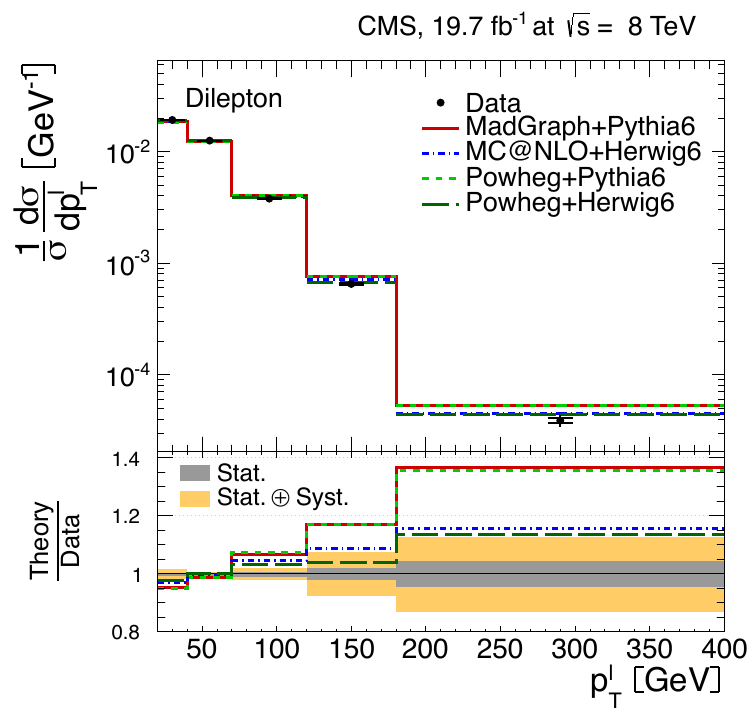}\hspace{3mm}
  \includegraphics[width=0.45\textwidth]{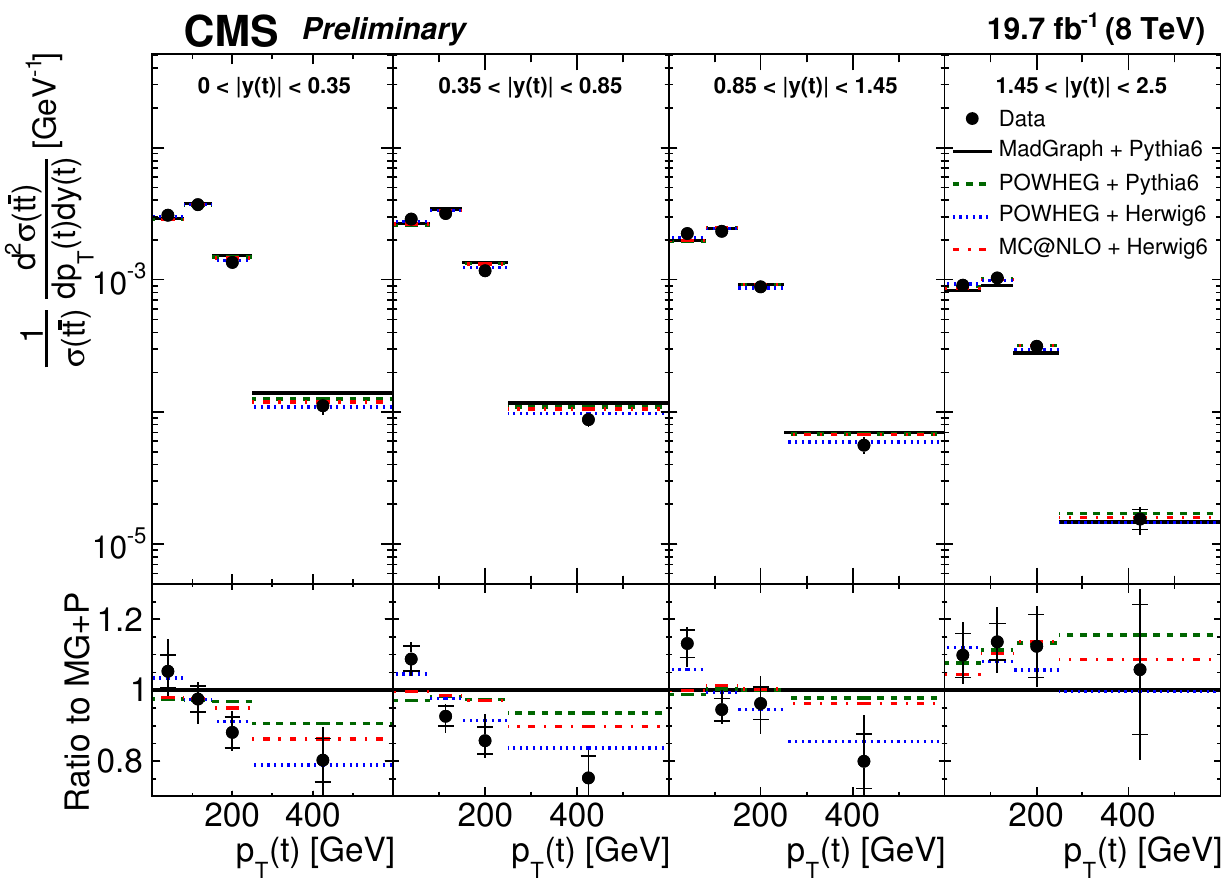}

  \caption{Examples of differential \xsec measurements from the
    Tevatron and the \ac{lhc}: 
    Top-quark transverse
    momentum at the Tevatron~\cite{Abazov:2014vga} (top left).
    Transverse momentum of the pseudo-top-quark
    pair at~$\sqrts=\SI{7}{TeV}$~\cite{Aad:2015eia} (top right).
    Transverse momentum of the lepton from the top-quark decay 
    at~$\sqrts=\SI{8}{TeV}$~\cite{Khachatryan:2015oqa} (bottom left).
    Transverse momentum of the top-quark for four different intervals
    of the top-quark rapidity at
    $\sqrts=\SI{8}{TeV}$~\cite{CMS:2016cue} (bottom right).
    The distributions are unfolded to particle or parton level and
    compared to predictions using recent \ac{mc} event generators
    and/or higher-order \ac{qcd} calculations.}
  \label{fig:differential}
\end{figure}

\paragraph{Tevatron Results}
The limited size of the \ttbar data samples at the Tevatron only
allowed for a small number of differential \xsec measurements. CDF
published the differential \ttbar \xsec as a function of the invariant
\ttbar mass~\cite{Aaltonen:2009iz}, which can also be interpreted as a
search for exotic particles decaying into \ttbar. D0 published a
comprehensive set of differential \xsec{}s using the full Tevatron
Run~II dataset~\cite{Abazov:2014vga}, see Fig.~\ref{fig:differential}.

\begin{figure}[!t]
   \centering
   \includegraphics[width=0.45\textwidth]{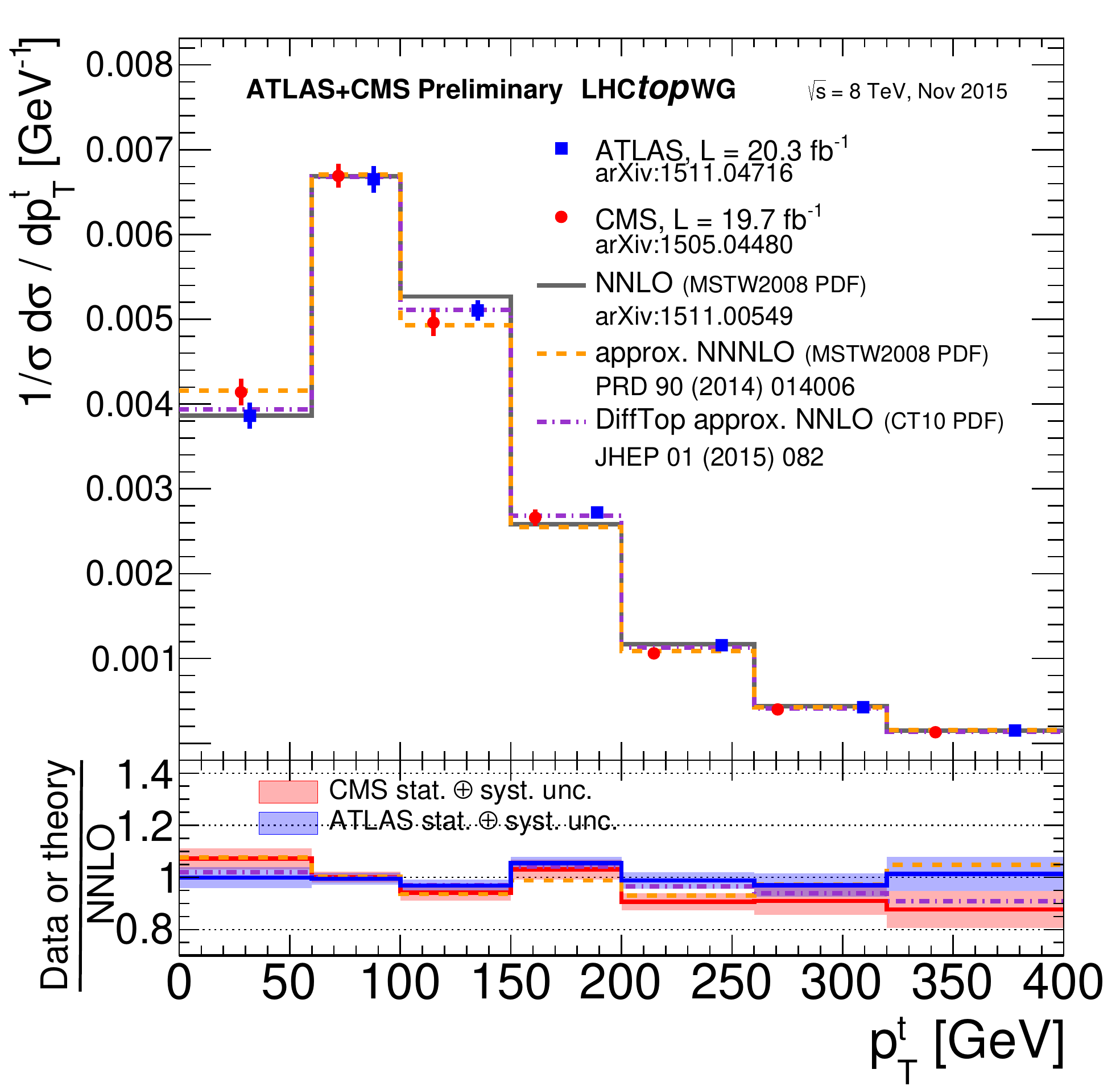}\hspace{3mm}
   \includegraphics[width=0.45\textwidth]{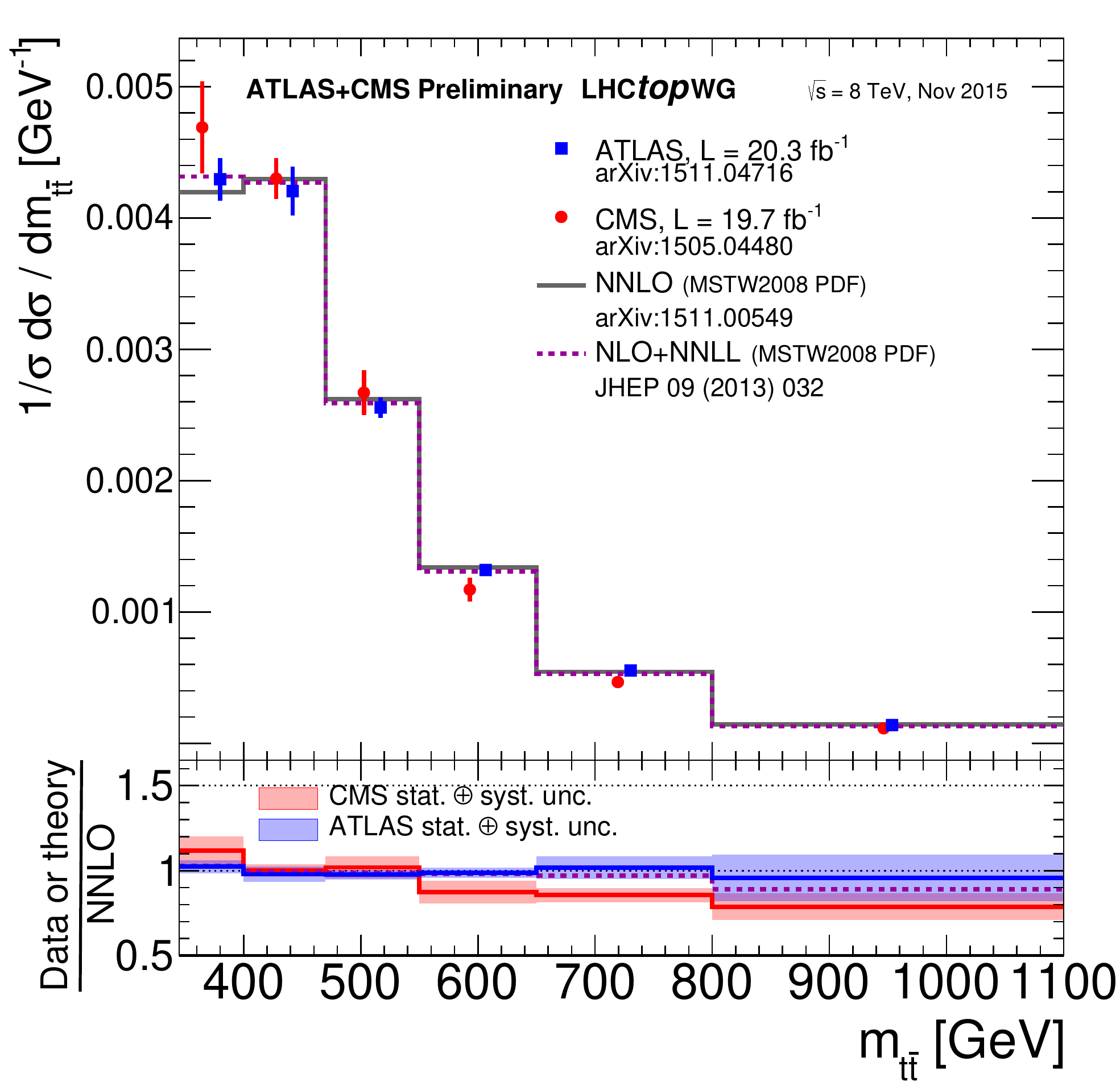}\\[3mm]
   \includegraphics[width=0.45\textwidth]{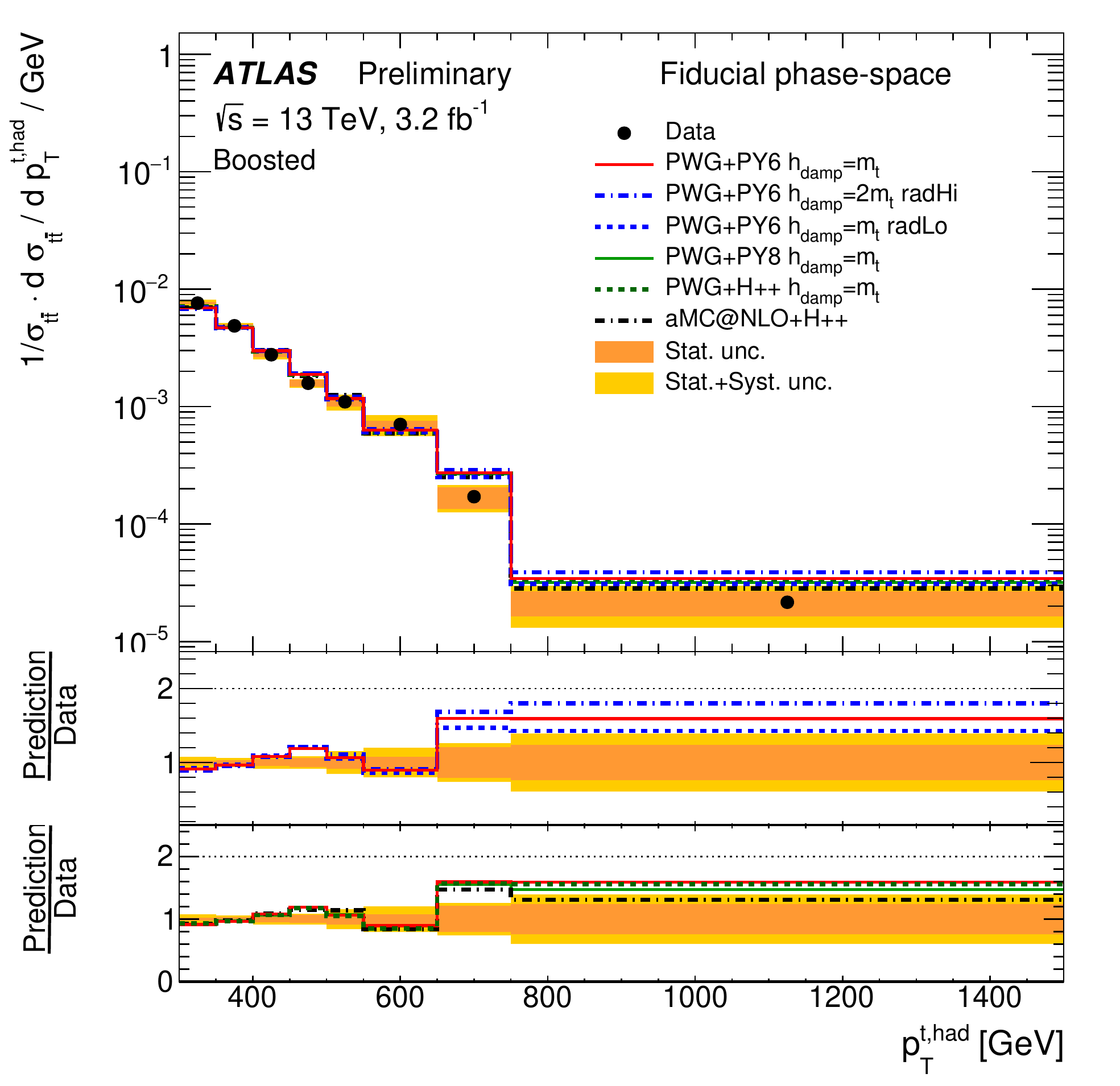}\hspace{3mm}
   \includegraphics[width=0.45\textwidth]{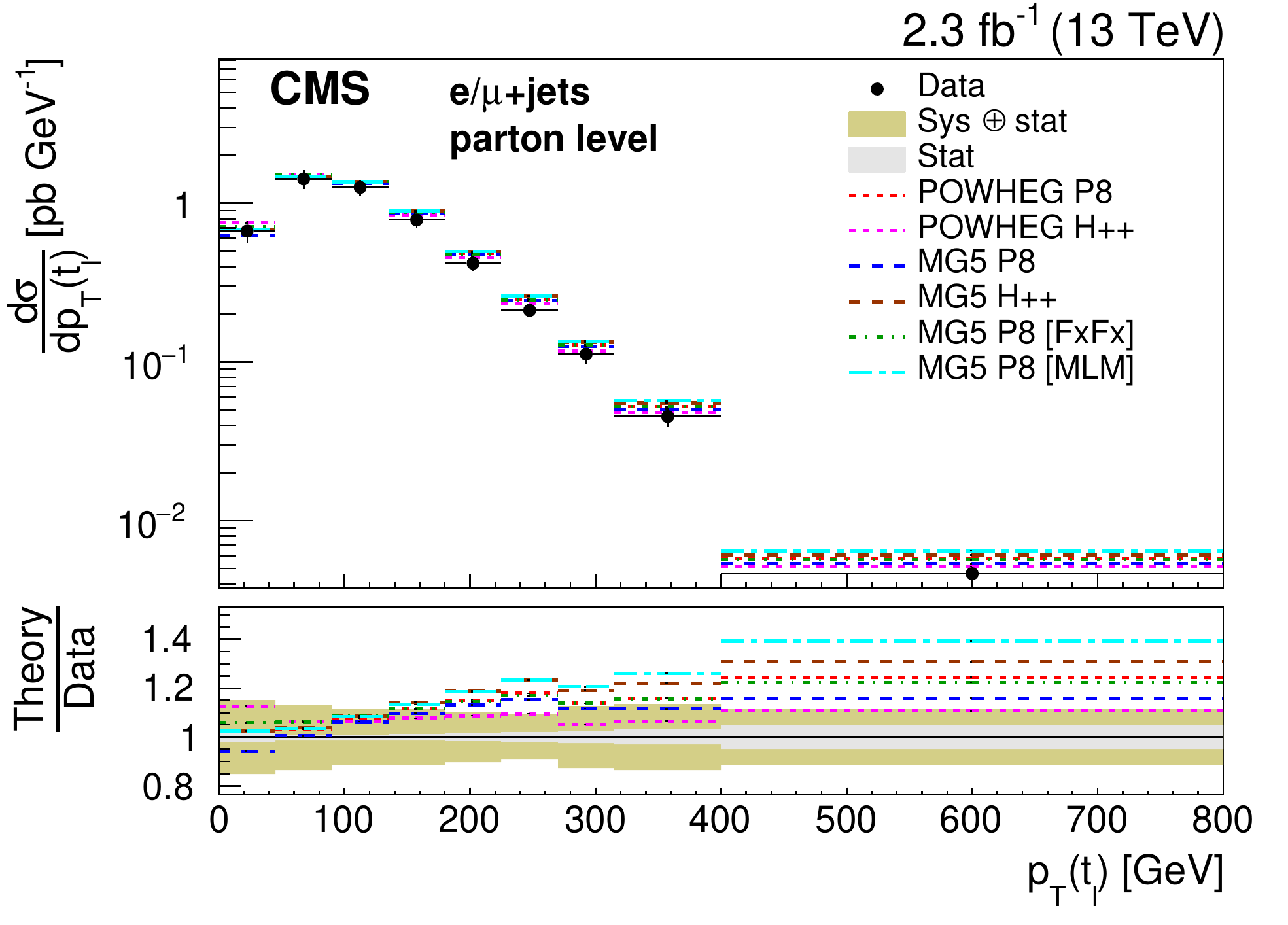}

   \caption{Examples of differential \xsec measurements from the
    \ac{lhc}: 
    Comparison of ATLAS and CMS measurements of the top-quark
    transverse momentum (top left) and the \ttbar invariant mass (top right)
    at~$\sqrts=\SI{8}{TeV}$~\cite{LHCTopWG}.
    Transverse momentum of the hadronically decaying top quark 
    at~$\sqrts=\SI{13}{TeV}$ using boosted-top
    reconstruction~\cite{ATLAS:2016soq} (bottom left).  
    Transverse momentum of the leptonically decaying top quark at
    $\sqrts=\SI{13}{TeV}$~\cite{Khachatryan:2016mnb} (bottom right).
    The distributions are unfolded to parton level and compared to
    predictions using recent \ac{mc} event generators and/or
    higher-order \ac{qcd} calculations.}
   \label{fig:differential2}
 \end{figure}

\paragraph{LHC Results}
The \ac{lhc} experiments have published a large number of differential
\ttbar \xsec measurements at $\sqrts=\SI{7}{TeV}$ and
\SI{8}{TeV}~\cite{Aad:2012hg,Aad:2014zka,Chatrchyan:2012saa,Aad:2015eia,Aad:2015mbv,Aaboud:2016iot,Khachatryan:2015oqa,CMS:2015tpz},
and recently also at
$\sqrts=\SI{13}{TeV}$~\cite{ATLAS:2016soq,CMS:2016uiq,CMS:2016xyh,Khachatryan:2016mnb}.
A small selection of the vast body of results is presented in
Figs.~\ref{fig:differential} and~\ref{fig:differential2}, including
both particle and parton level observables.  After unfolding, these
results are compared with the predictions of \ac{mc} generators at
particle and parton level, and \ac{sm} predictions at parton
level. Generally current \ac{mc} event generators as those introduced
in Section~\ref{sec:mc} describe the differential \xsec{}s well over a
wide kinematic range. In CMS, the measured \pt spectrum of top quarks
was found to be softer than most \ac{mc} predictions, while ATLAS
results are consistent with the predictions.  The recent full
\ac{nnlo} calculation of the top-quark \pt spectrum shows improved
agreement with the measured spectrum, compared to previous
calculations. Recently, due to the large \ttbar data samples at the
\ac{lhc}, also the first double differential \xsec{}s were published,
for example as a function of \pt and $y$ of the top
quark~\cite{CMS:2016cue}.  To study the production of top quarks with
large transverse momenta, differential \xsec measurements using
boosted-top reconstruction techniques (see Section~\ref{sec:boosted})
have been devised~\cite{Aad:2015hna,ATLAS:2016soq,Khachatryan:2016gxp,
  ATLAS:2016jct,CMS:2016rtp}. A first differential \ttbar \xsec
measurement as a function of the mass of boosted top quarks has been
performed as a proof of principle to measure \mt in boosted-top final
states~\cite{CMS:2016bnj}.

\afterpage{\clearpage}

Differential \ttbar production \xsec{}s measured as a function of
further event-level quantities provide additional insight into the
production mechanisms and are at the same time sensitive probes of
\ac{bsm} physics contributions to \ttbar production. Measurement of
the jet multiplicity and the number of additional jets, not coming
from the decay products of the \ttbar pair, in \ttbar events from
ATLAS~\cite{Aad:2014iaa,Aaboud:2016omn,ATLAS:2016soq,ATLAS-CONF-2015-065,
  Aaboud:2016xii} and
CMS~\cite{Chatrchyan:2014gma,Khachatryan:2015mva,CMS:2016ooc,CMS:2016uiq,Khachatryan:2016mnb}
probe the treatment of \ac{qcd} radiation in \ac{mc} event generators.
Of particular interest for rare \ac{sm} processes such as associated
\ttH production (with the decay \Hbb) as well as for the search for
\ac{bsm} physics is the production of \ttbar pairs with additional
\bjet{}s.  In measurements of these processes, the ratio of \xsec{}s
for $\ttbar$ production with two additional \bjet{}s and $\ttbar jj$
production, where $j$ is a jet of any flavor, has been used as a
robust observable sensitive to $\ttbar\,\bbbar$
production~\cite{CMS:2014yxa,Aad:2015yja,CMS:2016tlo}.

Differential \xsec{}s as a function of event-level observables such as
the missing transverse momentum (\met), the scalar sum of the jet
transverse momenta (\HT), or the scalar sum of the transverse momenta
of all physics objects (\ST) are sensitive to rare processes, \eg the
associated production of \ttbar and $W$, $Z$, or Higgs bosons, as well
as to \ac{bsm} physics processes with lepton+multijet
signatures~\cite{Khachatryan:2016oou,CMS:2015imk}.

\subsection{$\ttbar+X$ Production}
\label{sec:ttX}
The production of \ttbar pairs in association with ``something else''
($\ttbar+X$), where $X$ can be the electroweak gauge bosons $\gamma$,
$W$, and $Z$, or the Higgs boson, is predicted to be rare in the
\ac{sm}, with inclusive production \xsec{}s in $pp$ collisions at
$\sqrts=\SI{13}{TeV}$ predicted by \ac{nlo} \ac{qcd} below
\SI{1}{\pico\barn}. Measuring these processes gives access to the
coupling of the top quark to the electroweak bosons as well as to the
Yukawa coupling of the top quark, all of which could be modified by
\ac{bsm} physics effects. In particular \ttZ production gives direct
access to the $Ztt$ coupling, as in \ac{lo} the $Z$ boson is
predominantly emitted from a top (anti)quark.  These processes also
constitute backgrounds for many hypothetical \ac{bsm} processes with
high particle multiplicities and similar final states.

\paragraph{Electroweak Gauge Bosons}
Evidence for \ttg production was first reported by
CDF~\cite{Aaltonen:2011sp}. The process was first observed by
ATLAS~\cite{Aad:2015uwa} in the experiment's full dataset recorded at
$\sqrts=\SI{7}{TeV}$ and also measured by CMS in $\sqrts=\SI{8}{TeV}$
data~\cite{CMS:1900ipz}. The production \xsec{}, defined in a fiducial
region of phase space, is compatible with the \ac{sm} prediction at
\ac{nlo}. The challenge of these analyses lies in separating photons
from \ttg production from hadron decays into photon pairs and hadrons
and electrons misidentified as photons, which is done by studying the
photon isolation.

Measurements of \ttW and \ttZ production have only become feasible
with the large \ac{lhc} datasets. The processes feature very massive
final states of more than \SI{425}{GeV} and therefore profit a lot
from the increase in production \xsec{}s at \ac{lhc} Run~2 compared to
Run~1.  The \ac{lhc} experiments have seen evidence of these processes
in leptonic decays of the $W$ and $Z$ bosons in combination with
single-lepton and dilepton decays of the \ttbar pair, in events
containing multiple jets, \btagged jets and two to four charged
leptons~\cite{Aad:2015eua,Khachatryan:2015sha,Aaboud:2016xve,CMS:2016ium,CMS:2016dui}.
While the backgrounds in these events are generally low, it is
difficult to estimate the number of analysis objects wrongly
identified as leptons (``fake leptons'') precisely. The fake lepton
background is usually modeled from data events in control regions, as
described in Section~\ref{sec:background}. The measurements are
compatible with \ac{sm} predictions and are used to constrain \ac{bsm}
physics contributions to the $Ztt$
coupling~\cite{Khachatryan:2015sha}.

\paragraph{Higgs Boson}
Higgs-boson production in the associated \ttH channel is a process
that has not yet been established experimentally. The channel is
challenging due to the small production \xsec{}, approximately
\SI{0.5}{pb} at $\sqrts=\SI{13}{TeV}$~\cite{LHCHXSWG}, and large
irreducible backgrounds. In particular in the \Hbb decay channel, the
background from \ttbar\bbbar production is large and hard to
control. With sophisticated multivariate methods, both physics
motivated (\acl{mem}, see Section~\ref{sec:mem}) and from machine
learning, and the inclusion of final states with boosted top quarks,
so far only upper limits on the production \xsec have been
determined. Both ATLAS and CMS first conducted three independent
analyses in the \Hbb, \Hgg and multilepton\footnote{The multilepton
  channel summarizes all Higgs-boson decays with multiple leptons from
  the decay channels \HZZ, \HWW, and \Htt.}  decay channels that were
statistically combined in a second step.

The individual \ttH searches performed using the \ac{lhc} Run~1
dataset~\cite{Aad:2014lma,Aad:2015iha,Aad:2016zqi,Khachatryan:2014qaa,Khachatryan:2015ila}
were combined with other Higgs-physics results from ATLAS and CMS to
arrive at a significance of 4.4 standard deviations for \ttH
production, where only 2.0 standard deviations were expected. This
unexpected result is driven by a small excess of events in the
multilepton channel~\cite{Khachatryan:2016vau}. With the four-fold
increase in \ttH production \xsec at $\sqrt{s}=\SI{13}{TeV}$, the
\ac{lhc} experiments are expected to finally become sensitive to \ttH
production with \SIrange{30}{50}{\per\femto\barn} of luminosity. First
preliminary results using up to \SI{13}{\invfb} of data have already
been
presented~\cite{CMS:2016qwm,CMS:2016rnk,CMS:2016vqb,CMS:2016ixj,CMS:2016zbb,ATLAS:2016awy,ATLAS:2016nke,ATLAS:2016ldo,ATLAS:2016axz},
already with increased sensitivities compared to the Run-1 results.

\paragraph{Invisible Particles}
The production of \ttbar pairs in association with invisible particles
results in a significant amount of \ac{met}. Such processes have been
studied in the context of searches for \ac{bsm} physics, in particular
in the search for supersymmetric particles or more generically in
dark-matter searches. This will be discussed in Section~\ref{sec:dm}.

\subsection{Single Top-Quark Production}
\label{sec:singletop}

Electroweak \st production was first observed at the
Tevatron~\cite{Abazov:2009ii,Aaltonen:2009jj}. The expected inclusive
\xsec in \ppbar collisions at $\sqrts=\SI{1.96}{TeV}$ is small, of the
order of~\SI{3}{pb} adding all production
channels~\cite{Kidonakis:2010tc,Kidonakis:2011wy}. At the \ac{lhc}, \st
production in the \tch has a moderately large \xsec, of the order of
\SI{65}{pb} at $\sqrts=\SI{7}{TeV}$~\cite{Kidonakis:2011wy}, so that
\st production was established early in \ac{lhc}
Run~1~\cite{Chatrchyan:2011vp,Aad:2012ux}. In the following the most
precise measurements of \st production available from the Tevatron and
the \ac{lhc} are summarized. A more detailed account of \st production
at the \ac{lhc} can be found in two recent
reviews~\cite{Wagner-Kuhr:2016zbg,Giammanco:2015bxk}.

\begin{figure}[t]
  \centering
  \includegraphics[width=0.43\textwidth]{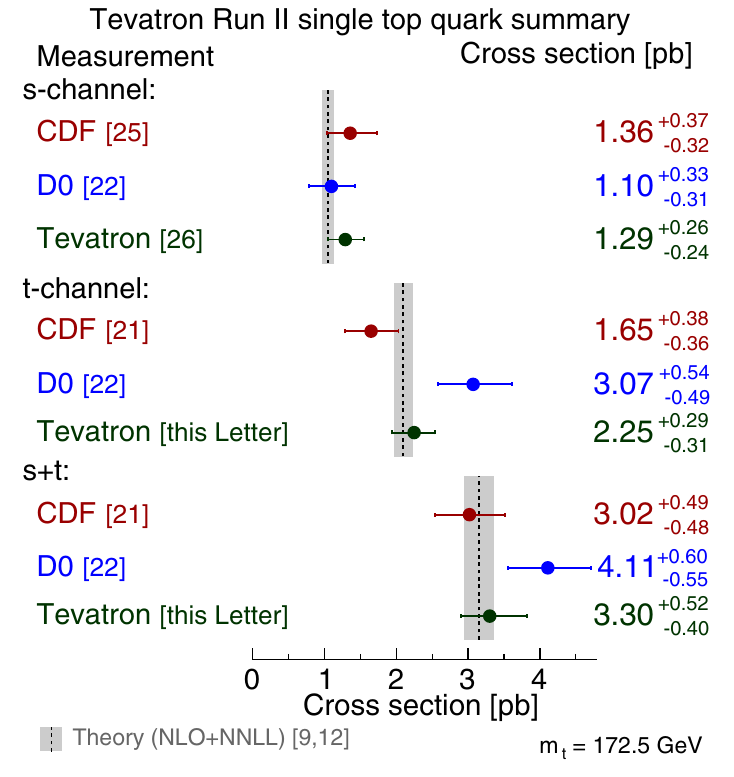}\hspace{5mm}
  \includegraphics[width=0.50\textwidth]{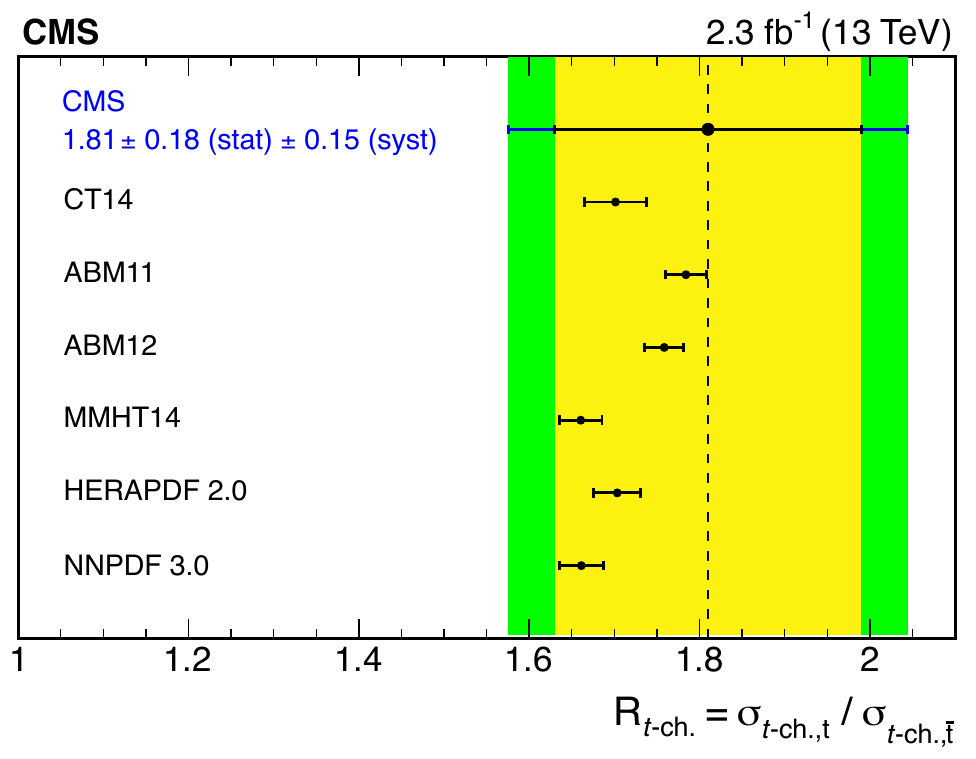}
  \caption{Compilation of Tevatron measurements of the \st \xsec in
    the \sch and the \tch as well as for both channels
    combined~\cite{Aaltonen:2015cra} and compared to \ac{sm}
    predictions at approximate
    \ac{nnlo}~\cite{Kidonakis:2010tc,Kidonakis:2011wy} (left).  Ratio
    of top quark and antiquark production \xsec{}s in the \tch at the
    \ac{lhc} compared to various \ac{pdf} sets~\cite{Sirunyan:2016cdg}
    (right).}
  \label{fig:sttev}
\end{figure}

\paragraph{Tevatron Results}
At the Tevatron, only \tch and \sch production were accessible. The
characteristic \tch signature of a semileptonically decaying top quark
and a spectator jet in forward direction is overwhelmed by background
mainly from \wjets production. This requires sophisticated
multivariate techniques to separate signal and background and profile
likelihood ratio fits to extract the production \xsec.  Both CDF and
D0 have published \st production measurements using the above
techniques with their full Run-II
datasets~\cite{Abazov:2013qka,Aaltonen:2014mza} as well as a
combination~\cite{Aaltonen:2015cra}, shown in Fig.~\ref{fig:sttev}
(left). The \sch, which has a smaller production \xsec and larger
backgrounds compared to the \tch, was only established as a separate
\st production channel by combining the full CDF and D0 Run-II
datasets~\cite{CDF:2014uma}. The combined Tevatron results are in good
agreement with \ac{sm} predictions at approximate
\ac{nnlo}~\cite{Kidonakis:2010tc,Kidonakis:2011wy}.

\paragraph{LHC Results}
Similar to the Tevatron, the \xsec measurements at the \ac{lhc} are
based on multivariate separation of signal and background and profile
likelihood ratio fits to extract the \xsec.  Precise measurements of
the \tch \st production \xsec have been performed at
$\sqrts=\SI{7}{TeV}$~\cite{Chatrchyan:2012ep,Aad:2014fwa},
\SI{8}{TeV}~\cite{Khachatryan:2014iya,ATLAS:2014dja}, and
\SI{13}{TeV}~\cite{Sirunyan:2016cdg,Aaboud:2016ymp}. As for \ttbar
production, modeling uncertainties have been reduced by reporting
fiducial \xsec{}s in addition to inclusive
\xsec{}s~\cite{ATLAS:2014dja,CMS:2015jca}. With the large datasets
available at the \ac{lhc}, also the first differential \xsec{}s for
\tch \st production as a function of the $t$ or \tbar transverse
momentum and rapidity became feasible~\cite{CMS:2014ika,CMS:2016xnv}.

An interesting observable in \tch \st production in $pp$ collisions is
the ratio of production rates for top quarks and antiquarks,
$\Rtch = \sigma_t / \sigma_\tbar$. While the top quark is produced
with an up-type quark (or down-type antiquark) in the initial state,
the top antiquark is produced with a down-type quark (or up-type
antiquark). Hence $\Rtch$ is sensitive to the ratio of \acp{pdf} for
up-type and down-type quarks (and down-type and up-type antiquarks),
with a naive expectation of $\Rtch=2$ for up and down valence quarks
only. Also anomalous $Wtb$ couplings as expected from \ac{bsm} physics
would modify $\Rtch$. Experimentally $\Rtch$ is a robust observable in
which many uncertainties cancel. Measurements of \Rtch from \ac{lhc}
Run~1~\cite{Aad:2014fwa, Khachatryan:2014iya} and
Run~2~\cite{Sirunyan:2016cdg} are compatible with the \ac{sm}
prediction, see Fig.~\ref{fig:sttev} (right). Another complementary
constraint on \acp{pdf} can be obtained from the ratio of \tch
\xsec{}s at $\sqrt{s}=\SI{7}{TeV}$ and
\SI{8}{TeV}~\cite{Khachatryan:2014iya}.

The $Wt$ associated production channel, whose \xsec was negligible at
the Tevatron, was observed for the first time at the
\ac{lhc}~\cite{Chatrchyan:2012zca,
  Aad:2012xca,Chatrchyan:2014tua,Aad:2015eto, ATLAS:2016lte}. The $Wt$
production \xsec{}s obtained by ATLAS and CMS at $\sqrts=\SI{8}{TeV}$
have recently been combined~\cite{CMS:2016ufa}.  At the \ac{lhc}, the
smallest \st production \xsec{} is expected in the \sch. First
evidence for this process has been reported in an ATLAS analysis at
$\sqrts=\SI{8}{TeV}$ using a sophisticated \ac{mem} technique (see
Section~\ref{sec:mem})~\cite{Aad:2015upn}, the corresponding CMS
search shows a slightly smaller significance~\cite{Khachatryan:2016ewo}.

\begin{figure}[t]
  \centering
  \includegraphics[width=\textwidth]{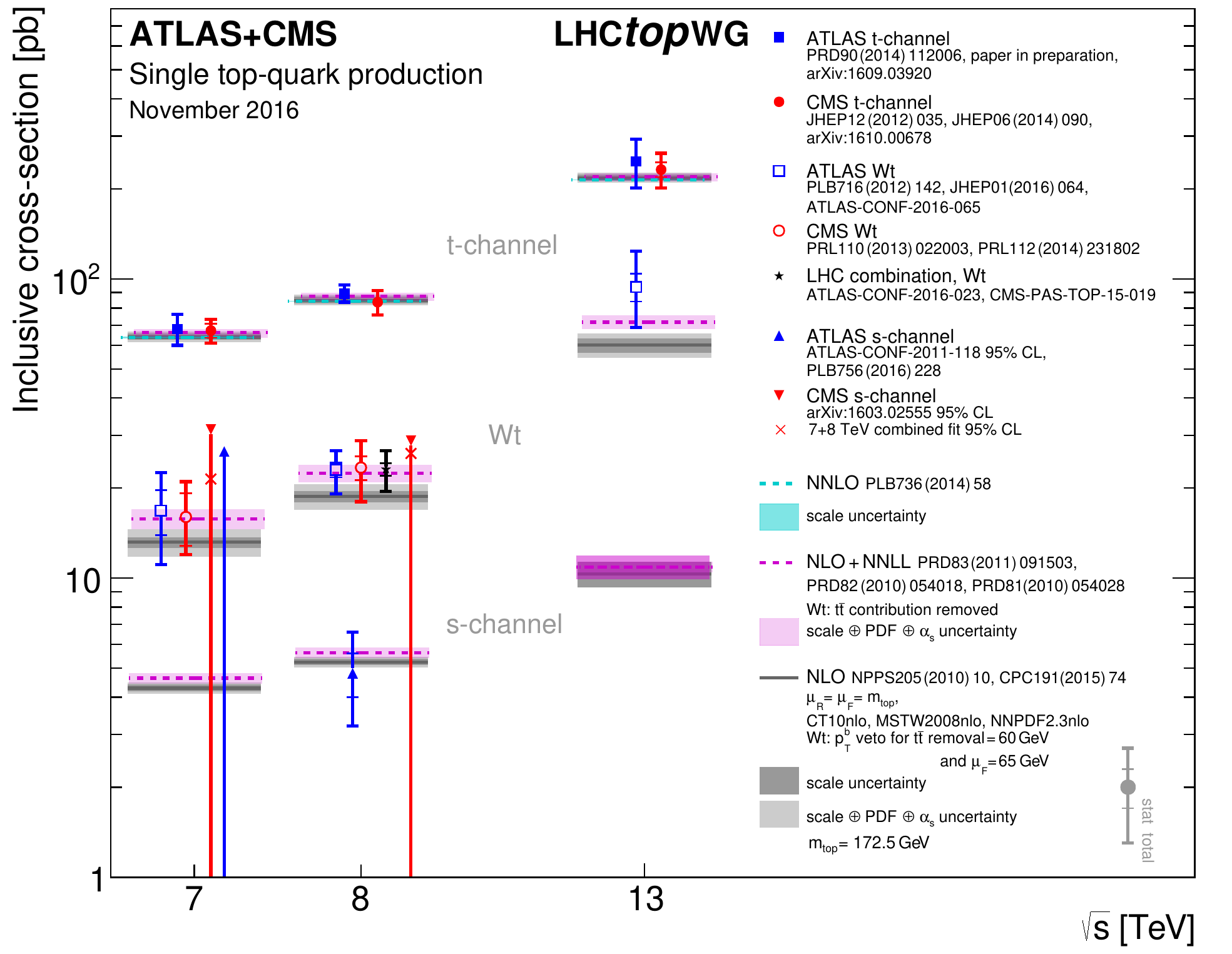}
  \caption{Compilation of measurements and \ac{sm} predictions of the
    inclusive \st \xsec for different \ac{lhc} center-of-mass
    energies~\cite{LHCTopWG}.}
  \label{fig:stxsec}
\end{figure}

A summary of inclusive \st \xsec measurements at the \ac{lhc} in all
production channels and for different center-of-mass energies is
presented in Fig.~\ref{fig:stxsec}. All measurements are in good
agreement with each other and with the \ac{sm} predictions.

\subsubsection{CKM Matrix Element \Vtb}
As the \st production \xsec is proportional to \Vtbsq, many of the
\xsec measurements at the Tevatron and the \ac{lhc} presented above
are also interpreted in terms of constraints on $|\Vtb|$. In such
analyses, it is usually assumed that \st production is only mediated
by $W$-boson exchange, that $\Vtbsq$ is much larger than the sum of
$|V_{ts}|^2$ and $|V_{td}|^2$, and that the $Wtb$ vertex is a
CP-conserving \VminusA coupling with a coupling strength modifier
$\mu_{V}^L$, with $\mu_{V}^L = 1$ in the \ac{sm}, see
Section~\ref{sec:decays}. No assumptions about the unitarity of the
\ac{ckm} matrix are made.

\begin{figure}[!t]
  \centering
  \includegraphics[width=0.9\textwidth]{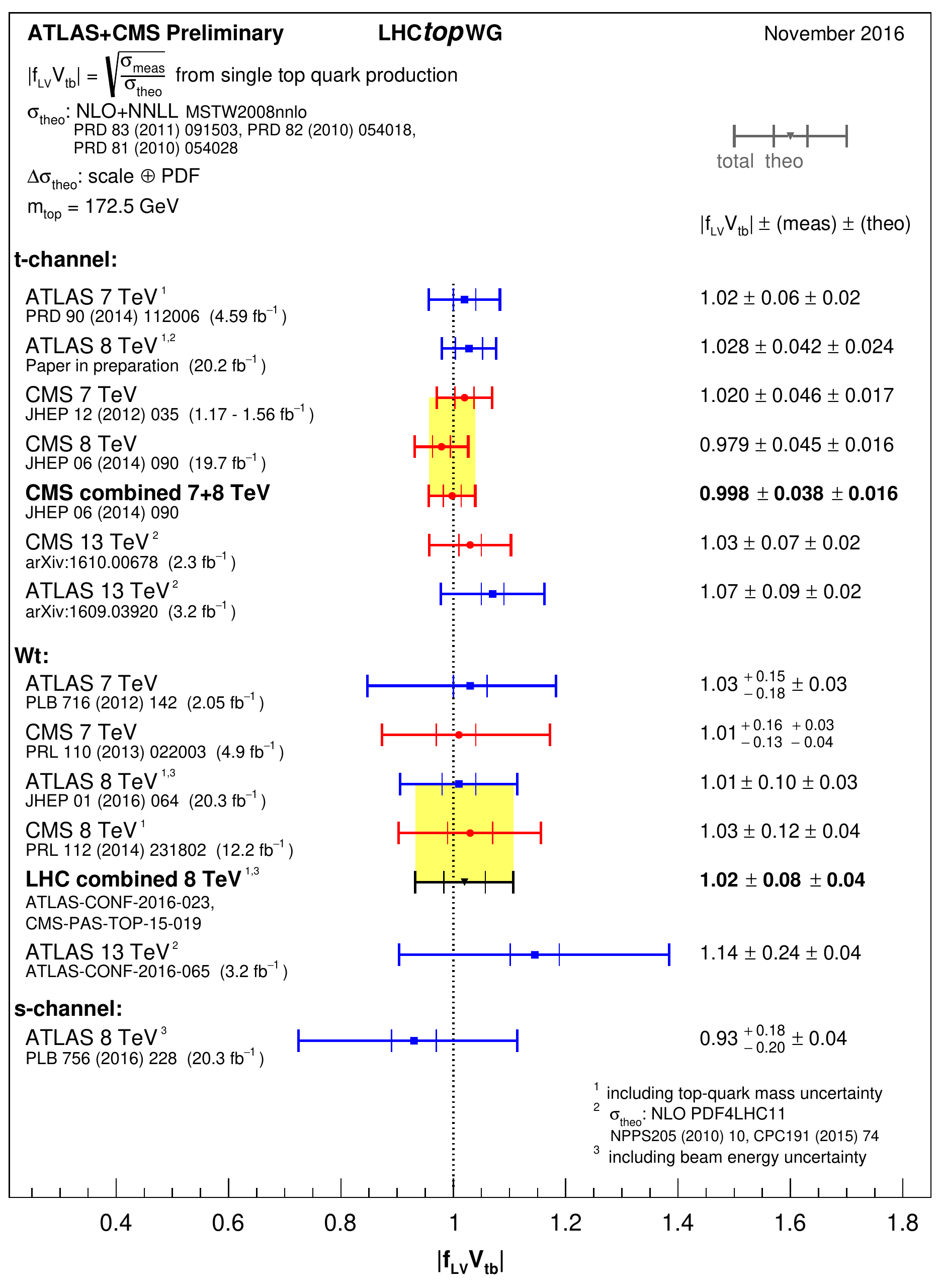}
  \caption{Compilation of \ac{lhc} measurements of
    $\left| \mu_{V}^L \Vtb \right|$ (denoted as $|f_{LV} \Vtb|$ in the
    figure)~\cite{LHCTopWG}.}
  \label{fig:Vtb}
\end{figure}

A compilation of \Vtb results from the \ac{lhc} is presented in
Fig.~\ref{fig:Vtb}.  The current most precise value is obtained from a
combination of CMS \tch \xsec measurements at $\sqrts=\SI{7}{TeV}$ and
\SI{8}{TeV}~\cite{Khachatryan:2014iya}:
\begin{equation}
  \left| \mu_{V}^L \Vtb \right| = 0.998 \pm 0.038\,\mathrm{(exp)} \pm 0.016\,\mathrm{(th)},
\end{equation}
where the first uncertainty originates from experimental and the
second from theoretical sources. A value for $|\Vtb|$ was also
extracted from the combination of $Wt$ production measurements:
$\left| \mu_{V}^L \Vtb \right|=1.02\pm0.09$~\cite{CMS:2016ufa}. 

\afterpage{\newpage}

\subsubsection{Single Top + X Production}
The associated production of \st{}s and additional particles has only
received little attention so far. The associated production of a
single top-quark with a Higgs boson ($tHq$ and $tHW$) is sensitive to
the sign of the top-quark's Yukawa coupling relative to the Higgs
coupling to vector bosons, while in \ttH production only the absolute
value of the Yukawa coupling is probed. The process is extremely rare
in the \ac{sm} due to destructive interference of the scattering
amplitudes for the Higgs boson coupling to the $W$ boson and the top
quark and would be enhanced significantly by \ac{bsm} physics. First
direct limits on the $tHq$ (and $tHW$) production \xsec in Higgs-boson
decays to $\gamma\gamma$, \bbbar, and multiple charged leptons using
\ac{lhc} data at $\sqrts=\SI{8}{TeV}$ and $\sqrts=\SI{13}{TeV}$ have
been published in~\cite{Khachatryan:2015ota,CMS:2016ygt}. Indirect
limits on the $Htq$ coupling were also obtained from the search for
\ttH production in the \Hgg decay channel~\cite{Aad:2014lma}.

\subsection{Summary}
The production of top quarks in the \ac{sm} is very well understood
both experimentally and theoretically. Most kinematic distributions
are well described by modern \ac{mc} event generators. This provides a
solid basis for studying the properties of the top quark as well as
for searches for \ac{bsm} physics in which top quarks are part of the
signal and/or of the background.  A possible new direction is
measuring the fiducial \xsec of the full process
$\pp \to W^+ b\,W^- \bbar + X$, which includes double-resonant
(\ttbar), single-resonant ($Wt$), as well as non-resonant
contributions.

\section{Top-Quark Mass}
\label{sec:mass}

The top-quark mass \mt is an important free parameter of the
\ac{sm}. Quantum corrections to certain \ac{sm} observables lead to
relations with other \ac{sm} parameters, such as the masses of the $W$
boson and the Higgs boson. Therefore precise measurements of \mt are
an important ingredient of precision tests of the \ac{sm}\footnote{The
  relation of \mt with the stability of universe, assuming that the
  \ac{sm} is valid up to very high energy scales, has been an
  interesting point of discussion in recent years, see
  \eg\cite{Degrassi:2012ry,Alekhin:2012py}.}.  At hadron colliders the
conventional way of measuring \mt relies on the kinematic
reconstruction of the \ttbar final state. Kinematic reconstruction is
the most precise method to determine \mt to date, with innovations
including the \acf{mem} and in-situ calibration of the jet energy
scale to increase the \mt sensitivity, see
Section~\ref{sec:history}. The different \ttbar decay channels and
different observables sensitive to \mt are subject to different
systematic uncertainties. If measurements from different channels and
with different observables are consistent, even higher precision is
obtained by combining them.

As discussed in Section~\ref{sec:mass_theory}, \mt measurements based
on the kinematic reconstruction of the \ttbar final state suffer from
ambiguities in the definition of \mt. Therefore alternative methods to
determine the top-quark mass at hadron colliders have been proposed
that are based either on different kinematic observables such as
endpoints of distributions or on (differential) \xsec measurements.
These methods have not reached the precision of the kinematic
reconstruction but provide valuable independent cross checks with
complementary systematic uncertainties. The perspectives of precision
\mt measurements, both at the \ac{lhc} and its upgrades, and at future
\epem colliders, will be discussed in Section~\ref{sec:outlook}.
Further details on measurements of the mass of the top quark can be
found in a recent review article~\cite{Cortiana:2015rca}.

\subsection{Kinematic Reconstruction}

\paragraph{Mass Determination Methods}
A straightforward way to determine the top-quark mass is to compare
the \mt distribution\footnote{In this chapter \mt may stand for any
  kinematic observable from which the top-quark mass may be inferred,
  \eg the invariant mass of the top-quark's decay products.}  as
reconstructed from the data with a set of \ac{mc}-simulated \mt
distributions (``templates'') with different values of the top-quark
mass parameter in the simulation.  Alternatively, each event can be
assigned an \mt-dependent likelihood. This event-level likelihood is
composed of process-level likelihoods that included the
hard-scattering matrix elements of the signal and the most important
background processes (\ac{mem}, see
Section~\ref{sec:mem}~\cite{Kondo:1988yd,Abazov:2004cs}). The ideogram
method~\cite{Mulders:2001mp, Abazov:2007rk} represents an alternative
to the \ac{mem} that is less computing-intensive. In the ideogram
method the likelihood for each process is assumed to factorize into an
\mt-independent factor depending only on the event topology and an
\mt-dependent factor depending on the event kinematics.


\paragraph{In-Situ Jet Energy Scale Calibration}
A major limitation on the precision of kinematic methods is the
limited resolution of \mt when reconstructed (partly) from jet
energies. In addition, a miscalibrated jet-energy scale would lead to
shifts in the reconstructed \mt similar to shifts caused by a
different \mt value. This correlation between \mt and the \ac{jes} can
be exploited by reconstructing hadronic decays of other particles with
known mass in the same dataset. Hadronically decaying $W$ bosons,
available in single-lepton and all-hadronic \ttbar events, are the
particles of choice: by reconstructing \mt and \mW simultaneously and
constraining \mW to the known value the precision of \mt is
significantly improved~\cite{Arguin:2005kx}. This method of
calibrating the \ac{jes} in-situ can be applied to template, \ac{mem},
and ideogram methods alike. Strictly speaking the in-situ \ac{jes}
calibration is only applicable to the same composition of light-flavor
quark jets the $W$ boson decays to. On the other hand, the \bjet{}s
from top-quark decays fragment differently from light-flavor jets and
could need a separate calibration (``$b$-\ac{jes}''). This is either
dealt with by assigning a systematic uncertainty to the residual
difference between \bjet{}s and light-flavor jets or by determining
both \ac{jes} and $b$-\ac{jes} in situ.

\paragraph{Tevatron and LHC Results}
The most precise measurements of the top-quark mass currently
available from the Tevatron and \ac{lhc} experiments are summarized in
Table~\ref{table:mass}. The CDF result~\cite{Aaltonen:2012va} is
obtained using a template method, while D0 utilizes a
\ac{mem}~\cite{Abazov:2014dpa,Abazov:2015spa}.  The CMS result---the
single most precise \mt measurement to date---employs the ideogram
method~\cite{Khachatryan:2015hba}. All measurements presented above
were performed in the single-lepton channel and used an in-situ
calibration of the \ac{jes} scale. The most precise ATLAS
result~\cite{Aaboud:2016igd} is based on a template method in the
dilepton channel.

\begin{table}[t]
  \centering
  \small
  \caption{Summary of the most precise individual measurements of the 
    top-quark mass \mt from the Tevatron and the \ac{lhc} experiments as 
    of November 2016, together with their statistical and systematic 
    uncertainties as well as their total relative uncertainties.}
  \label{table:mass}
  \vspace{1mm}

  \begin{tabular}{lccccl}
    \toprule
    Experiment & \mt (\si{GeV}) & stat (\si{GeV}) & 
    syst (\si{GeV}) & total (\%) & Ref.\\
    \midrule
    CDF   & 172.85 & 0.71 & 0.85 & 0.65 & \cite{Aaltonen:2012va}\\
    D0    & 174.98 & 0.58 & 0.49 & 0.43 & \cite{Abazov:2014dpa,Abazov:2015spa}\\
    \midrule
    ATLAS & 172.99 & 0.41 & 0.74 & 0.49 & \cite{Aaboud:2016igd}\\
    CMS   & 172.35 & 0.16 & 0.48 & 0.29 & \cite{Khachatryan:2015hba}\\
    \bottomrule
  \end{tabular}
\end{table} 

\begin{figure}[t]
  \centering
  \includegraphics[width=0.41\textwidth]{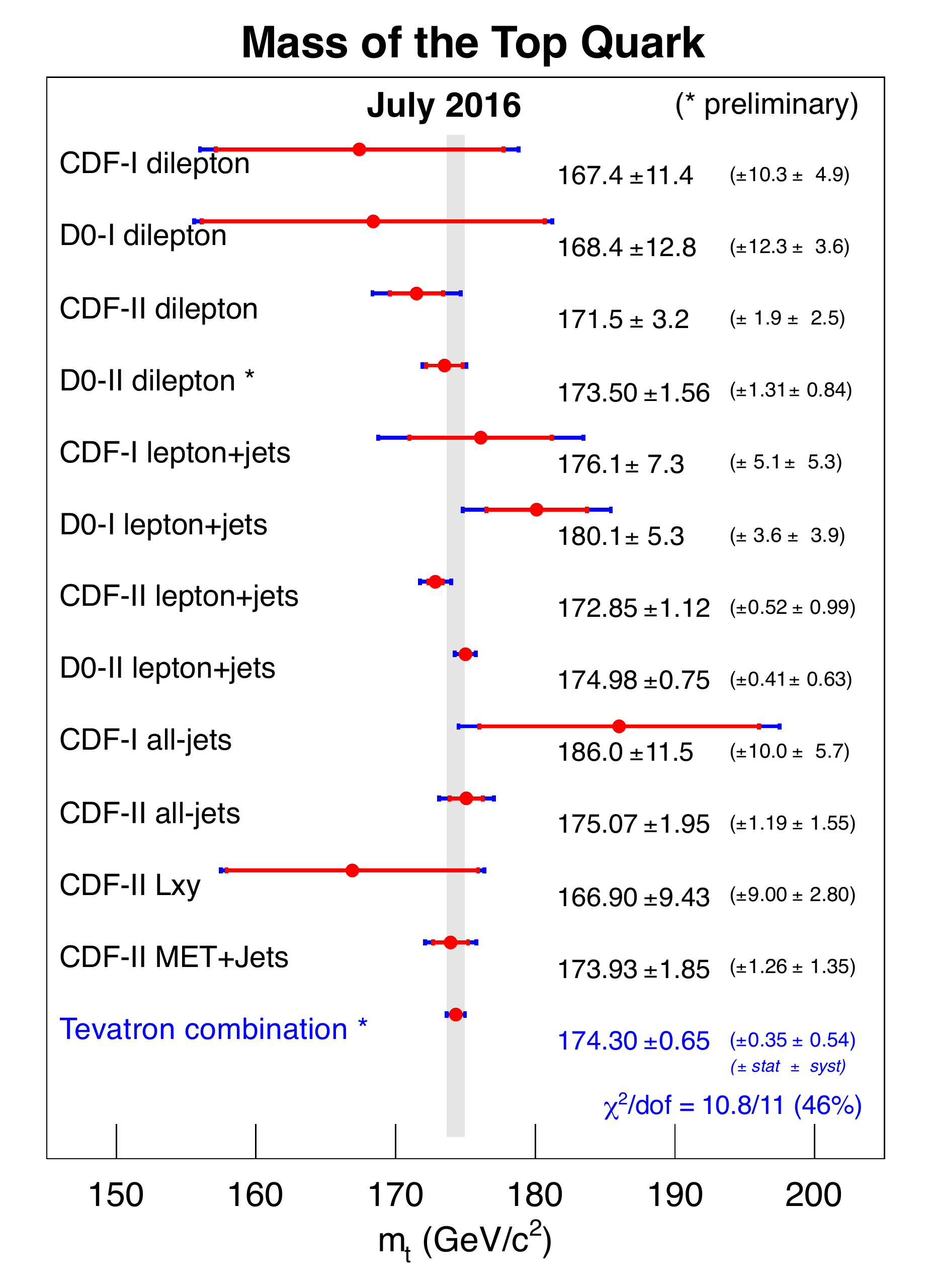}
  \includegraphics[width=0.58\textwidth]{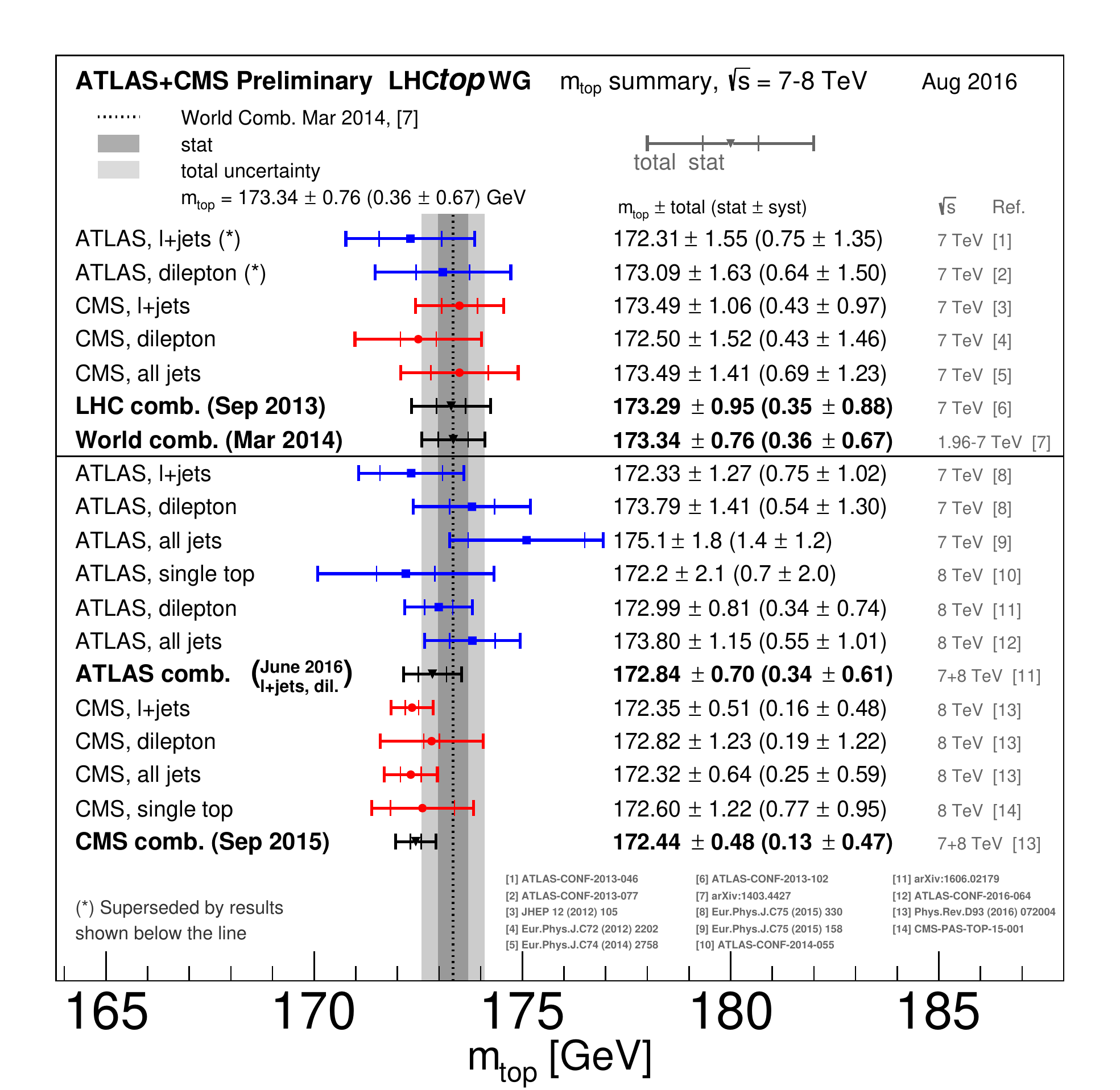}
  \caption{Compilation of recent top-quark mass measurements at the
    Tevatron~\cite{TevatronElectroweakWorkingGroup:2016lid} (left) and
    at the LHC~\cite{LHCTopSummary} (right).}
  \label{fig:mass}
\end{figure}

While kinematic \mt measurements in other \ttbar decay channels are
less precise, they still provide valuable cross-checks, as they
are subject to different systematic uncertainties and
may be influenced by \ac{bsm} physics differently.  The Tevatron
experiments have provided a full suite of further measurements, mostly
with the full Run-II dataset, in the single-lepton
channel~\cite{Abazov:2014dpa}, the dilepton
channel~\cite{Aaltonen:2015hta,D0:2015dxa,D0:2016ull}, for fully
hadronic \ttbar decays~\cite{Aaltonen:2014sea}, and for events with
jets and \acl{met}~\cite{Aaltonen:2013aqa}.  With the large \ac{lhc}
datasets and better understanding of systematic effects, also the
precision of \mt measurements in these channels is improving, as
seen from the \ac{lhc} results in the single and dilepton
channel~\cite{Aad:2015nba,Chatrchyan:2012cz,Chatrchyan:2012ea,Khachatryan:2015hba}
and the fully hadronic
channel~\cite{Aad:2014zea,ATLAS:2016byb,Chatrchyan:2013xza,Khachatryan:2015hba}.
As shown in Fig.~\ref{fig:mass}, all results are compatible with each
other within uncertainties.

\subsection{Alternative Methods to Extract The Top-Quark Mass}
Various alternative methods to extract \mt from experimental data have
been proposed. These methods may be separated into alternative
kinematic methods with systematic uncertainties that are partly
uncorrelated to those in standard methods and into methods that
extract \mt from other physics observables such as \xsec{}s and thus
avoid the ambiguities in the definition of \mt.

\subsubsection{Kinematic Methods}
One class of alternative methods is designed to avoid calorimetric jet
reconstruction and its limited resolution and scale uncertainties but
rather to use purely track-based observables to determine \mt. The
most important uncertainties when applying these methods are the
modeling of the kinematic properties of the top quark and of the
$b$-quark fragmentation.

\paragraph{Decay Length and Lepton Transverse Momentum}
As virtually all top-quark decays involve a \bjet, the \bjet
properties can be used to determine \mt.  In a method known as the
``\Lxy method''\footnote{The \Lxy method is named after the
  two-dimensional projection of the distance between the primary
  vertex and the \bhad decay vertex in the transverse plane, usually
  denoted by \Lxy.}---pioneered at CDF---the correlation of \mt and
the \bhad decay length is
exploited~\cite{Hill:2005zy,Aaltonen:2009hd}. The decay length is
determined from track information only, and the main systematic
uncertainties of the method originate from the \ac{mc} modeling of the
track multiplicity in \bhad decays. Another measure of \mt exploited
in~\cite{Aaltonen:2009hd} is the transverse momentum of the charged
lepton ($e$ or $\mu$) from the leptonic decay of a $W$ boson coming
from the top quark. The combination of both leads to a 4\% uncertainty
on \mt, which is dominated by the limited size of the CDF Run-II
dataset. The \Lxy method has also been applied to the CMS dataset at
$\sqrt{s}=\SI{8}{TeV}$ in a preliminary study~\cite{CMS:2013cea},
achieving a precision of 1.9\% on \mt. The dominant systematic
uncertainties for the \Lxy and lepton transverse momentum methods come
from the lepton moment scale and the \ac{mc} modeling of signal,
background, and the track multiplicity in \bhad decays.  A further CMS
analysis exploiting \bjet{}s from top-quark decays extracts \mt from
the peak position in the energy spectrum of \bjet{}s in the laboratory
frame---a method proposed in~\cite{Agashe:2012bn}---with a precision
of 1.7\% on the same dataset~\cite{CMS:2015jwa}.

\paragraph{Invariant Mass of Final-State Particle Combinations}
Another mass-dependent kinematic observable used to determine \mt that
is based only on the momenta of charged particles is the invariant
mass of the charged lepton from a leptonic $W$ boson decay and the
secondary vertex of a \bhad decay, both from a top-quark decay. In a
CMS measurement, a precision of 0.9\% has been
achieved~\cite{Khachatryan:2016wqo}. With the large \ac{lhc} datasets
also a more exclusive \ttbar decay is accessible, in which the \bhad
decays according to $B\to\jpsi+X\to\mumu+X$. As a proxy for \mt, the
invariant mass of the charged lepton from a leptonic $W$ boson decay
and the \jpsi is employed~\cite{Kharchilava:1999yj}. A first result
using this technique has been published recently based on the full CMS
dataset at $\sqrt{s}=\SI{8}{TeV}$~\cite{Khachatryan:2016pek}. The
precision of the result is 1.8\%, limited by the available statistics.

The reconstruction of \mt in the dilepton channel is kinematically
underconstrained due to the two undetected neutrinos in the final
state. In this case kinematic observables are explored whose shapes,
peaks, edges, or endpoints are sensitive to \mt. One example is the
invariant mass of the \bjet and the charged lepton from a top-quark
decay, $m_{b\ell}$.  Also observables initially developed for \ac{bsm}
physics searches with semi-invisible final states can be employed,
such as ``stransverse mass''
$m_{T2}$~\cite{Lester:1999tx,Aaltonen:2009rm,ATLAS:2012poa}. A recent
CMS measurement using $m_{b\ell}$ and $m_{T2}$ reports a precision on
\mt of 0.6\%~\cite{CMS:2016kgk}, becoming competitive with results
from standard kinematic methods.

In the dilepton channel, \mt can also be determined from leptonic
observables alone~\cite{Frixione:2014ala}. For example, the transverse
momentum of the \lplm pair, $\pt(\lplm)$ turns out to be sensitive to
\mt and robust against modeling uncertainties. Sensitivities better
than 2\% are obtained from the shape of the $\pt(\lplm)$ distribution
and its first and second moments~\cite{CMS:2016xfv}.

Further kinematic methods to measure \mt are based on the ratio of the
three-jet to the two-jet invariant mass, where one of the three jets
is a \bjet~\cite{CMS:2015agz}. In this method, the shape of the
combinatorial background from wrong assignments of jets to the
hadronic \ttbar decay products, is determined by mixing jets from
different events. A precision of 0.6\% is achieved based on the full
CMS dataset at $\sqrt{s}=\SI{8}{TeV}$.

\subsubsection{Top-Quark Mass from Single-Top Quark Events}
While traditionally the top-quark mass has been extracted from \ttbar
events only, the large \st datasets at the \ac{lhc} also allow for \mt
measurements based on the kinematic reconstruction of \tch \st
events. The top-quark mass has been extracted from fits to the
invariant mass distribution of the charged lepton, the neutrino, and
the \bjet from the top-quark decay, $m_{\ell\nu b}$, with a precision
of up to 0.7\%~\cite{ATLAS:2014baa,CMS:2016fdm}.

\subsubsection{Cross-Section Methods}
\label{sec:xsecmass}
\begin{figure}[t]
  \centering
  \includegraphics[width=0.5\textwidth]{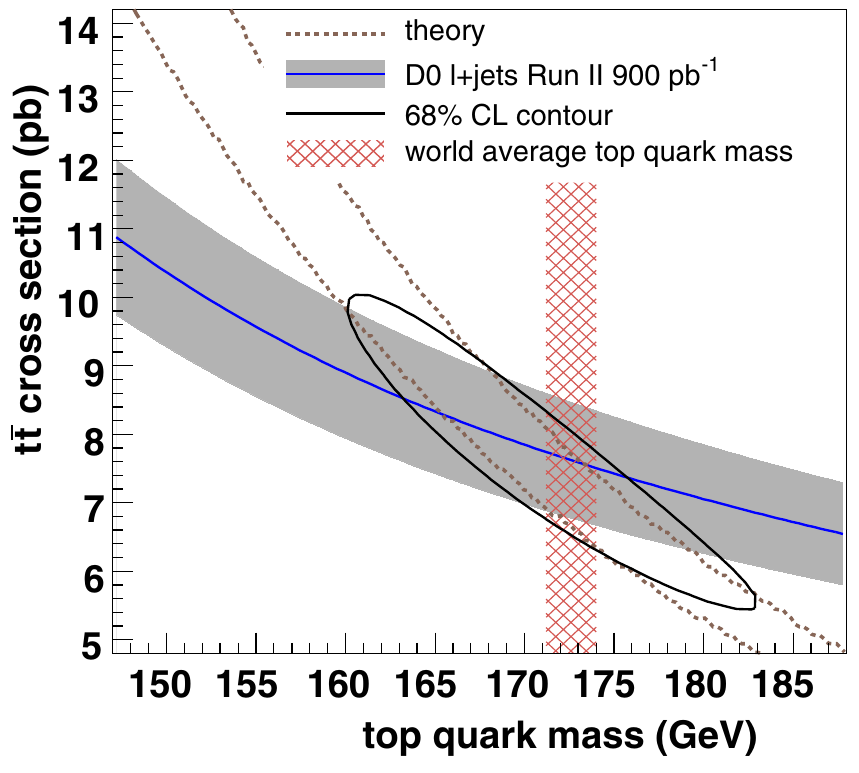}
  \caption{First determination of the top-quark mass from the \ttbar
    production \xsec. The \xsec is obtained from the intersection of
    the measured and the theoretical \ttbar \xsec as a function of
    \mt\cite{Abazov:2008gc}.}
  \label{fig:mass_xsec}
\end{figure}

The inclusive \ttbar production \xsec{} predicted by perturbative
\ac{qcd} is a steeply falling function of \mt.
 In a given renormalization scheme, \eg the
on-shell scheme or the \MSbar scheme, the \mt parameter in the \ttbar
\xsec{} is defined unambiguously. The measured inclusive \xsec also
has a weak dependence on the \mt parameter used in the \ac{mc}
simulation: Because of the larger momenta transferred to the \ttbar
decay products with larger \mt, the acceptance for \ttbar events
increases slightly with \mt, hence the measured \xsec decreases, see
Eq.~(\ref{eq:xsec}). The top-quark mass can be determined from the
intersection of the curves describing the \mt dependence of the
theoretical and the measured cross section, as illustrated in
Fig.~\ref{fig:mass_xsec}. This method has been pioneered by
D0~\cite{Abazov:2008gc} and has been applied both at the Tevatron and
at the \ac{lhc}. The results have reached a precision on the pole mass of the
top-quark below 2\% at the Tevatron~\cite{Abazov:2016ekt} and approximately
1\% at the \ac{lhc}~\cite{Aad:2014kva,Khachatryan:2016mqs}.

Also differential \xsec{}s can be used to extract \mt. In \ttbar
events with one additional jet, the normalized differential \xsec as a
function of the observable $\rho_s = 2m_0 / \sqrt{s_{\ttbar j}}$ is
sensitive to the top-quark pole mass \mtpole~\cite{Alioli:2013mxa}:
\begin{equation}
\mathcal{R}(\mtpole,\rho_s) = 
\frac{1}{\sigma_{\ttbar j}}
\frac{\mathrm{d}\sigma_{\ttbar j}(\mtpole,\rho_s)}{\mathrm{d}\rho_s},
\end{equation}
where $\sigma_{\ttbar j}$ is the inclusive or fiducial \xsec for
$\ttbar$+1\,jet production, $m_0$ is a mass scale of the order of \mt
and $\sqrt{s_{\ttbar j}}$ is the invariant mass of the \ttbar+1\,jet
system in the final state. The \ac{lhc} experiments have presented
first \mtpole measurements based on the observable
$\mathcal{R}$~\cite{Aad:2015waa,CMS:2016khu}, which reach a precision
of up to 1.3\%.

\subsection{Combinations of Top-Quark Mass Results}

For an observable like \mt additional precision may be gained by
combining measurements from different \ttbar decay channels within the
same experiment, and from different experiments, that are at least
partially uncorrelated, see Sections~\ref{sec:combination}
and~\ref{sec:statcomb}. Combined results are also considered more
robust if they contain channels with complementary systematic
uncertainties. At the Tevatron the most precise measurements of \mt
from the full Run~I and Run~II datasets of CDF and D0 have recently
been combined~\cite{TevatronElectroweakWorkingGroup:2016lid}. The
\ac{lhc} experiments have provided combinations of their results
individually~\cite{Aaboud:2016igd,Khachatryan:2015hba}. Tevatron
results on \mt have been combined with results from the \ac{lhc} in
the first (and so far only) \mt ``world combination'' in
2014~\cite{ATLAS:2014wva}.  All CMS top-quark mass measurements using
alternative methods have also been combined
recently~\cite{CMS:2016nxz}. The value of \mt obtained in this
combination is in excellent agreement with, but less precise than, the
earlier CMS combination of high-precision kinematic \mt
measurements~\cite{Khachatryan:2015hba}.

The \mt combination results are summarized in
Table~\ref{table:masscombi}. The central \mt values and the
uncertainties are usually driven by the one or two most precise
measurements, but show a moderate reduction of total uncertainty with
respect to individual results.

\begin{table}[t]
  \centering
  \small
  \caption{Recent \mt combinations at the Tevatron and the
    \ac{lhc}. Shown are the central values together with the
    statistical and systematic uncertainties and the total relative
    uncertainty of the combinations.}
  \label{table:masscombi}
  \vspace{1mm}

  \begin{tabular}{lccccl}
    \toprule
    Experiment & \mt (\si{GeV}) & stat (\si{GeV}) & 
    syst (\si{GeV}) & total (\%) & Ref.\\
    \midrule
    CDF + D0        & 174.30 & 0.27 & 0.71 & 0.44 & \cite{TevatronElectroweakWorkingGroup:2016lid}\\
    ATLAS           & 172.84 & 0.34 & 0.61 & 0.40 & \cite{Aaboud:2016igd} \\
    CMS             & 172.44 & 0.13 & 0.47 & 0.28 & \cite{Khachatryan:2015hba} \\
    CMS Alternative & 172.58 & 0.21 & 0.72 & 0.43 & \cite{CMS:2016nxz}\\
    \midrule
    World 2014 & 173.34 & 0.36 & 0.67 & 0.44 & \cite{ATLAS:2014wva}\\
    \bottomrule
  \end{tabular}
\end{table}

\subsection{Top Quark-Antiquark Mass Difference}
A measurement of the mass difference between top quarks and
antiquarks, $\Delta\mt$, is a test of the invariance of the \ac{sm}
under the simultaneous transformations of charge conjugation, parity,
and time reversal (CPT). Such measurements have been conducted both at
the Tevatron~\cite{Abazov:2011ch,Aaltonen:2012zb} and at the
\ac{lhc}~\cite{Chatrchyan:2012uba,Aad:2013eva,Chatrchyan:2016mqq}. The
analyses begin by performing a kinematic fit of the \ttbar system in
single-lepton events, where the top quark and antiquark are
distinguished by the charge of the lepton\footnote{The probability to
  misreconstruct the charge of a high-\pt electron is very small and
  negligible for the charge of a high-\pt muon.}  and the fit does not
assume \mt and \mtbar to be equal. In the next step methods similar to
\mt measurements are applied, \eg the ideogram method in CMS and an
unbinned \ac{ml} fit to the reconstructed $\Delta\mt$ distribution in
ATLAS.  All results obtained so far are consistent with the top quark
and antiquark masses being the same, as required by CPT
invariance. The uncertainty of the most precise measurement of
$\Delta\mt$ to date is~\SI{210}{MeV}~\cite{Chatrchyan:2016mqq}.

\subsection{Summary}
From measurements of \mt at the Tevatron and the \ac{lhc}, the
top-quark mass has been determined with excellent precision, rendering
the top-quark mass the most precisely known quark mass. The most
precise methods are based on the kinematic reconstruction of the final
state, with a current precision of approximately \SI{500}{MeV} or
0.3\%. Recently results obtained from a large number of alternative
methods to extract the top-quark mass have become available, with
complementary systematic uncertainties and reaching a precision up to
\SI{1}{GeV} or 0.6\%. The prospects of these methods with increasing
datasets at the \ac{lhc} will be discussed in
Section~\ref{sec:hl-lhc}.

\section{Top Quark Properties}
\label{sec:properties}
In the \ac{sm} the properties of the top quark are well defined, as
outlined in Section~\ref{sec:topsm}. However, in \ac{bsm} physics
models, various deviations from the \ac{sm} expectations are
predicted. Therefore measurements of the properties of the top quark
constitute tests of the \ac{sm} and often provide constraints on
\ac{bsm} physics models at the same time. There is a wide range of
top-quark properties to be tested: Basic properties include the
electric charge, the mass (already discussed in
Chapter~\ref{sec:mass}), and the decay width of the top quark. Further
insight into the production and decay properties of top quarks is
gained by studying production asymmetries, spin observables, and the
top quark's coupling structure in general. 

In many \ac{bsm} models, top quarks are preferred decay products of
new heavy particles, leading to observable resonances in invariant
mass spectra, or are produced in association with new particles, for
example dark-matter candidates.

In kinematic regions where the \ac{sm} predictions are known to
describe experimental data well, top quarks may also be used as a
powerful tool, for example to extract \btagging efficiencies,
\aclp{pdf}, or the strong coupling constant.


\subsection{Basic Top-Quark Properties}
\subsubsection{Top-Quark Electric Charge}
The electric charge of the top quark, which is $Q_t=2/3$ in the
\ac{sm}, can be determined from its coupling strength to the photon or
from the charge of its decay products. 

The coupling strength of the $\gamma tt$ vertex is related to the
\xsec for associated \ttg production, which has been studied both at
the Tevatron~\cite{Aaltonen:2011sp} and the
\ac{lhc}~\cite{Aad:2015uwa}, see Section~\ref{sec:ttX}. However, \ttg
final states can not only be produced by photons coupling to top
quarks, but also by photons coupling to other charged particles in the
initial and final state of the process. The corresponding scattering
amplitudes interfere, such that the interpretation as the top-quark
charge would required more sophisticated techniques, such as an
angular analysis of the final state, and therefore has not been
attempted yet.

Charge conservation in the decay $\ttbar \to W^+ W^- b \bbar$ allows
exotic heavy quarks with $Q=-4/3$ decaying to $W^- b$ instead of the
top quark decaying to $W^+ b$. From the combined charge of the decay
products $W$ boson and $b$ quark, the charge of the mother particle
can be inferred. In leptonic $W$-boson decays the $W$-boson charge
sign can be determined with great confidence from the charge sign of
the charged lepton. However the $b$ quark is a colored particle whose
charge information is diluted during hadronization. Experimentally the
charge sign can only be determined on a statistical basis by
constructing observables that infer the $b$ quark charge from the
charges of all particles in the corresponding \bjet after
hadronization, such as the \jetq observable~\cite{Field:1977fa}:
\begin{equation}
  \jetq = \frac{\sum_\mathrm{tracks}(\vec{p}_\mathrm{track} \cdot
    \vec{p}_\mathrm{jet})^\kappa \,Q_\mathrm{track}}
  {\sum_\mathrm{tracks}(\vec{p}_\mathrm{track} \cdot
    \vec{p}_\mathrm{jet})^\kappa},
\end{equation}
where the charge of each particle in the \bjet, $Q_\mathrm{track}$,
is weighted with the particle's momentum $\vec{p}_\mathrm{track}$
derived from the track, projected on the jet momentum axis
$\vec{p}_\mathrm{jet}$. The exponent $\kappa$ is a free parameter that
has been optimized for \ttbar events to be around $\kappa=0.5$.  The
two charge hypotheses $Q=2/3$ and $Q=-4/3$ can then be compared in a
statistical hypothesis test, typically with the product of \jetq and
the lepton charge as the test statistic. Measurements of the top-quark
charge based on \jetq have been performed both at the
Tevatron~\cite{Aaltonen:2013sgl,Abazov:2014lha} and at the
\ac{lhc}~\cite{Aad:2013uza}. The hypothesis that all $W^+ W^- b \bbar$
final states stem from exotic quarks with $Q=-4/3$ has been excluded
with a significance of more than eight standard
deviations~\cite{Aad:2013uza}. In the same publication, the charge of
the top quark is determined as
\begin{equation}
  Q_t = 0.64 \pm 0.02\,\mathrm{(stat)} \pm 0.08\,\mathrm{(syst)},
\end{equation}
compatible with the \ac{sm} expectation of $Q_t = 2/3$.

\subsubsection{Top-Quark Width and Lifetime} 
The total decay width of the top quark $\Gamma_t$, see
Eq.~(\ref{eq:width}), and its inverse, the top-quark lifetime
$\tau_t = 1/\Gamma_t$ can be determined indirectly from a combination
of two measurements: $\Gamma_t$ is proportional to
$\Vtbsq/\mathcal{B}(t\to Wb)$. The CKM matrix element factor \Vtbsq
can be accessed by comparing the measured value of the \xsec for \st
production in the \tch, $\sigma_{t\text{-ch}}$, to the theory
expectation $\sigma_{t\text{-ch}}^\mathrm{theo}$, see
Section~\ref{sec:singletop}. The ratio of branching fractions
$\mathcal{R}=\mathcal{B}(t\to Wb)/\sum_q\mathcal{B}(t\to Wq)$, where
the sum is over the down-type quarks, $q=d$, $s$, $b$, can be measured
from the number of \bjet{}s in \ttbar events. Assuming
$\sum_q\mathcal{B}(t\to Wq)=1$, \ie
$\mathcal{R}=\mathcal{B}(t\to Wb)$, the top-quark width is given by
\begin{equation}
  \Gamma_t =
  \frac{\sigma_{t\text{-ch}}}{\sigma_{t\text{-ch}}^\mathrm{SM}}
  \cdot
  \frac{\Gamma(t\to Wb)^\mathrm{SM}}{\mathcal{B}(t\to Wb)},
  \label{eq:topwidth}
\end{equation}
where $\Gamma(t\to Wb)^\mathrm{SM}\approx \SI{1.35}{GeV}$ is the
\ac{sm} expectation for the $t\to Wb$ partial decay width.
Measurements of $\Gamma_t$ based on Eq.~(\ref{eq:topwidth}) have been
performed both at the Tevatron~\cite{Abazov:2012vd} and at the
\ac{lhc}~\cite{Khachatryan:2014nda}. The more precise \ac{lhc} result
is based on a profile likelihood ratio fit to the \bjet multiplicity
in \ttbar dilepton final states. The fits leaves $\Gamma_t$ as a free
parameter and treats the \btagging and mistagging efficiencies as well
as the uncertainties of $\sigma_{t\text{-ch}}$ and
$\sigma_{t\text{-ch}}^\mathrm{SM}$ as nuisance parameters. The
resulting top-quark width of
\begin{equation}
  \Gamma_t = \left( 
    1.36 \pm 0.02\,\mathrm{(stat)} ^{+0.14}_{-0.11}\,\mathrm{(syst)}
  \right)\si{GeV}
\end{equation}
is in very good agreement with the \ac{sm} expectation.

The width of the top-quark can also be determined directly from the
kinematic reconstruction of its decay products, as performed by
CDF~\cite{Aaltonen:2013kna} and CMS~\cite{CMS:2016hdd}. Similar to
measurements of \mt, an observable sensitive to the top-quark width is
built from reconstructed quantities. In the more recent CMS analysis,
the observable is the invariant mass of charged lepton--\bjet pairs in
dilepton \ttbar events. In a series of binary hypothesis tests the
\ac{sm} value of $\Gamma_t$ is probed against different non-\ac{sm} width
hypotheses to extract a 95\%~\ac{cl}~central confidence interval of
$\SI{0.6}{GeV}<\Gamma_t<\SI{2.4}{GeV}$~\cite{CMS:2016hdd}. The
sensitivity of this direct method is lower than the sensitivity of the
indirect method described above.



\subsection{\ttbar Production Asymmetries}
\begin{figure}[t]
  \centering
  \includegraphics[width=0.6\textwidth]{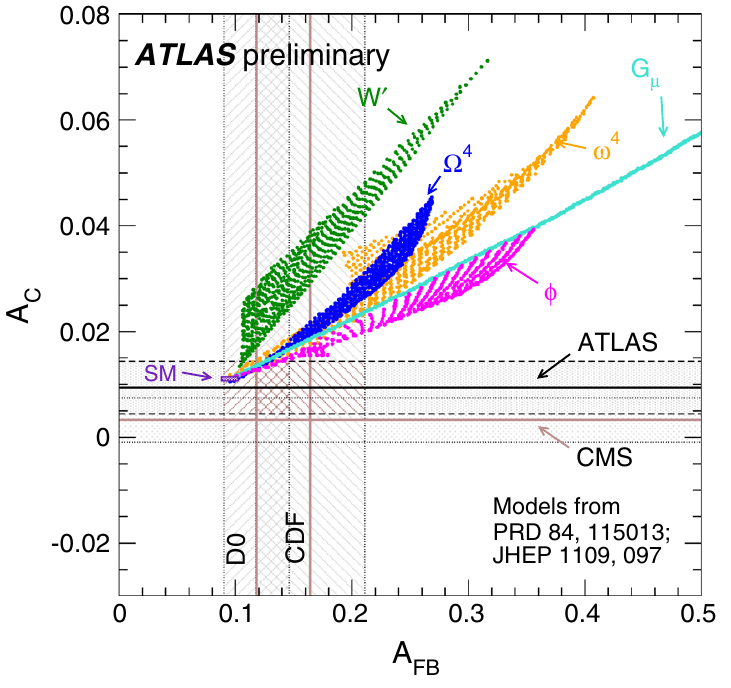}
  \caption{Measurements of \Afb at the Tevatron (vertical lines and
    bands) and \Ac at the \ac{lhc} (horizontal lines and bands)
    compared to the \ac{sm} prediction as well as several classes of
    \ac{bsm} models: heavy gauge boson ($W'$), heavy gluon with axial
    couplings $(G_\mu$), scalar isodoublet ($\phi$), color-triplet
    scalar ($\omega^4$), and color-sextet scalar ($\Omega^4$). The
    bands represent the 68\% \ac{cl} confidence
    intervals of the measurements~\cite{Aad:2015noh}.}
  \label{fig:Afb_vs_Ac}
\end{figure}

At \ac{lo} in \ac{qcd} perturbation theory, \ttbar pair production is
symmetric under the exchange of $t$ and \tbar. While the production
process $gg\to\ttbar$ remains symmetric also at \ac{nlo}, the process
$\qqbar\to\ttbar$ shows a small asymmetry~\cite{Kuhn:1998jr}.  The
asymmetry is caused by the interference of tree-level and one-loop
contributions to the squared amplitude for \ttbar production, which is
antisymmetric unter the exchange of $t$ and \tbar. Additional small
asymmetries arise from electroweak
corrections~\cite{Hollik:2011ps}. The observation of large \ttbar
production asymmetries would be a sign of \ac{bsm} physics. The
different initial states (\ppbar vs.\ $pp$) and the different
fractions of $gg$-initiated and $\qqbar$-initiated \ttbar production
at the Tevatron and the \ac{lhc} lead to different asymmetry
observables, the forward-backward asymmetry \Afb at the Tevatron and
the charge asymmetry \Ac at the \ac{lhc}, as defined
below. Fig.~\ref{fig:Afb_vs_Ac} shows that \ac{bsm} physics
contributions would influence \Afb and \Ac simultaneously but in
different ways depending on the model.

The expected magnitude and sign of the asymmetries depend on the
kinematic region considered; therefore, with increasingly large
datasets, differential asymmetries as a function of kinematic
observables with increasingly fine binning are being reported in
addition to inclusive asymmetries.  Further details on the physics of
\ttbar production asymmetries can be found \eg in a recent review
article~\cite{Aguilar-Saavedra:2014kpa}.

\subsubsection{Forward-Backward Asymmetries at the Tevatron}
At the Tevatron the \ttbar production asymmetry manifests itself as a
forward-backward asymmetry, generally defined as
\begin{equation}
  \Afb = \frac{N_F - N_B}{N_F + N_B},
\end{equation}
where $N_F$ is the number of forward events and $N_B$ is the number of
backward events. The forward-backward asymmetry in \ttbar production
is usually defined at parton level. The observable of choice is the
rapidity difference of the top quark and antiquark,
$\Delta y = y_t - y_\tbar$, which is invariant under Lorentz boosts in
the beam direction. The kinematics of the top quark and antiquark are
measured and then corrected back to the parton level using unfolding
techniques. The uncertainties on \Afb are dominated by the limited
size of the Tevatron datasets. The largest systematic uncertainties
originate from the background estimation and the \ac{mc} modeling of
hadronization.

Early measurements of \Afb at the Tevatron showed discrepancies to
\ac{nlo} \ac{qcd} preditions, in particular in events with large
\ttbar invariant masses~\cite{Aaltonen:2011kc}. However, the CDF and
D0 results with the full Run-II
dataset~\cite{Aaltonen:2012it,Aaltonen:2016bqv,Abazov:2014cca,Abazov:2015fna}
and the most recent \ac{sm} predictions with \ac{qcd} corrections up
to \ac{nnlo} ~\cite{Czakon:2014xsa} or approximate
\ac{n3lo}~\cite{Kidonakis:2015ona}, both with \ac{nlo} electroweak
corrections, are compatible within less than 1.5~standard deviations,
as shown in Table~\ref{table:Afb}. Also the differential \Afb
measurements show agreement with the \ac{sm} prediction at the level
of two standard deviations or
better~\cite{Aaltonen:2016bqv,Czakon:2016ckf}.

\begin{table}
  \centering
  \small
  \caption{Inclusive \ttbar production asymmetry results from the
    Tevatron compared to the most recent \ac{sm} predictions. The
    asymmetries are quoted together with their combined statistical and
    systematic uncertainties.}
  \label{table:Afb}
  \vspace{1mm}

  \begin{tabular}{lcl}
    \toprule
    Source & \Afb & Ref. \\
    \midrule
    CDF Combination & $0.160 \pm 0.045$ & \cite{Aaltonen:2016bqv}\\
    D0 Combination  & $0.118 \pm 0.028$ & \cite{Abazov:2015fna}\\
    \midrule
    NNLO \ac{qcd} + NLO electroweak             & $0.095 \pm 0.007$ & \cite{Czakon:2014xsa}\\
    approx.\ N$^3$LO \ac{qcd} + NLO electroweak & $0.100 \pm 0.006$ & \cite{Kidonakis:2015ona}\\
    \bottomrule
  \end{tabular}
\end{table}

A complementary approach to determine \ttbar production asymmetries is
to measure the charge asymmetry of leptons from \ttbar decays. These
may be defined as a function of the product of charge and
pseudorapidity of the leptons, $Q_\ell\cdot\eta_\ell$, or as the
pseudorapidity difference of the leptons in dilepton events,
$\Delta\eta = \eta_{\ell^+}-\eta_{\ell^-}$. \ac{sm} predictions for
the lepton asymmetries are available with \ac{nlo} \ac{qcd} and electroweak
corrections~\cite{Bernreuther:2012sx} and include cuts on the lepton
acceptance, resulting in a very small model dependence. The
experimental results from the full Tevatron Run-II
dataset~\cite{Abazov:2013wxa,Abazov:2014oea,Aaltonen:2014eva} are in
good agreement with the \ac{sm} predictions. A summary is given in
Fig.~\ref{fig:Afb} (left). With a full set of measurements using the
full Tevatron Run-II dataset and \ac{sm} predictions including
corrections beyond \ac{nlo}, no strong hints of \ac{bsm} physics in
\ttbar production asymmetries remain.

\subsubsection{Charge Asymmetries at the LHC}
The initial deviations from the \ac{sm} expectation for \Afb observed
during Tevatron Run~II also triggered an extensive measurement program
at the \ac{lhc}. Due to the symmetric $pp$ initial state at the
\ac{lhc}, \ttbar production asymmetries do not manifest themselves as
forward-backward asymmetries like in \ppbar collisions. Instead a
charge asymmetry \Ac can be observed, where top antiquarks from the
process $\qqbar \to\ttbar$ show a narrower rapidity distribution
compared to the top quarks from the same process. The process
$gg \to\ttbar$ remains charge-symmetric.  The charge asymmetry \Ac is
defined in terms of the difference of absolute rapidity,
$\Delta|y| = |y_t| - |y_\tbar|$:
\begin{equation}
  \Ac= \frac{N(\Delta|y|>0) - N(\Delta|y|<0) }
  {N(\Delta|y|>0) + N(\Delta|y|<0) }.
\end{equation}
In the \ac{sm}, the inclusive charge asymmetry is expected to be
small; the expectation at \ac{nlo} \ac{qcd} with electroweak
corrections at $\sqrt{s}=\SI{8}{TeV}$ amounts to
$\Ac = 0.0111 \pm 0.0004$~\cite{Bernreuther:2012sx}.  In many \ac{bsm}
physics models that predict larger \Afb compared to the \ac{sm}, also
significant deviations in \Ac are
expected~\cite{AguilarSaavedra:2011hz,AguilarSaavedra:2011ug}, see
also Fig.~\ref{fig:Afb_vs_Ac}.

Measurements of \Ac have been presented both at $\sqrts=\SI{7}{TeV}$
and at $\sqrts=\SI{8}{TeV}$. The sensitivities are similar, as the
smaller expected \Ac due to the larger fraction of
$gg$-initiated \ttbar events at $\sqrts=\SI{8}{TeV}$ is compensated by
the four-fold increase in dataset size. In the single-lepton channel,
ATLAS and CMS have measured inclusive asymmetries and asymmetries
differential in the invariant mass and transverse momentum of the
\ttbar system at parton
level~\cite{Chatrchyan:2012cxa,Khachatryan:2015oga,Aad:2013cea,Aad:2015noh}.
The inclusive \Ac measurements using data taken by ATLAS and CMS at
$\sqrts=\SI{7}{TeV}$ have been combined in the context of the
\ac{lhctopwg}~\cite{CMS:2014jua}.

As the charge asymmetry expected at the \ac{lhc} is small, much care
has been taken in unfolding the data, using a regularized matrix
unfolding technique in CMS and fully Bayesian unfolding in ATLAS, see
Section~\ref{sec:unfolding}.  This included detailed studies of the
correlations between the bins of the unfolded differential
asymmetries. Statistical uncertainties dominate the total uncertainty
for the measurements at $\sqrts=\SI{7}{TeV}$ while their size becomes
similar to the size of the systematic uncertainties for
$\sqrts=\SI{8}{TeV}$. The dominant systematic uncertainties stem from
the incomplete knowledge of the jet energy scale and resolution as
well as from \ac{mc} signal modeling, in particular for differential
asymmetries where migrations occur between bins of the differential
distributions within the systematic uncertainties.

To increase the \Ac sensitivity for invariant \ttbar masses above
\SI{750}{GeV}, where the Tevatron \Afb measurements hinted at tensions
with the \ac{sm}, ATLAS has also applied boosted top-quark
reconstruction techniques in an \Ac measurement~\cite{Aad:2015lgx}. In
addition to \Ac measurements based on unfolding, CMS has applied a
template method in an inclusive \Ac analysis.  The template method
results in smaller statistical uncertainties compared to unfolding, at
the expense of a larger model dependence~\cite{Khachatryan:2015mna}.

As for the \Afb measurements from the Tevatron, leptonic asymmetries
with reduced model dependence are accessible in \ttbar dilepton events
at the \ac{lhc}~\cite{Chatrchyan:2014yta, Khachatryan:2016ysn,
  Aad:2015jfa, Aad:2016ove}. The uncertainties of these measurement
are dominated by statistical uncertainties, followed by signal
modeling uncertainties.

\begin{figure}[t]
  \centering
  \begin{minipage}[b]{0.44\textwidth}
    \includegraphics[width=\textwidth]{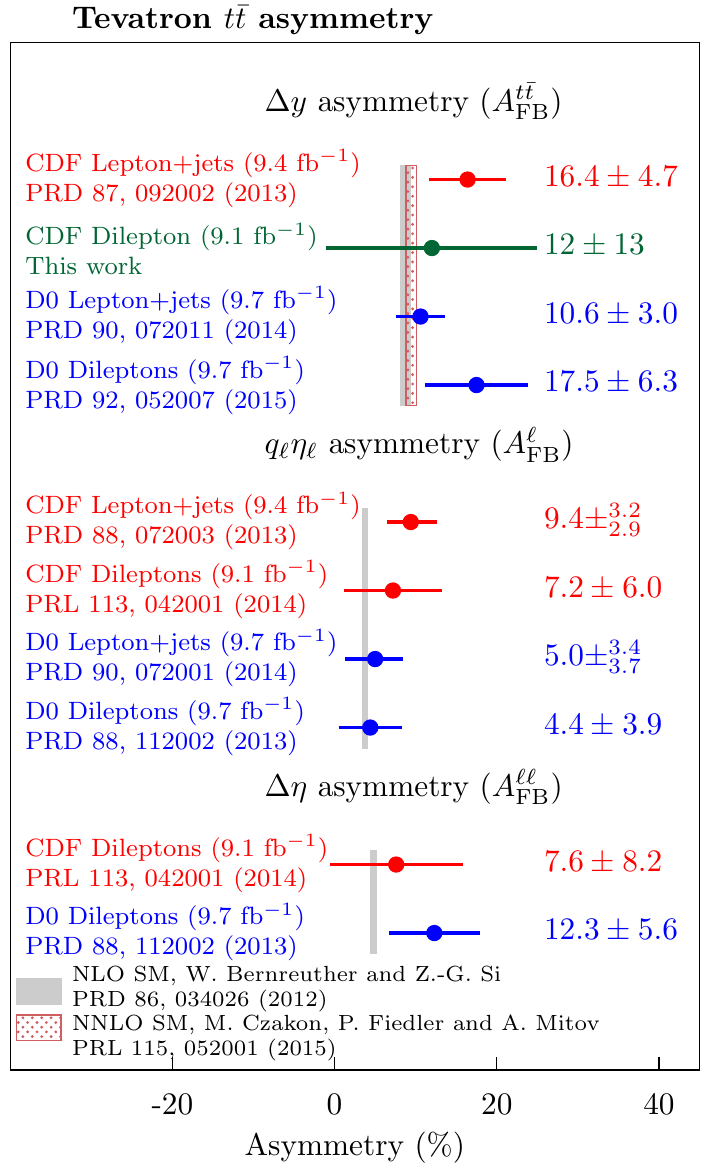}
  \end{minipage}
  \hspace{3mm}
  \begin{minipage}[b]{0.51\textwidth}
    \includegraphics[width=\textwidth]{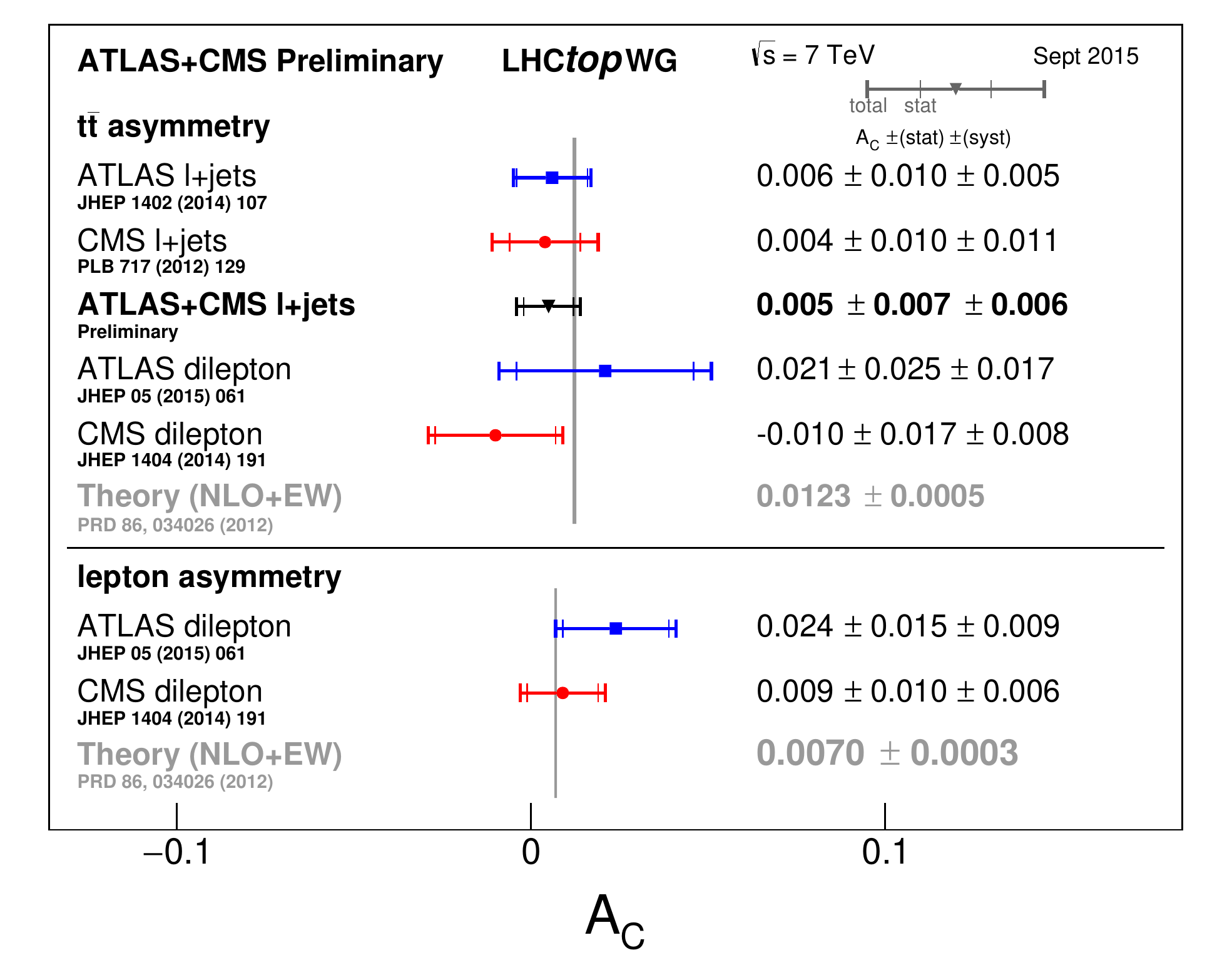}\\
    \includegraphics[width=\textwidth]{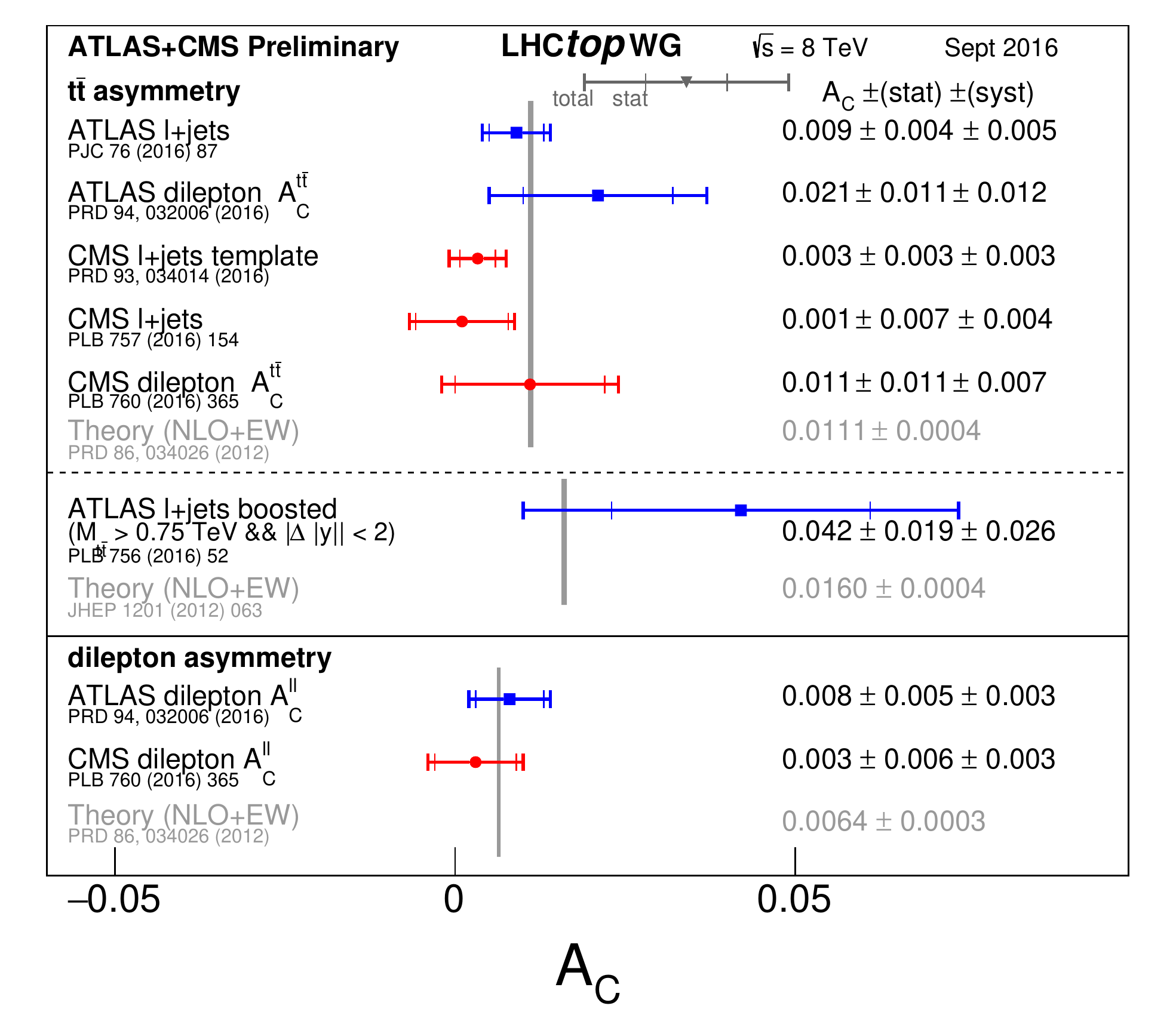}
  \end{minipage}
  \caption{Compilation of inclusive \ttbar forward-backward asymmetry
    results from the Tevatron~\cite{Aaltonen:2016bqv} (left) and of
    charge asymmetry results from the \ac{lhc}~\cite{LHCTopSummary}
    (right) compared to \ac{sm} predictions. Note the different \Ac
    axis scales in the \ac{lhc} summary plots for
    $\sqrt{s}=\SI{7}{TeV}$ (top right) and $\sqrt{s}=\SI{8}{TeV}$
    (bottom right).}
  \label{fig:Afb}
\end{figure}

All inclusive, differential, and leptonic asymmetry measurements at
the \ac{lhc} agree well with \ac{sm} predictions with \ac{nlo}
\ac{qcd} and electroweak corrections~\cite{Bernreuther:2012sx}. This
is illustrated in Fig.~\ref{fig:Afb} (right). It remains to be seen if
the even larger datasets, counteracted by the smaller expected \Ac due
to the larger fraction of $gg$ initial states at
$\sqrt{s}=\SI{13}{TeV}$, will allow for more stringent tests of \Ac at
\ac{lhc} Run~2.

\afterpage{\clearpage}


\subsection{Spin Observables in Events with Top-Quarks}
As introduced in Sections~\ref{sec:ttbarprod}
and~\ref{sec:singletopprod}, top quarks and antiquarks from \ttbar
production are expected to be essentially unpolarized in the \ac{sm},
but the $t$ and \tbar spins are correlated. On the other hand, single
top quarks produced via a $Wtb$ vertex are 100\% polarized. Likewise
$W$ bosons from the decay $t\to Wb$ are polarized. Measurements of
polarization observables sensitive to the above effects are tests of
\ac{sm} predictions for the top-quark couplings and may reveal
\ac{bsm} physics contributions to top-quark production or decay, or
both.

\subsubsection{$W$-Boson Polarization in Top-Quark Decays}

\paragraph{Observables}
The polarization of $W$~bosons stemming from top-quark decays is
measured as a differential \xsec in the observable \cosths.  The angle
$\theta^*$ is defined as the angle between the charged lepton or the
down-type quark from the $W$-boson decay and the $t$ or \tbar boost
direction\footnote{Most measurements consider both the $t$ and the
  \tbar, regardless of their decay mode, to exploit the full
  polarization information from the two $W$~bosons in each \ttbar
  event.} in the $W$-boson rest frame. As the above definition of
$\theta^*$ relies on parton-level information, a $W$-boson
polarization measurement requires either folding the polarization
effects into reconstructed observables or unfolding of the
reconstructed observables. Usually the first step to compute \cosths
is to reconstruct both the $t$ and the $\tbar$, from the \bjet and the
$W$-boson decay products, either $\ell \nu$ or the jets from
$q\qbar'$, employing a kinematic fit.

Assuming that the top quark and antiquark in a \ttbar event are
unpolarized, the polarization of each $W$~boson is decoupled from the
rest of the event and can be studied separately. The differential
production \xsec can then be expressed as a function of the fractions
of left-handed polarization ($F_L$), longitudinal polarization
($F_0$), and right-handed polarization ($F_R$) of the $W$~bosons
introduced in Section~\ref{sec:topsm}:
\begin{equation}
\frac{1}{\sigma}\frac{\mathrm{d}\sigma}{\mathrm{d}\cosths} =
\frac{3}{8}\left( 1 - \cosths\right )^2 F_L +
\frac{3}{4}\left( 1 - \cos^2\theta^*\right ) F_0 +
\frac{3}{8}\left( 1 +\cosths\right )^2 F_R.
\end{equation}
Measurements of the $W$-boson polarization fractions can be compared
with the \ac{sm} predictions directly and also be interpreted as
limits on anomalous $Wtb$ couplings, which will be discussed in
Section~\ref{sec:anomalousWtb}. The $W$-boson polarization has been
measured in \ttbar events with single-lepton decays both at the
Tevatron and the \ac{lhc} in fits that determine $F_L$ and $F_0$
simultaneously, taking into account the correlations between these
polarization fractions, and derive $F_R$ from the constraint
$F_L + F_0 + F_R = 1$\footnote{In the publications the polarization
  fractions are also quoted for the case when one fraction is fixed to
  the \ac{sm} expectation.}.

\paragraph{Tevatron Results}
Using the first \SIrange{2.7}{5.4}{\invfb} of Tevatron Run~II data,
the $W$-boson polarization has been measured using the \acf{mem} and
template-fit techniques. In the CDF \ac{mem} analysis, a likelihood
ratio discriminant is constructed from the \ac{lo} \ttbar production
matrix element parameterized as a function of the $W$-boson
polarization fractions~\cite{Aaltonen:2010ha}.  In the D0 template-fit
analysis, independent templates for the three polarization states as a
function of the reconstructed \cosths distribution are constructed
from simulated data and fitted to the data~\cite{Abazov:2010jn}.  The
results of these measurements have been
combined~\cite{Aaltonen:2012rz} to arrive at a relative uncertainty on
$F_L$ and $F_0$ of 11\%. From the full Run~II CDF dataset a relative
uncertainty of 13\% is achieved, again employing a
\ac{mem}~\cite{Aaltonen:2012lua}. Within their uncertainties the
$W$-boson polarization results obtained at the Tevatron are compatible
with the \ac{sm} predictions.

\begin{figure}[t]
  \centering
  \includegraphics[width=0.48\textwidth]{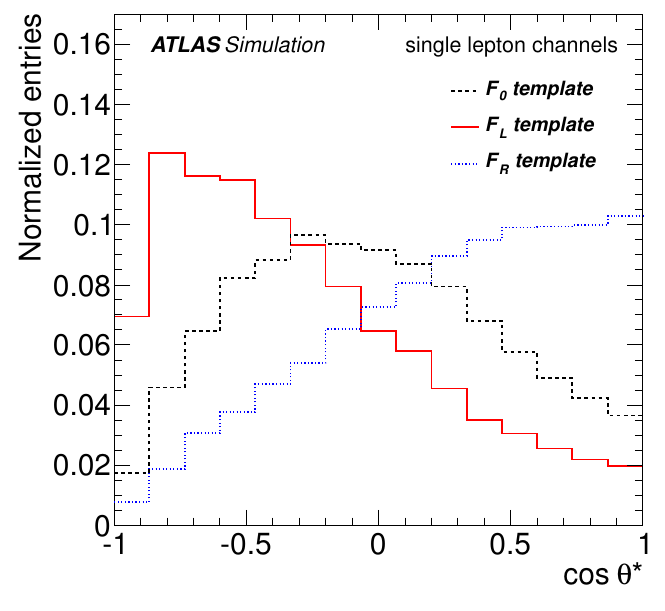}
  \hspace{3mm}
  \includegraphics[width=0.48\textwidth]{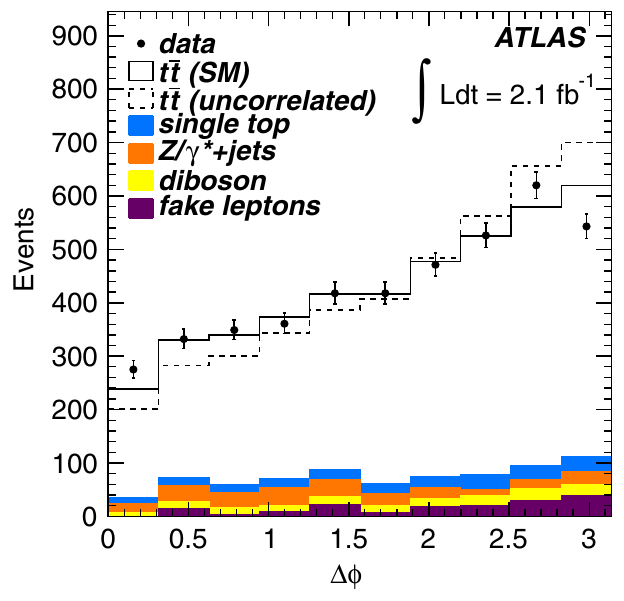}
  \caption{Simulated signal templates for longitudinal ($F_0$),
    left-handed ($F_L$), and right-handed ($F_R$) $W$ boson
    polarization as a function of \cosths~\cite{Aad:2012ky} (left).
    Reconstructed distribution of the angle $\Delta\phi$ between the
    two charged leptons in \ttbar dilepton decays in data compared to
    simulated \ttbar and background distributions (right). The
    simulated \ttbar distributions are shown for the case of spin
    correlations as expected in the \ac{sm} (solid line) and for the
    case of no spin correlations (dashed line)~\cite{ATLAS:2012ao}.}
  \label{fig:spincorr}
\end{figure}

\paragraph{LHC Results}
The ATLAS and CMS collaborations have performed their first set of
measurements of the $W$-boson polarization using \ac{lhc} Run~1 data
at $\sqrts=\SI{7}{TeV}$. CMS has extracted the $W$-boson polarization
from a template fit technique~\cite{Chatrchyan:2013jna}. ATLAS has
determined the $W$-boson polarization from a template fit and
additionally from a complementary set of observables that is based on
asymmetries in \cosths~\cite{Aad:2012ky}. An example of \cosths
templates is shown in Fig.~\ref{fig:spincorr}~(left).  The dominant
systematic uncertainties on these measurements are due to the modeling
of the \ttbar signal and the determination of the dominant \wjets
background, which is a source of unpolarized $W$~bosons.  Based on the
above individual ATLAS and CMS measurements, a combination was
performed in the context of the
\ac{lhctopwg}~\cite{ATLAS:2013tla}. The relative uncertainties on
$F_L$ and $F_0$ obtained in the combination are around 10\%. Improved
measurements have been presented using the $\sqrts=\SI{8}{TeV}$
dataset: the uncertainties on $F_L$ and $F_0$ were further reduced to
below 5\%~\cite{Khachatryan:2016fky}.

\begin{figure}[t]
  \centering
  \includegraphics[width=\textwidth]{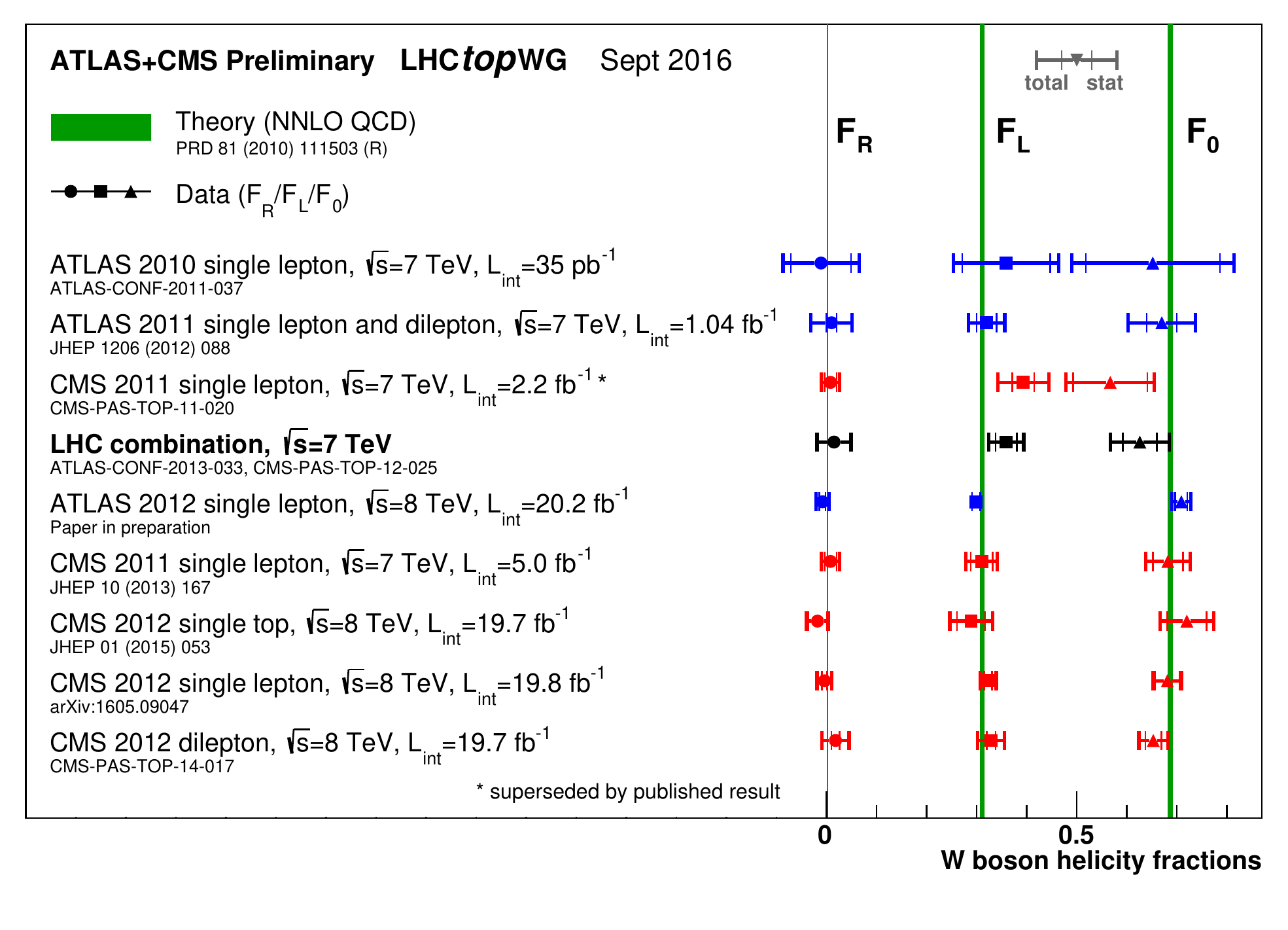}
  \caption{Compilation of \ac{lhc} results on the $W$-boson
    polarization in top-quark decays, expressed as the polarization
    fractions $F_R$, $F_L$, and $F_0$~\cite{LHCTopSummary}. The
    experimental results are compared to \ac{sm} predictions at
    \ac{nnlo}~\cite{Czarnecki:2010gb}.}
  \label{fig:Wpol}
\end{figure}

Also events with dilepton \ttbar decays~\cite{CMS:2015fja} as well
as events with a single reconstructed top
quark~\cite{Khachatryan:2014vma,ATLAS:2016dcn} have been used to
measure the $W$-boson polarization, albeit with larger uncertainties
than using single-lepton \ttbar events. The \ac{lhc} results on the
$W$-boson polarization are summarized in Fig.~\ref{fig:Wpol}. They are
compatible with the \ac{sm} predictions within their uncertainties.

\subsubsection{Top Quark Polarization and \ttbar Spin Correlations}
The top-quark polarization in \ttbar events is predicted to be very
small in the \ac{sm}~\cite{Bernreuther:2015yna}; the exact value
depends on the choice of the spin quantization axis, as discussed in
Section~\ref{sec:decays}. Because of the different initial states at
the Tevatron (\ppbar) and at the \ac{lhc} ($pp$), different degrees of
top-quark polarization are expected for a given set of quantization
axes.

In the dilepton decay channel, a clean way to extract the polarization
is to measure double-differential distributions of the polar angles
$\theta^+$ and $\theta^-$ for the positively and negatively charged
lepton with respect to a given set of spin quantization axes:
\begin{equation}
  \frac{1}{\sigma}
  \frac{\mathrm{d}\sigma}{\mathrm{d}\cos\theta_+\,\mathrm{d}\cos\theta_-}
  = \frac{1}{4}
  \left(
    1 + P_+\kappa_+ \cos\theta_+ + P_-\kappa_-\cos\theta_- - C\cos\theta_+\cos\theta_-
  \right).
\end{equation}
Here $P_\pm$ are the polarizations of the charged leptons and
$\kappa_\pm$ are their spin analyzing power, while $C$ is the spin
correlation coefficient. Instead of analyzing the full angular
distribution, the more robust spin asymmetry observable $A_s$ can
constructed. The spin asymmetry is related with the polarization and
the spin analyzing power via $A_{s,\pm} = P_\pm\kappa_\pm/2$. 

\paragraph{Top-Quark Polarization}
Top-quark polarization measurements from the Tevatron have been
presented first simultaneously with the leptonic and inclusive
forward-backward asymmetries~\cite{Abazov:2012oxa,Abazov:2015fna} with
which the polarization is anticorrelated. Also a dedicated D0
measurement of the top-quark polarization along several quantization
axes, including its transverse polarization, is
available~\cite{Abazov:2016tba}.  Direct measurements of the top-quark
polarization have also been presented by the \ac{lhc}
experiments~\cite{Aad:2013ksa,Khachatryan:2016xws,ATLAS-CONF-2016-099}.
These measurements start from the kinematic reconstruction of the
\ttbar event and proceed by unfolding distributions sensitive to the
top-quark polarization to parton or particle level. Earlier
measurements also used template fits to the reconstructed
distributions.  The most important uncertainties include those
originating from the \ttbar modeling; for some polarization
observables also the \ac{jes} uncertainty becomes relevant. Within the
current uncertainties, all top-quark polarization observables in
\ttbar events are compatible with the \ac{sm} expectation.

The top-quark polarization times spin analyzing power has also been
determined in single-top quark events. The main observable is the
angle~$\theta_\ell$ between the top-quark spin quantization axis and
the charged lepton from the top decay. Unfolding this observable to
parton level, values of $P\kappa\approx 0.9$ are
obtained~\cite{Khachatryan:2015dzz,ATLAS:2016dcn}, compatible with the
\ac{sm} expectation.

\paragraph{Spin Correlations}
The correlation between the spins of the \ttbar pair predicted by the
\ac{sm} is another interesting spin observable to be tested.  The
different dominant \ttbar production mechanisms at the Tevatron and
the \ac{lhc} make measurements of \ttbar spin correlations at the two
colliders complementary.  Recent \ttbar spin correlation measurements
have also be interpreted as limits on top-squark pair production in
supersymmetric models. The measurement is complementary to other
top-squark searches in that it probes top-squark masses close to \mt.

A very good observable to measure \ttbar spin correlations is the
difference $\Delta\phi$ in azimuthal angle of the two leptons in the
dilepton channel in the laboratory frame. From the $\Delta\phi$
distribution, the asymmetry in $\Delta\phi$ can be extracted as a
measure of spin correlations.  The spin correlations can be extracted
either from a template fit to the reconstructed data or from the
unfolded distribution. The hypotheses of fully correlated spins and
uncorrelated spins are then tested against each other in a hypothesis
test.  The first spin correlation measurements have been presented at
the Tevatron~\cite{Aaltonen:2010nz,Abazov:2011gi,Abazov:2015psg};
however, with significances for correlated spins below five standard
deviations. The first observation of \ttbar spin correlations with
more than five standard deviations was reported by
ATLAS~\cite{ATLAS:2012ao}, illustrated in Fig.~\ref{fig:spincorr}
(right).  More precise ATLAS and CMS measurements both at
$\sqrts=\SI{7}{TeV}$~\cite{Aad:2014pwa,Aad:2015bfa,Chatrchyan:2013wua}
and $\sqrts=\SI{8}{TeV}$~\cite{Aad:2014mfk} with similar analysis
strategies followed. A different approach is taken
in~\cite{Khachatryan:2015tzo}, where the single-lepton \ttbar decay
channel is considered instead of the dilepton channel.  A hypothesis
test is constructed from a likelihood ratio test statistic, for which
the \ac{lo} \ttbar matrix elements with and without spin correlations
are compared. The data do not show a clear preference for either
hypotheses, and from a template fit to the test statistic, the
fraction of \ttbar pairs with correlated spins is determined. Spin
correlations have also been observed with more than five standard
deviations significance in a recent top-quark polarization
measurement~\cite{ATLAS:2016dcn}.


\subsection{Anomalous Top-Quark Couplings}
Top-quark couplings can also be analyzed in a more general context:
The $Wtb$ vertex structure can be studied to constrain anomalous
couplings, as they occur in \ac{bsm} physics models. The most general
$Wtb$ coupling contains CP-conserving as well as CP-violating
contributions. Processes that change a quark's flavor without changing
its charge, known as \acf{fcnc} interactions, are forbidden in the
\ac{sm} at tree level and heavily suppressed at the level of quantum
corrections. Strong enhancement of \ac{fcnc} interactions in top-quark
production or decay would be a clear sign of \ac{bsm} physics. Due to
the special role the top quark is expected to play in many \ac{bsm}
physics models, it is plausible to assume that hypothetical new
particles with masses at the \si{TeV} scale have significant couplings
to the top quark. Therefore top quarks are preferred decay products of
heavy particles in many \ac{bsm} models. The large Yukawa coupling of
the top quark may also indicate a relation to \ac{dm} that can be
studied in the associated production of \ttbar pairs with \ac{dm}. In
the absence of new heavy resonances accessible experimentally, the
top-quark couplings may be studied in an \ac{eft} approach, in which
all heavy \ac{bsm} particles are ``integrated out'' and their effect
at energies accessible experimentally is parameterized in a
comprehensive set of effective couplings.

\subsubsection{Anomalous $Wtb$ Couplings and CP Violation}
\label{sec:anomalousWtb}
In \ac{bsm} physics models, the \ac{sm} $Wtb$ vertex may be modified.
The Lagrangian density corresponding to the most general $Wtb$ coupling 
structure extends Eq.~(\ref{eq:Wtbsm}) to read:
\begin{align}
\mathcal{L}_{Wtb} =& - \frac{g}{\sqrt{2}}\, \bbar\,\gamma^\mu
\left( f_V^L P_L + f_V^R P_R \right) t\, W_\mu^-\nonumber\\
&-  \frac{g}{\sqrt{2}}\, \bbar\, \frac{i\sigma^{\mu\nu} q_\nu
  }{\mW}
\left( f_T^L P_L + f_T^R P_R \right) t\, W_\mu^-
+\text{h.c.},
\label{eq:Wtbfull}
\end{align}
where $P_{L,R} = (1\mp\gamma_5)/2$,
$\sigma_{\mu\nu} = i [\gamma_\mu,\gamma_\nu]/2$, and $q_\nu$ is the
four-momentum of the $W$ boson. The Lagrangian contains left-handed
and right-handed vector and tensor couplings, expressed through the
complex coupling constants $f_V^{L,R}$ and $f_T^{L,R}$. In the \ac{sm}
at \ac{lo}, the only non-vanishing constant is $f_V^L=V_{tb}$, giving
rise to a purely left-handed \VminusA coupling structure. Non-zero
imaginary parts of the couplings could be either due to final-state
interactions or to CP violation, see
\eg\cite{Bernreuther:2008us}. 

Limits on the coupling constants in Eq.~(\ref{eq:Wtbfull}) have been
derived \eg in~\cite{Fabbrichesi:2014wva}.  There are also software
tools available to extract the coupling constants from fits to data:
The {\sc TopFit} program code~\cite{AguilarSaavedra:2010nx,TopFit} is
specialized to the $Wtb$ vertex, and {\sc
  EFTfitter}~\cite{Castro:2016jjv} is a more general software tool to
perform fits to arbitrary coupling structures, showcasing the above
$Wtb$ coupling model in the publication.

Anomalous $Wtb$ couplings have been studied in \tch \st production, by
interpreting measurements of the $W$-boson polarization in the
framework of
Eq.~(\ref{eq:Wtbfull})~\cite{ATLAS:2014dja,Khachatryan:2014vma}, by
measuring the differential production \xsec as a function of angular
variables~\cite{Aad:2015yem}, and by constructing asymmetries in
various angular distributions~\cite{ATLAS:2013ula,ATLAS:2016dcn}.
Major systematic uncertainties arise from the \ac{jes} calibration and
the \st signal modeling.  When interpreting the results, it should be
noted that the tightness of the constraints on the individual coupling
constants depends on the assumptions on the other constants. For
example, when fixing $f_V^L=1$ and $f_V^R=0$, the real part of $f_T^R$
can be constrained to better than $\pm0.08$, while the ratios of the
real and imaginary part of $f_T^R/f_V^L$ are much less constrained, of
the order of 0.2 to 0.3~\cite{Aad:2015yem}.  Within the current
measurement precision, all results from \ac{lhc} Run~1 agree with the
\ac{sm} predictions and limits on anomalous $Wtb$ couplings have been
set. The sensitivity of searches for anomalous $Wtb$ couplings is
expected to increase with the increased size of the data samples at
\SI{13}{TeV} in \ac{lhc} Run~2 compared to Run~1.

CP-violating observables can be constructed from the \ttbar decay
products in a framework with CP-violating
operators~\cite{Atwood:2000tu,Gupta:2009wu,Gupta:2009eq}. CP
asymmetries in four of these observables have been studied for the
first time at CMS~\cite{Khachatryan:2016ngh}. No signs of CP violation
in \st production or \ttbar decay have been found yet.

As \bhad{}s produced directly from \bbbar pairs or in hadronic
interactions, also those from top-quark decays undergo mixing and
decay. In $t\to Wb$ decays, the $b$ quark's charge sign at production
time can be determined from the charge sign of the lepton from the
$W$-boson decay. The charge sign at decay time can be obtained from
the soft lepton in a semileptonic \bhad decay. Based on the
measurements of these two charges, various charge asymmetries
sensitive to CP violation can be constructed. These charge asymmetries
are compatible with the \ac{sm} expectation within current
uncertainties~\cite{Aaboud:2016bmk}.

\subsubsection{Flavor-Changing Neutral Currents}
\ac{fcnc} top-quark interactions are interactions with a transition of
a top quark into another up-type quark $q$ ($q=u,c$) by coupling to a
neutral gauge boson ($\gamma$, $Z$, or $g$) or the Higgs boson.  In
the \ac{sm}, \ac{fcnc} interactions are forbidden at tree level. They
can occur via higher-order corrections, but are strongly suppressed
due to destructive interference effects in loop corrections, a variant
of the GIM mechanism~\cite{Glashow:1970gm}. For example the \ac{sm}
prediction for the \ac{fcnc} decay $t\to Zc$ is far below current
detection limits, $\mathcal{B}(t\to Zc)\approx\num{e-14}$.  On the
other hand, in many \ac{bsm} physics models the expected rates of
\ac{fcnc} processes are increased by several orders of magnitude,
see~\cite{Agashe:2013hma} for a recent review.

Searches for \ac{fcnc} interactions in the top-quark sector can be
pursued by searching for either production or decay channels in
addition to the channels predicted by the \ac{sm}, governed by one of
the above (effective) interactions.

\paragraph{FCNC Top-Quark Production}
The LEP experiments have set first limits on anomalous \st production
in the process $\epem\to tq$, with sensitivity to $\gamma tu$ and
$Ztu$
couplings~\cite{Heister:2002xv,Abdallah:2003wf,DELPHI:2011ab,Achard:2002vv,Abbiendi:2001wk}.
While the \xsec for \ac{sm} \st production at the $ep$ collider HERA
was too small to be detected, the HERA data were used to search for
anomalous \st production via the same $\gamma tu$ and $Ztu$ vertices
relevant at LEP~\cite{Chekanov:2008gn,Aaron:2009wp,
  Aaron:2009ab,Aaron:2009vv,Abramowicz:2011tv}. A concise review of
the HERA results on \ac{fcnc} top-quark production can be found
in~\cite{Behnke:2015qja}.

Hadron-collider searches for \ac{fcnc} processes mediated by $gtq$
vertices are performed best as searches for anomalous \st production,
as the decay $t\to gq$ is overwhelmed by \ac{qcd} multijet background.
Similar to the $Wtb$ vertex, the flavor-changing $gtq$ vertex can be
parameterized in the most general way as
\begin{equation}
\mathcal{L}_{gtq} = \frac{\kappa_{gtq}}{\Lambda}\, g_s\, \qbar\, \sigma^{\mu\nu}
\frac{\lambda^a}{2} t\,G_{\mu\nu}^{a}, 
\label{eq:gtq}
\end{equation}
where $\kappa_{gtq}$ is the dimensionless coupling constant of the
interaction, $\Lambda$ is the expected \ac{bsm} physics scale,
$g_s = \sqrt{4\pi\alphaS}$ is the \ac{qcd} coupling, $\lambda^a$ are
the Gell-Mann matrices and $G_{\mu\nu}^a$ is the gluon field strength
tensor. 

\ac{fcnc} searches via anomalous \st production have been conducted
both at the Tevatron~\cite{Aaltonen:2008qr,Abazov:2010qk} and with
\ac{lhc} Run-1
data~\cite{Aad:2012gd,Aad:2015gea,Khachatryan:2016sib}. Similar to
measurements of the \ac{sm} \st production \xsec, these searches
employ multivariate methods to separate signal from background; hence
they share similar systematic uncertainties. However the searches are
performed in kinematic regions different from those of \st production
and the multivariate methods are optimized for the separation of a
possible \ac{fcnc} signal from the \ac{sm} background, including \st
production.  The current best 95\% \ac{cl} limits on the branching
fractions for $t\to gq$ are
$\mathcal{B}(t\to gu)<\num{2e-5}$~\cite{Khachatryan:2016sib} and
$\mathcal{B}(t\to gc)<\num{2e-4}$~\cite{Aad:2015gea}. In the framework
of Eq.~(\ref{eq:gtq}) these limits can also be expressed as limits on
$\kappa_{gtq}/\Lambda$.  In anomalous \st production associated with a
photon, also the $\gamma tq$ vertex can be
probed~\cite{Khachatryan:2015att}.

\paragraph{FCNC Top-Quark Decays}
Studying top-quark decays at the Tevatron and the \ac{lhc}, the
\ac{fcnc} $\gamma tq$, $Ztq$, and $Htq$ couplings can be probed. The
Tevatron experiments have searched for decays governed by these
couplings, for example for the decay $t\to Zc$ in events with two or
three charged leptons and jets. Template fits were performed to
observables sensitive to the final state of the \ac{fcnc} interaction.
In the absence of a significant signal, limits on the \ac{fcnc}
branching fraction were derived. For $\mathcal{B}(t\to Zq)$, the
Tevatron limits are of the order of a few
percent~\cite{Aaltonen:2008ac,Abazov:2011qf}, where the sensitivity
was limited by the dataset size and irreducible backgrounds, for
example \zjets production.

The much larger datasets recorded at \ac{lhc} Run~1 allow for more
stringent \ac{fcnc}
limits~\cite{Aad:2012ij,Chatrchyan:2012hqa,Aad:2015uza,Chatrchyan:2013nwa},
while using analysis techniques very similar to the Tevatron
experiments.  For example, a limit of
$\mathcal{B}(t\to Zq)<\num{5e-4}$ is obtained from the CMS data taken
at $\sqrts=\SI{7}{TeV}$ and
$\sqrts=\SI{8}{TeV}$~\cite{Chatrchyan:2013nwa}. The main systematic
limitations of these searches are uncertainties in the modeling of the
\ac{sm} \ttbar background and from the \ac{jes} calibration.

\begin{figure}[t]
  \centering
  \includegraphics[width=0.7\textwidth]{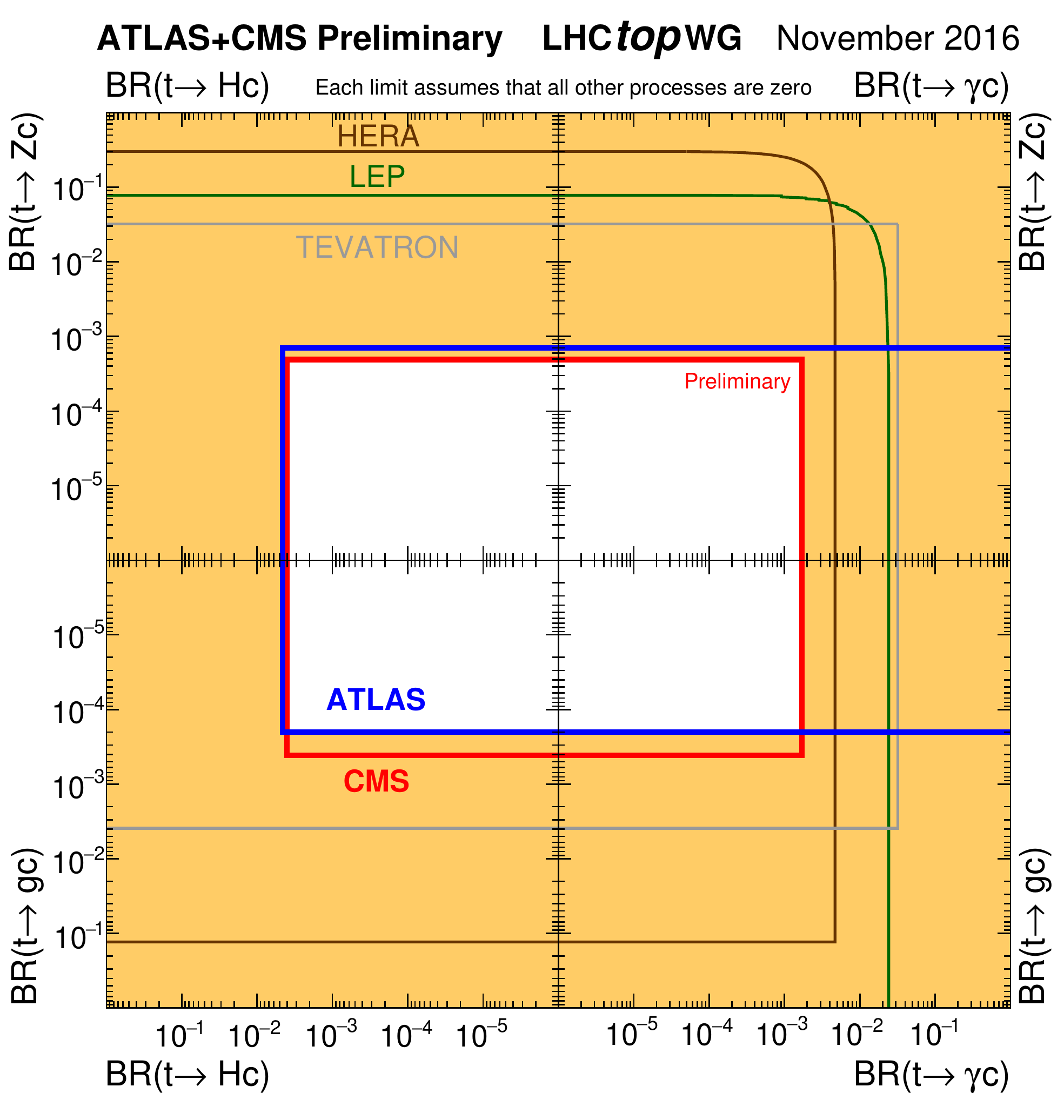}
  \caption{Summary of 95\% \ac{cl} limits on the branching fractions of
    \ac{fcnc} processes $t\to Xc$, where
    $X=g,Z,\gamma,H$~\cite{LHCTopWG}.}
  \label{fig:fcnc}
\end{figure}

At the \ac{lhc} also flavor-changing top-Higgs couplings ($Htq$) have
been studied~\cite{Aad:2015pja,Khachatryan:2016atv}. A search in final
states with two leptons with the same charge sign and with three
leptons is sensitive to Higgs-boson decays into $WW$, $ZZ$, and
$\tautau$ final states. In addition final states with a photon pair
and with a charged lepton and a \bjet are studied, to cover the decays
\Hgg and \Hbb. The current best 95\% CL limit of
$\mathcal{B}(t\to Hc)<\num{4.3e-3}$ is derived from a simultaneous fit
to suitable kinematic observables in all decay
channels~\cite{Khachatryan:2016atv}. A recent summary of limits on
\ac{fcnc} interactions is given in Fig.~\ref{fig:fcnc}.

\afterpage{\clearpage}

\subsubsection{Heavy-Particle Decays to Top Quarks}
Many \ac{bsm} models predict heavy particles that have top quarks
among their decay products. In order not to depend on a specific
\ac{bsm} model, heavy particles can be classified according to their
generic decay signatures that in turn depend on their quantum numbers
and couplings. Electrically neutral particles may decay into \ttbar
pairs, while charged particles may decay into a single top quark plus
another particle. While the larger center-of-mass energy makes
\ac{lhc} searches for heavy particles generally more sensitive for
masses above \SI{1}{TeV}, the Tevatron experiments were often able to
add complementary sensitivity to lower-mass particles. The large
Lorentz boost of the decay products of heavy particle with masses well
above \SI{1}{TeV} makes heavy-particle searches a prime field to apply
the boosted top-quark reconstruction techniques introduced
in~Section~\ref{sec:boosted}. 

\paragraph{Neutral Heavy Particles}
The signature of neutral heavy particle decays $X\to\ttbar$ is
resonances in the invariant mass spectrum of the \ttbar pair, \mtt. As
the ``true'' \mtt is only accessible at parton level, usually a proxy
for \mtt is computed from the four-momenta of reconstructed objects.
The decay width of a heavy resonance depends on the underlying
\ac{bsm} physics model, and narrow resonances lead to different
experimental signatures than wide resonances. Narrow resonances
feature decay widths of the order of a few percent of their mass,
comparable with the detector resolution. They are often represented by
a benchmark model with a ``leptophobic'' $Z'$ boson, a heavy neutral
gauge boson that only shows weak couplings to leptons (or else the
resonance would have been discovered in searches for $Z'\to\lplm$
decays already). Such $Z'$ bosons occur for example in
topcolor-assisted technicolor (TC2) models~\cite{Hill:1994hp} and are
often assumed to have a relative width of 1\% or 1.2\%. Wide
resonances show a decay width of 10\% of their mass or
above. Representative of wide resonances are Kaluza-Klein (KK) gluons
or gravitons, as they are predicted in Randall-Sundrum models of
warped extra dimensions with \ac{sm} particles propagating in the
five-dimensional bulk (RS2)~\cite{Randall:1999ee}.

\begin{figure}[t]
  \centering
  \includegraphics[width=0.465\textwidth]{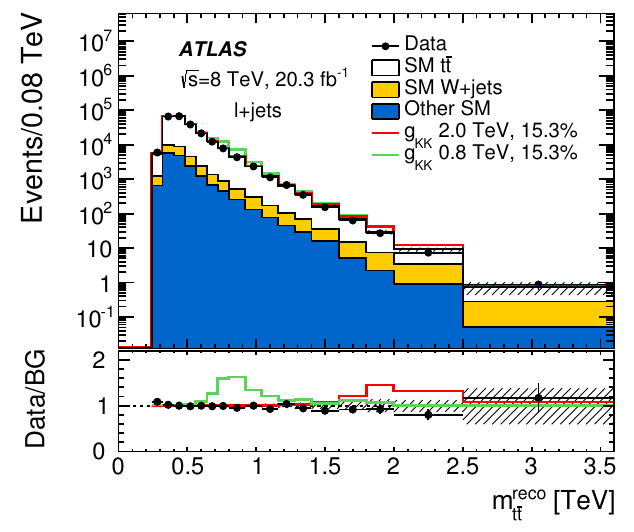}
  \includegraphics[width=0.525\textwidth]{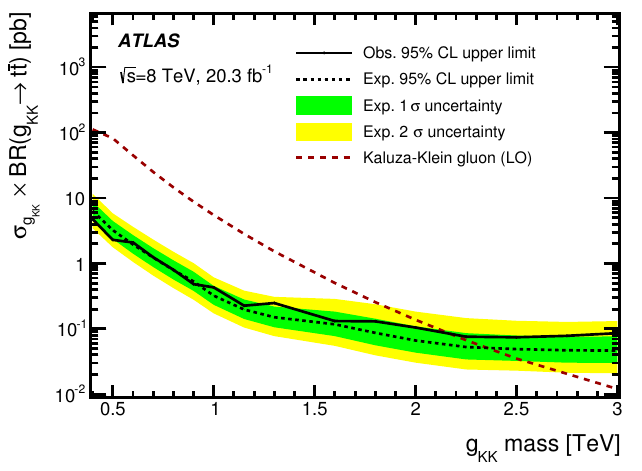}
  \caption{Reconstructed \mtt distribution for both boosted and
    resolved \ttbar decay channels (left). The distribution is overlaid with
    the expected distributions of hypothetical KK gluons with masses
    of \SI{0.8}{TeV} and \SI{2.0}{TeV} and a relative width of 15.3\%.
    Observed and expected 95\% \ac{cl} upper limits on the
    production \xsec times branching fraction to \ttbar final states
    of a KK gluon (right). The data exclude KK gluon masses between
    \SI{0.4}{TeV} and \SI{2.2}{TeV}. Taken from~\cite{Aad:2015fna}.}
  \label{fig:kkgluon}
\end{figure}

At the Tevatron, a CDF search excluded a narrow $Z'$ resonance in the
\mtt spectrum up to masses of \SI{915}{GeV} at
95\%~\ac{cl}~\cite{Aaltonen:2012af}, while D0 reports a slight excess
around $\mtt=\SI{950}{GeV}$, leading to a weaker
limit~\cite{Abazov:2011gv}.  The \ac{lhc} experiments have published
\ttbar resonance searches in both boosted and resolved decay channels
with the full dataset at
$\sqrts=\SI{7}{TeV}$~\cite{Aad:2013nca,Aad:2012ans,
  Aad:2012dpa,Chatrchyan:2012cx,Chatrchyan:2012yca} and
$\sqrts=\SI{8}{TeV}$~\cite{Aad:2015fna,Khachatryan:2015sma}.  The
searches target single-lepton, dilepton, and fully hadronic \ttbar
final states. The object reconstruction is adapted to the \mtt range
considered, for example with narrower fat jets and less isolated
charged leptons at large \mtt, well above \SI{1}{TeV}. The dominant
uncertainties are related to the modeling of the non-resonant \ttbar
background, the \ac{jes} of fat jets, and the \acp{pdf}.  As a result
of the \ac{lhc} searches for heavy resonances decaying into \ttbar
pairs, leptophobic $Z'$ bosons were excluded up to masses
of~\SI{2.4}{TeV} at 95\% \ac{cl}. If the data are interpreted in RS2
models~\cite{Agashe:2006hk}, KK gluons up to masses of~\SI{2.8}{TeV}
can be excluded at 95\% \ac{cl}. An example of a reconstructed \mtt
spectrum and its interpretation in terms of KK gluons is shown in
Fig.~\ref{fig:kkgluon}.  At the time of writing this review results of
\ttbar resonance searches with the first \ac{lhc} Run~2
from~2015~\cite{ATLAS-CONF-2016-014,CMS:2016zte,CMS:2016ehh} arrived
at sensitivities similar to Run~1. With more data at
$\sqrts=\SI{13}{TeV}$ improved sensitivities are expected.

In \ac{bsm} models with an extended Higgs sector, such as two-Higgs
doublet models, heavy scalar or pseudoscalar particles may decay to
\ttbar pairs. A search for a model in the single-lepton channel that
takes into account the interference with the \ac{sm} process
$gg\to\ttbar$ for the first time, is presented
in~\cite{ATLAS:2016pyq}. Due to the interference a peak-dip structure
is expected instead of a simple resonance peak, leading to reduced
sensitivity of ``bump hunt''-style analyses. The sensitivity is restored if
the interference effects are accounted for in the fit model.

\paragraph{Charged Heavy Particles}
A charged heavy gauge boson ($W'$) features a different set of decay
channels than a $Z'$ boson, \eg with decays into $tb$ ($t\bbar$ and
$\tbar b$). In contrast to the \ac{sm} $W$ bosons, arbitrary
combinations of left-handed and right-handed couplings to fermions are
allowed for $W'$ bosons.  For example, in a CDF
search~\cite{Aaltonen:2015xea} neural networks are trained to separate
a hypothetical $W'$-boson signal from background due to \ac{qcd},
\vjets and \ttbar processes. The data are fitted to a combined
neural-network discriminant to extract a limit on the $W'$-boson
mass. The uncertainties are dominated by the limited knowledge of $W$
and $Z$ boson production in association with heavy-flavor jets.  In a
scenario with purely right-handed $W'$ couplings the Tevatron
experiments have excluded $W'$ masses up
to~\SI{885}{GeV}~\cite{Abazov:2011xs,Aaltonen:2015xea}.

The \ac{lhc} experiments also used multivariate methods to select
$W'\to tb$ decays. The analyses arrived at 95\%-\ac{cl} limits of up
to~\SI{2.15}{TeV} for the same $W'$ model with only right-handed
couplings using data from \ac{lhc}
Run~1~\cite{Aad:2014xea,Chatrchyan:2014koa,Khachatryan:2015edz}.  With
the large center-of-mass energy available at \ac{lhc} Run~2, a simpler
approach was sufficient to supersede the Run-1 limits, a search for
bumps in the invariant mass spectrum of the $tb$ system, as
reconstructed from a charged lepton, jets and \acf{met}. No deviation
from the \ac{sm} prediction was observed and a 95\%-\ac{cl} limit
of~\SI{2.38}{TeV} was derived~\cite{CMS:2015aal}.

\paragraph{Vector-Like Quarks}
In addition to heavy bosons, heavy colored fermions are predicted in
many \ac{bsm} models. Heavier quarks with the same chiral couplings as
the six \ac{sm} quarks (``fourth-generation quarks'') have been
searched for at the Tevatron~\cite{Aaltonen:2011vr, Aaltonen:2011rr,
  Aaltonen:2011na, Aaltonen:2011tq, Abazov:2011vy}. However, a fourth
quark generation has been excluded by the fact that Higgs boson
production and decay rates are compatible with the \ac{sm}
prediction~\cite{Djouadi:2012ae}\footnote{Both the production channel
  $gg\to H$ and the decay channel \Hgg are mediated by fermionic
  triangle diagrams which are dominated by heavy quarks.}.  An
attractive alternative are \acp{vlq}, colored fermions that have
left-right symmetric couplings and do not have Yukawa couplings to
acquire their mass, see~\cite{Aguilar-Saavedra:2013qpa} for a recent
theory review.  As an example searches for vector-like heavy $T$
quarks with charge $Q=2/3$ will be discussed below. Vector-like $T$
quarks mix with the \ac{sm} top quark and decay into $Ht$, $Zt$, and
$Wb$ final states.  Other \acp{vlq} considered at the \ac{lhc} are
heavy $B$ quarks with charge $Q=-1/3$, and heavy $X$ quarks ($Q=5/3$)
and $Y$ quarks ($Q=-4/3$).

At hadron colliders, \acp{vlq} can be produced either singly
or in pairs, similar to top quarks. The \ac{lhc} experiments have
conducted searches for \acp{vlq} with Run-1 and in Run-2 data and in
various final states. The searches in \TTbar production
\cite{Aad:2014efa,Aad:2015kqa,Aad:2015gdg, ATLAS-CONF-2016-013,
  ATLAS:2016sno,ATLAS:2016qlg,
  ATLAS:2016cuv,ATLAS-CONF-2016-104,Chatrchyan:2013uxa,Khachatryan:2015axa,Khachatryan:2015oba,
  CMS:2016ete} generally aim at reconstructing an invariant mass
spectrum sensitive to resonances due to \acp{vlq}. For low invariant
masses regular resolved jets are used in the reconstruction, while for
high invariant masses, boosted-jet techniques are employed that also
allow for \btagging in the dense environment of fat jets.  No signs of
significant resonance peaks were observed and lower limits on the
$T$-quark mass \mT were derived, for example $\mT>\SI{1.10}{TeV}$ at
95\% \ac{cl} assuming
$\mathcal{B}(T\to Zt)=1.0$~\cite{ATLAS:2016qlg}\footnote{Note that
  depending on the \ac{bsm} model and the masses of the \acp{vlq} also
  other $T$ decays may be kinematically allowed, including decays into
  other \acp{vlq}.}.  In single $T$ quark production, stronger
exclusion limits can be
obtained~\cite{Aad:2016qpo,ATLAS:2016ovj,CMS:2016ccy,Sirunyan:2016ipo,CMS:2016usi,CMS:2016onb},
but these limits rely on assumptions on the $WTb$ coupling. Searches
for heavy $B$, $X$, and $Y$ quarks have also been conducted at the \ac{lhc},
see \eg\cite{ATLAS:2016sno,CMS:2016ete}.

\paragraph{Composite Top Quarks}
A feature of \ac{bsm} models with composite instead of fundamental top
quarks is excited top quarks ($t^*$). In~\cite{Chatrchyan:2013oba} a
search for $t^*\tbar^*\to tg\,\tbar g$ is documented. The final state
is reconstructed using a kinematic fit and the $tg$ invariant mass
distribution is scanned for resonances. With the analysis, excited
$t^*$ quarks with masses below \SI{803}{GeV} are excluded at 95\%
\ac{cl} in one specific Randall-Sundrum model~\cite{Hassanain:2009at}.

\paragraph{Summary on Heavy Resonances}
So far none of the searches for heavy particles decaying into top
quarks has provided a significant excess in the data compared to the
\ac{sm} prediction. The current best lower limits on the masses of
these heavy particles are summarized in Table~\ref{table:heavy}. With
additional Run-2 data from 2016 and beyond included in the analyses,
these limits are expected to improve significantly.  A comprehensive
review of \ac{lhc} searches for exotic new particles, including the
ones mentioned above, can be found in~\cite{Golling:2016thc}.
 
\begin{table}
  \centering
  \small
  \caption{Summary of the most stringent 95\% \ac{cl} limits on
    the mass of heavy particles decaying into top quarks.}
  \vspace{1mm}

  \begin{tabular}{lD{.}{.}{-1}ll}
    \toprule
    Heavy Particle & \multicolumn{1}{c}{95\%-\ac{cl} Mass Limit (TeV)} & 
    Experiment & Reference \\
    \midrule     
    Leptophobic $Z'$   & >2.4  & CMS \SI{8}{TeV} & \cite{Khachatryan:2015edz}\\
    Kaluza-Klein gluon & >2.8  & CMS \SI{8}{TeV} & \cite{Khachatryan:2015edz}\\
    Right-handed $W'$  & >2.38 & CMS \SI{8}{TeV} & \cite{CMS:2015aal}\\
    \midrule
    \ac{vlq} $T \to Zt$     & >1.10 & ATLAS \SI{13}{TeV} & \cite{ATLAS:2016qlg}\\
    \bottomrule
  \end{tabular}
  \label{table:heavy}
\end{table} 

\subsubsection{Top Quarks and Dark Matter}
\label{sec:dm}
In general, \ac{dm} can be searched for either in direct-detection
experiments, in which \ac{dm} particles scatter off baryonic matter,
or indirectly in \ac{sm} signatures of pair annihilation of \ac{dm}
particles in the universe, or in pair production under laboratory
conditions at colliders. \ac{dm} searches at colliders have the
distinct advantage that their interpretation is independent of
astrophysical input.  Classic \ac{dm} searches require at least a
single detectable \ac{sm} object recoiling against the undetected
\ac{dm} particles; the most well-known signature is a single high-\pt
jet (``monojet'') in association with a significant amount of
\ac{met}. Events with a \ttbar pair or a single top quark and large
\ac{met} ($\ttbar+\met$, $t+\met$) are also among the attractive
signatures studied in the quest for \ac{dm} at colliders.

\ac{dm} searches at the \ac{lhc} have been interpreted first in
\ac{eft}-based models~\cite{Aad:2014vea,Khachatryan:2015nua} and
limits on the \ac{dm}-nucleon \xsec as a function of the \ac{dm} mass
were derived. As a consequence of a coordinated effort between the
\ac{lhc} experiments for Run~2 (``Dark Matter Forum''), the focus
shifted to simplified models with defined benchmark
points~\cite{Abercrombie:2015wmb}.

In simplified models the interaction between top quarks and a
fermionic \ac{dm} candidate is mediated by a scalar or pseudoscalar
mediator particle. A recent comprehensive study of the collider
phenomenology of top-philic \ac{dm} is presented
in~\cite{Arina:2016cqj}. The $\ttbar+\met$ signature is also employed
in searches for pairs of top squarks in supersymmetric models. From
such top-squark searches with \ac{lhc} Run-2 data, the most sensitive
limits on the mass of the \ac{dm} and the mediator particles so far
were derived~\cite{ATLAS:2016ljb, ATLAS:2016xcm, ATLAS:2016jaa,
  CMS:2016mxc}.

Monotop signatures occur in models in which a new scalar resonance
decays into a top-quark and a colored ``dark'' fermion or in \ac{fcnc}
interactions producing a ``dark'' vector boson. The \ac{lhc}
experiments have searched for this signature with data taken at
$\sqrts=\SI{8}{TeV}$~\cite{Aad:2014wza,CMS:2016wdk} and recently also
in the $\sqrts=\SI{13}{TeV}$ data and using boosted-top
techniques~\cite{CMS:2016flr}. All measurements agree with the \ac{sm}
expectation and lower limits on the masses of the scalar resonance and
the ``dark'' vector boson have been placed, of the order of \SI{3}{TeV} at 95\%
\ac{cl} in a model in which the top quark and the \ac{dm} particle
originate from a heavy resonance.

\subsubsection{Top Couplings in an Effective Field Theory Approach}
\label{sec:effective}
In view of the null results of the searches for new heavy particles
decaying into top quarks, a comprehensive \acl{eft} approach to study
top-quark couplings becomes attractive. In such an approach the
indirect effects of \ac{bsm} physics on the top-quark couplings are
treated in a consistent way, by constructing a full set of effective
operators that mediate top-quark couplings with mass-dimension
six~\cite{Buchmuller:1985jz}:
\begin{equation}
\mathcal{L}_\mathrm{eff} = \mathcal{L}_\mathrm{SM} + 
\sum_i\frac{C_i^{(6)} O_i^{(6)}}{\Lambda^2} + 
\mathcal{O}(\Lambda^{-4}).
\end{equation}
In the above equation, the effective Lagrangian density
$\mathcal{L}_\mathrm{eff}$ is given by the \ac{sm} Lagrangian
$\mathcal{L}_\mathrm{SM}$ and a sum of dimension-six operators
$O_i^{(6)}$, each weighted with a Wilson coefficient $C_i^{(6)}$,
calculable in perturbation theory~\cite{Grzadkowski:2010es}. These
interactions are suppressed by the square of the new-physics scale
$\Lambda$. The operators relevant for top-quark interactions have been
worked out \eg in~\cite{AguilarSaavedra:2008zc,Zhang:2010dr}.

Compared to the anomalous coupling approaches discussed before the
\ac{eft} is the more comprehensive description of the top couplings in
a gauge-invariant and renormalizable way that respects all \ac{sm}
symmetries. Confronting the \ac{eft} approach with data requires a
global fit to Tevatron and \ac{lhc} data on differential cross
sections. A first global fit at \ac{lo} has been performed with the
{\sc TopFitter} software tool~\cite{Buckley:2015nca,Buckley:2015lku},
where the complementarity between \ac{lhc} and Tevatron measurements
has been demonstrated~\cite{Rosello:2015sck} and boosted-top final
states have been included~\cite{Englert:2016aei}. To match the
experimental precision, \ac{nlo} corrections to the \ac{eft} are being
worked out, with complications such as mixing of the operators in the
renormalization group evolution.  A first \ac{nlo} analysis of
\ac{fcnc} interactions in the top-quark sector in an \ac{eft}
framework has been presented in~\cite{Durieux:2014xla}.

\subsection{Top Quarks as a Tool}
\label{sec:tool}
Given the excellent understanding of the properties of the top quark,
the top quark is more and more considered a ``standard candle'' within
the \ac{sm}, similar to the role of the $W$ and $Z$ bosons at the
Tevatron and the \ac{lhc} so far. Events containing top quarks
can be used as a calibration source or as a reference for other
measurements. Top-quark production can also be used to better
constrain proton \acfp{pdf} and to measure the strong coupling constant~\alphaS.

\paragraph{$B$-Tagging Efficiency}
One unique property of \ttbar events is that they contain at least two
$b$-flavored quarks in the partonic final state. This can be exploited
in measurements of the \btagging efficiency \epsb in a ``busy''
environment with several jets and charged leptons, similar to the
signal region of many \ac{bsm} physics searches. Often the observable
of interest is not \epsb itself, but the \btagging scale factor \SFb,
defined as the ratio of \epsb obtained in a given data sample and in
an equivalent simulated data sample. One way of measuring \SFb is a
profile likelihood ratio fit in \ttbar dilepton candidate events,
similar to what is used in \xsec measurements presented in
Chapter~\ref{sec:production}.  The data sample is split into
categories according to the number of jets and the number of \bjet{}s
in the event and \SFb is extracted from the event counts in these
categories with a precision of up to
3\%~\cite{Chatrchyan:2012jua,Aad:2015ydr}. Another method to extract
\SFb is to apply a tag-and-probe technique to \bjet{}s in \ttbar
events.

\paragraph{Strong Coupling Constant and Parton Distribution Functions}
The strong coupling constant has been extracted from the \ttbar
production \xsec together with the pole mass of the top
quark~\cite{Chatrchyan:2013haa}. While \ac{nnlo} computations of
\ac{qcd} jet production at hadron colliders have only arrived in late
2016~\cite{Currie:2016bfm}, \ttbar production is already known to
\ac{nnlo} precision since 2013. Therefore the extraction of \alphaS
from \ttbar production constitutes the first \ac{nnlo} measurement of
\alphaS at a hadron collider. The resulting value of \alphaS evaluated
at the energy scale \mZ is
\begin{equation}
  \alphaS(\mZ) = 0.1151^{+0.0028}_{-0.0027},
\end{equation}
which tends to be lower than \alphaS values from other
sources.  The value is the only hadron collider result included into
the most recent world average value of
$\alphaS(\mZ) = 0.1181\pm0.0011$~\cite{Olive:2016xmw}.

Precision measurements and \ac{nnlo} calculations of the differential
\xsec for \ttbar production, see Section~\ref{sec:differential}, can
be used to include \ttbar data into proton \ac{pdf} fits, together
with other data, \eg from the HERA $ep$ collider. Including \ttbar
production improves the precision of the gluon \ac{pdf} at large
longitudinal momentum fractions $x$~\cite{Guzzi:2014wia}. As mentioned
in Section~\ref{sec:singletop}, also \st production can be used to
constrain \acp{pdf}. The ratio of single $t$ to single \tbar
production in the \tch is a measure of the ratio of the $u$ and $d$
quark
\acp{pdf}~\cite{Aad:2014fwa,Khachatryan:2014iya,Sirunyan:2016cdg};
however, this method is not yet precise enough to contribute to
\ac{pdf} fits significantly.

Also more general properties of hadron collision events can be
measured in \ttbar production. The color flow in \ttbar events has
been studied by measuring the pull angle between pairs of jets, which
is different for jet pairs coming from decays of color singlet and
color octet states~\cite{Aad:2015lxa}.  Another example is the
underlying event, defined as any hadronic activity not attributed to
the particles coming from the hard
scattering~\cite{CMS:2013mfa,CMS:2015usp}.

\section{Future Top-Quark Physics}
\label{sec:outlook}

Top physics has come a long way from discovery and first measurements
at the Tevatron in the 1990s to more and more sophisticated analyses
using Tevatron Run~II and \ac{lhc} data. In the top-quark physics
community the perspectives for the field are being evaluated, both for
the high-luminosity upgrade of the \ac{lhc} and for future lepton and
hadron colliders. As it is notoriously difficult to predict future
improvements and novel ideas, all of the projections presented below
should be taken with a grain of salt: all studies are only valid in a
context in which their underlying assumptions are valid as well.

\subsection{Towards the High-Luminosity LHC}
\label{sec:hl-lhc}
At the time of writing this review, Run~2 of the \ac{lhc} is in full
swing, with data-taking expected until 2018. Together with Run~3
(2021--2023) the \ac{lhc} will have recorded data corresponding to
about \SI{0.3}{\invab} of integrated luminosity. After that the ATLAS
and CMS detectors will undergo major upgrades, before the \ac{hllhc}
era will commence in 2026. At the end of data-taking at the \ac{hllhc}
in the late 2030s, integrated luminosities of the order of
\SI{3}{\invab} are expected.

The \ac{lhc} experiments have carried out some studies of the
top-quark physics potential at the \ac{hllhc} with projections of key
results, most prominently the expected uncertainty on the top-quark
mass. In the absence of obvious \ac{bsm} physics signals at the
\ac{lhc} so far, precision modeling of \ac{sm} backgrounds, often
including top quarks, is essential. The couplings of the top quark as
well as rare processes such as \acp{fcnc} may reveal deviations from
the \ac{sm}. The large top-quark datasets at high center-of-mass
energies will also allow investigation into new corners of the
kinematic phase space, such as very high invariant \ttbar masses, to
search for particles that decay into top quarks.

\begin{figure}[t]
  \centering
  \includegraphics[width=0.5\textwidth]{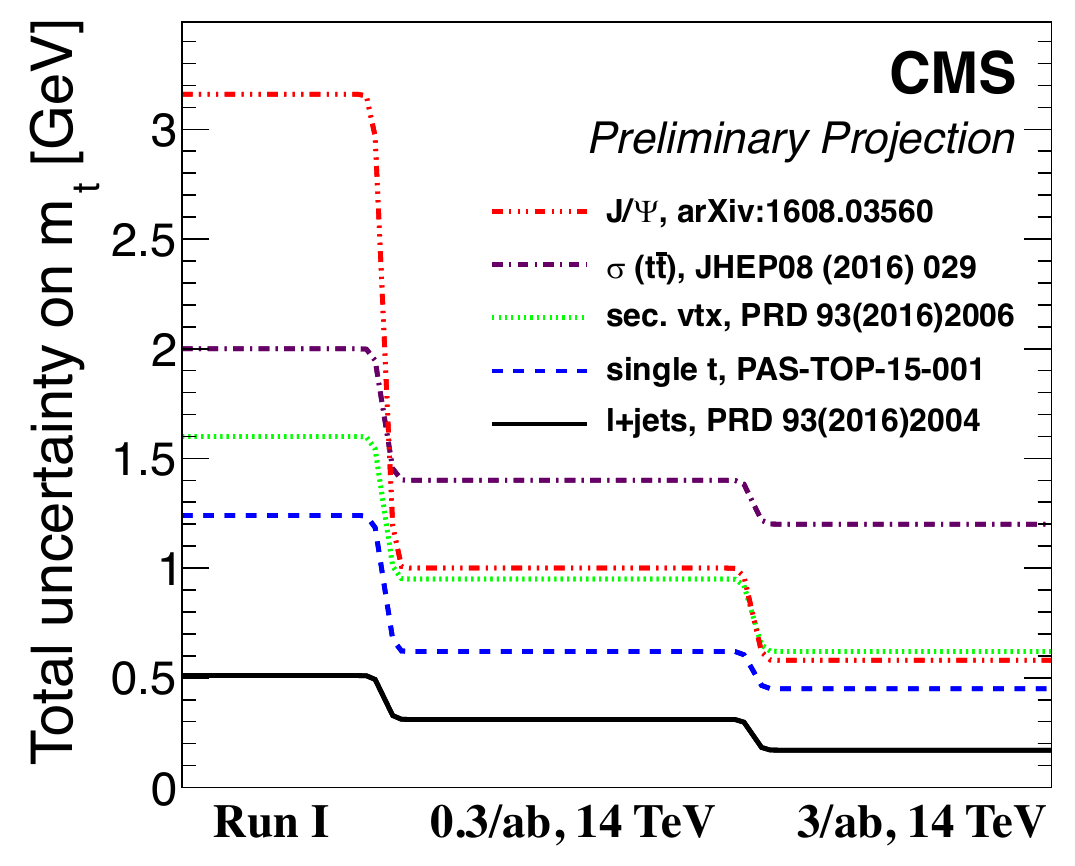}
  \caption{Total uncertainty on the mass of the top quark with
    different measurement methods comparing the uncertainty achieved
    in \ac{lhc} Run~1 with projections for integrated luminosities of
    \SI{0.3}{\invab} and \SI{3}{\invab} at
    $\sqrts=\SI{14}{TeV}$~\cite{CMS-DP-2016-064}.}
  \label{fig:massprojection}
\end{figure}

A recent CMS projection of the expected uncertainty on the top-quark
mass \mt is displayed in Fig.~\ref{fig:massprojection}. The projection
assumes that the upgraded CMS detector will maintain the same physics
performance as the current detector, that the trigger efficiency may
be reduced by up to a factor of three, and that the understanding of
many systematic uncertainties can be improved. The study shows that
kinematic methods to determine \mt will continue to be the most
precise, with rather optimistic uncertainty estimates below
\SI{200}{MeV}, or 0.12\%,  using the full \ac{hllhc} dataset. Methods based on the
reconstruction of exclusive final states, \eg \jpsi from the
hadronization of the $b$ quark, profit most from the increased dataset
sizes, arriving at uncertainties below
\SI{600}{MeV}~\cite{CMS-DP-2016-064}.

Projections have also been performed for \ac{fcnc} searches. As an
example, depending on the assumptions on the systematic uncertainties,
improvements by factors of two to six are expected for the limit on
$\mathcal{B}(t\to Zq)$~\cite{ATL-PHYS-PUB-2016-019}.

The \ttbar charge asymmetry, while reduced at \SI{14}{TeV} compared to
\ac{lhc} Run~1 center-of-mass energies due to the larger fraction of
$gg$-initiated \ttbar events, will profit from new analysis ideas
based the large number of events with boosted top quarks. It is
expected that \Ac can be measured very precisely also at the
\ac{hllhc}~\cite{Agashe:2013hma}.  Also the precision with which the
couplings of the top quarks can be determined in the future is
expected to improve significantly with the large data samples at the
\ac{hllhc}. This includes knowledge about the $Wtb$ vertex, as well as
the electroweak couplings of the top-quarks. The expected precision on
the \ac{sm} $\gamma tt$ ($Ztt$) coupling with the full \ac{hllhc}
dataset is 1.4\% (17\%)~\cite{Agashe:2013hma}, which at the same time
increases the sensitivity for anomalous top-quark couplings.  Also the
sensitivity to the Yukawa couplings of the top-quark to the Higgs
boson will be improved significantly.  An uncertainty of 10\% to 15\%
on the \ttH signal strength is expected for the full \ac{hllhc}
dataset~\cite{CMS:2013xfa,ATL-PHYS-PUB-2014-016}.

The large \ac{hllhc} datasets will also allow exploring
\ttbar production in final states with many jets, very high \ac{met}
and large \ttbar or $t+X$ invariant masses. Such measurements will
improve the modeling of \ac{sm} processes with top quarks in extreme
corners of phase space. This is a prerequisite for measurements of
rare \ac{sm} processes, such as \ttH production, and of \ac{bsm}
searches, \eg for supersymmetry, heavy resonances decaying into top
quarks, and associated production of top quarks and \ac{dm}.  As an
example, searches for heavy-resonance decays $Z'\to\ttbar$ and
$W'\to tb$ are expected to improve, with sensitivities to $Z'$ and
$W'$ masses up to \SI{4}{TeV}~\cite{CMS-DP-2016-064}.

\subsection{Top-Quark Physics at Future Lepton and Hadron Colliders}
\label{sec:future}
Plans for future particle colliders include both lepton (\epem)
colliders and hadron ($pp$) colliders, which are at different stages
of their planning.  Lepton collider projects include the
linear-collider projects \ac{ilc}~\cite{Behnke:2013xla}, with Japan as
the proposed host country, and \ac{clic}~\cite{Aicheler:2012bya} at
CERN, the circular collider projects
\ac{cepc}~\cite{CEPC-SPPCStudyGroup:2015csa,
  CEPC-SPPCStudyGroup:2015esa} in China, and the \epem option of the
CERN Future Circular Collider (FCC-ee). While the \ac{cepc}
center-of-mass energy will be too small for \ttbar pairs to be
produced, the option of the \ac{ilc} with $\sqrts=\SI{0.5}{TeV}$, all
\ac{clic} options, and the FCC-ee are intended to operate above the
\ttbar production threshold of approximately \SI{350}{GeV}. Plans for
future hadron colliders include the High-Energy \ac{lhc} (HE-LHC), an
upgraded version of the \ac{lhc} with very high-field magnets and
center-of-mass energies of up to \SI{33}{TeV}, the Chinese Super
Proton-Proton Collider (SppC) with up to $\sqrts=\SI{70}{TeV}$ as part
of the \ac{cepc} project, and the hadron collider option of the FCC
(FCC-hh) with up to $\sqrts=\SI{100}{TeV}$, all of which are circular
storage rings.

At future \epem colliders, the exact knowledge of the initial state,
in particular the center-of-mass energy, can be exploited to determine
the top-quark mass via a scan of the \ttbar production threshold at
$\sqrt{s}=2\mt$. The beam energies are varied such that \sqrts is
around $2\mt$ and the \ttbar production \xsec is measured as a
function of \sqrts. From the characteristic shape of the \xsec
turn-on, which has been computed including corrections up to either
\ac{n3lo}~\cite{Beneke:2015kwa} or \ac{nnll}~\cite{Hoang:2013uda}
accuracy, \mt can be determined with an expected total uncertainty of
\SI{100}{MeV} or below, without the ambiguities of the kinematic
reconstruction. The top-quark mass can also be determined from the
kinematics of the \ttbar decay products.  The excellent expected
precision of future \epem colliders will also allow for more precise
studies of the \ac{qcd} and electroweak couplings of the top quark in
an \ac{eft} framework. The top-quark's Yukawa coupling is expected to
be known to around 4\%. A recent summary of top-quark physics at
future \epem colliders can be found in~\cite{Vos:2016til}.

At a hadron collider with $\sqrts=\SI{100}{TeV}$ the \xsec for \ttbar
production will increase by a factor of almost 40 compared to the
current 13-\si{TeV} \ac{lhc}. The mass reach of all searches for
\ac{bsm} physics with top quarks will be extended significantly. Also
the role of the top quark will change: the top quark will become a
``light'' quark compared to the available collision energy. This will
have an impact on \ttbar production, which will prefer forward
rapidity, similar to \bbbar production at the \ac{lhc}, and much
higher boosts of the top quarks. In addition proton \ac{pdf} sets will
likely have to include top quark (and $W$ and $Z$ boson)
\acp{pdf}. However at this point it seems very difficult to scale the
\ac{lhc} expectations for systematic uncertainties to very high
luminosities at future hadron colliders.  Reviews discussing the
perspectives of \ac{sm} and \ac{bsm} physics at the FCC-hh can be
found in in~\cite{Mangano:2016jyj,Golling:2016gvc}.

\section{Conclusions}
\label{sec:conclusions}

Studying the physics of the top quark, the heaviest particle of the
\acl{sm} of particle physics, is an important, and very interesting,
task. Since the discovery of the top quark, more than 20~years of
research went into establishing its properties. With more and more
sophisticated analysis methods, the top quark properties are
remarkably well understood. In addition, the top-quark is considered a
possible stepping stone to physics beyond the \ac{sm}, both as part of
a signal and as a major background. All \ac{sm} measurements and
searches for \ac{bsm} physics so far are compatible with the
expectations of the \ac{sm} and contribute significantly to
constraining possible \ac{bsm} physics models. Top-quark physics will
remain important after the upcoming upgrades of the \ac{lhc}
experiments, and experiments at future colliders may take the quest
for the top to the next level.

\section*{Acknowledgment}
It is my pleasure to thank Alison Lister for her valuable comments on
this article.

\clearpage
\small

\end{document}